\documentclass[11pt, oneside]{article}   	
\usepackage{jheppub}
\usepackage{amsmath}
\usepackage{amssymb}
\usepackage{amsfonts}
\usepackage{amsthm}
\usepackage{amsbsy}
\usepackage{array}
\usepackage{mathtools}
\usepackage[dvipsnames]{xcolor}
\usepackage{subcaption}
\usepackage{dsdshorthand}
\usepackage{graphicx}
\usepackage[vcentermath]{youngtab}
\usepackage{hyperref}
\def\bpm{\begin{pmatrix}}
\def\epm{\end{pmatrix}}
\usepackage{tikz}
\usepackage{comment}
\usepackage[compat=1.1.0]{tikz-feynman}

\renewcommand\vol{\mathop{\mathrm{vol}}}

\newcommand\vx{{\vec x}}
\newcommand\vk{{\vec k}}
\newcommand\vy{{\vec y}}
\newcommand\vz{{\vec z}}

\newcommand\tn{{\tilde{n}}}
\newcommand\tj{{\tilde{j}}}
\newcommand{\Op}{\mathcal{O}}

\newcommand{\NO}[1]{(#1)}
\newcommand{\xx}{\vec{x}}

\makeatletter
\def\@fpheader{\ }
\makeatother

\title{Detectors in Nonrelativistic Conformal Field Theories}
\author{Cyuan-Han Chang$^{1}$, Subham Dutta Chowdhury$^{2}$, Ian Moult$^{3}$, and Dam Thanh Son$^{1}$}
\affiliation{$^1$Leinweber Institute for Theoretical Physics, University of Chicago, Chicago, Illinois 60637, USA}
\affiliation{$^2$The Abdus Salam ICTP, Strada Costiera 11, 34151, Trieste, Italy}
\affiliation{$^3$Department of Physics, Yale University, New Haven, CT 06511, USA}
\emailAdd{cchang10@uchicago.edu}
\emailAdd{sdutta\_c@ictp.it}
\emailAdd{ian.moult@yale.edu}
\emailAdd{dtson@uchicago.edu}

\date{}
\abstract{Correlations in asymptotic fluxes provide one of the primary experimental means of studying physical systems.
In the context of relativistic QFT, the formalization of such experiments in terms of detector operators has led to significant progress. Motivated by these successes, in this paper we initiate a study of detector operators in nonrelativistic conformal field theories (NRCFTs). In NRCFTs, fluxes are distributed in both angle and velocity, making the natural detector operator the energy flux differential in velocity, $\mathcal{E}_v(\hat n)$. We provide non-perturbative definitions of these operators in NRCFTs, and study their symmetries and algebra.
We calculate one-point functions of these detector observables in a variety of few and many body states of representative NRCFTs, including free fermions, as well as fermions at unitarity. 
In interacting three-body states, they exhibit a remarkably rich behavior reflective of the Efimov three-body wave-function, including the formation of an enhanced ``jet" structure due to the presence of resonant interactions. 
We use these detectors to derive a nonrelativistic generalization of the Hofman-Maldacena conformal collider bounds by arguing for positivity of $\mathcal{E}_v(\hat n)$, and imposing this in spin 1 states. 
We outline a number of future directions for the study of detectors in NRCFTs, and discuss how they can be measured in a variety of experimental platforms from cold atoms to low energy neutron experiments.}

\begin{document}

\maketitle
\pagenumbering{roman}
\setcounter{page}{2}
\pagenumbering{arabic}
\setcounter{page}{1}

\section{Introduction} 

In the context of relativistic quantum field theories (QFTs), in particular conformal field theories (CFTs), there has been tremendous recent progress in formalizing idealized theoretical abstractions of collider physics, often referred to as ``conformal colliders" \cite{Hofman:2008ar}. Key to this is the notion of detector operators, which implement the measurement of asymptotic fluxes in terms of QFT operators. 
The most famous example of a detector operator is the energy flow operator, defined in the seminal works of Sterman \cite{Sterman:1975xv} and Korchemsky, Oderda, and Sterman \cite{Korchemsky:1997sy}
\begin{equation}\label{eq:ANEC_op}
\mathcal{E}(\hat n) = \lim_{r\to \infty}  \int\limits_0^\infty\! dt\, r^2 n_i T_{0i}(t,r \hat n)\,.
\end{equation}
In the case of a CFT, where all radiation goes to null infinity, this operator can be expressed in terms of the average null energy (ANE) operator 
\begin{equation}
\mathcal{E}(\hat n)=\lim_{v\to \infty} 2 v^2 \!\int\limits_{-\infty}^\infty\! du\, T_{uu}(u,v,\vec 0)\,,
\end{equation}
where $u$ and $v$ denote light-cone coordinates in the null directions $(1,\pm \hat n)$ defined with respect to $\hat n$. In a CFT, the ANE can also be put on a null-sheet at a generic position in the bulk of spacetime by performing a null inversion \cite{Hofman:2008ar}, showing that this limit is well defined. Such a construction has been generalized to define a wide range of operators localized along null rays, referred to as ``light-ray operators" \cite{Kravchuk:2018htv}, which formalize the notion of the measurement of different fluxes.

\begin{figure}
\begin{center}
\includegraphics[width=0.35\linewidth]{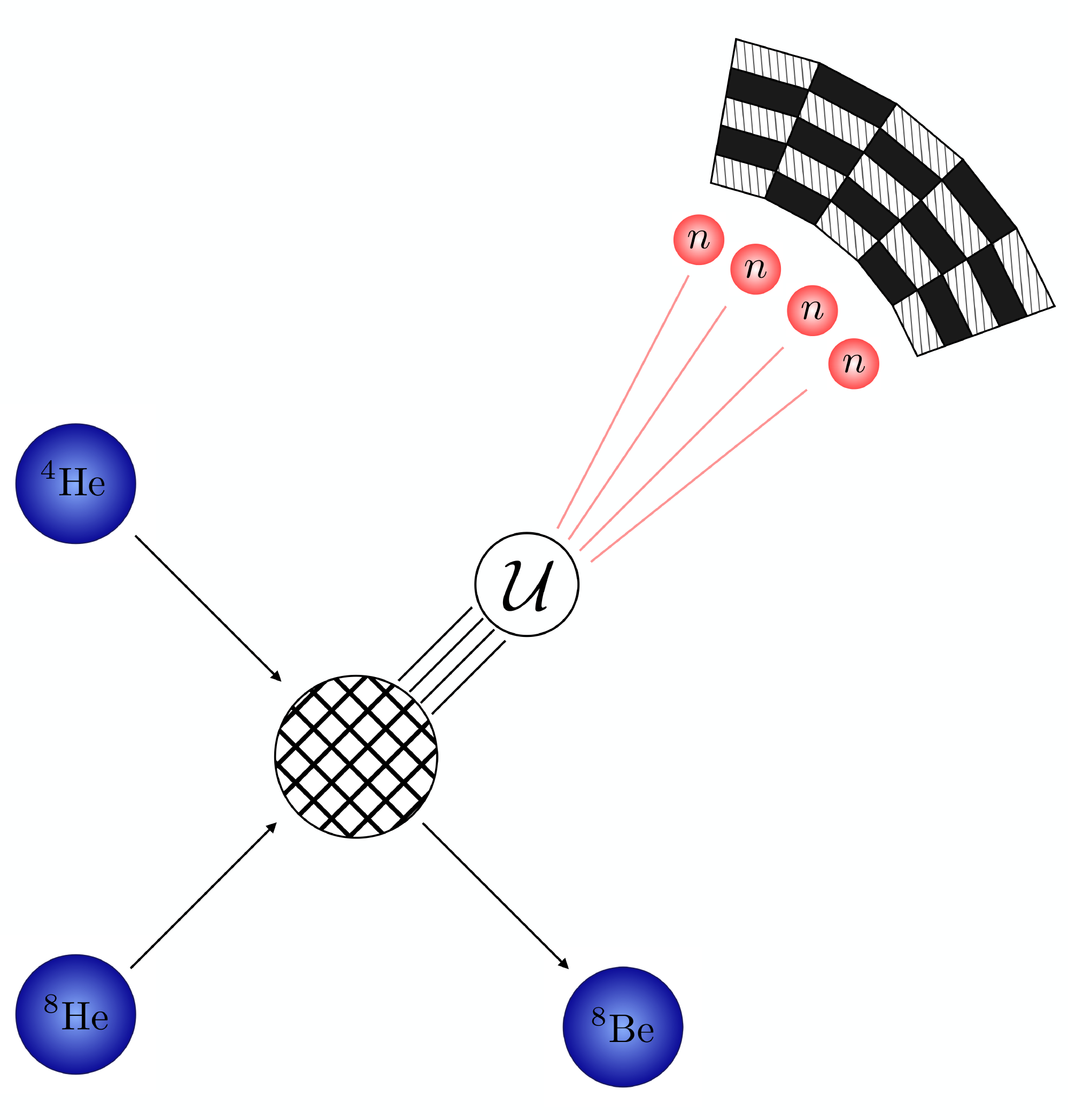} \qquad
\includegraphics[width=0.5\linewidth]{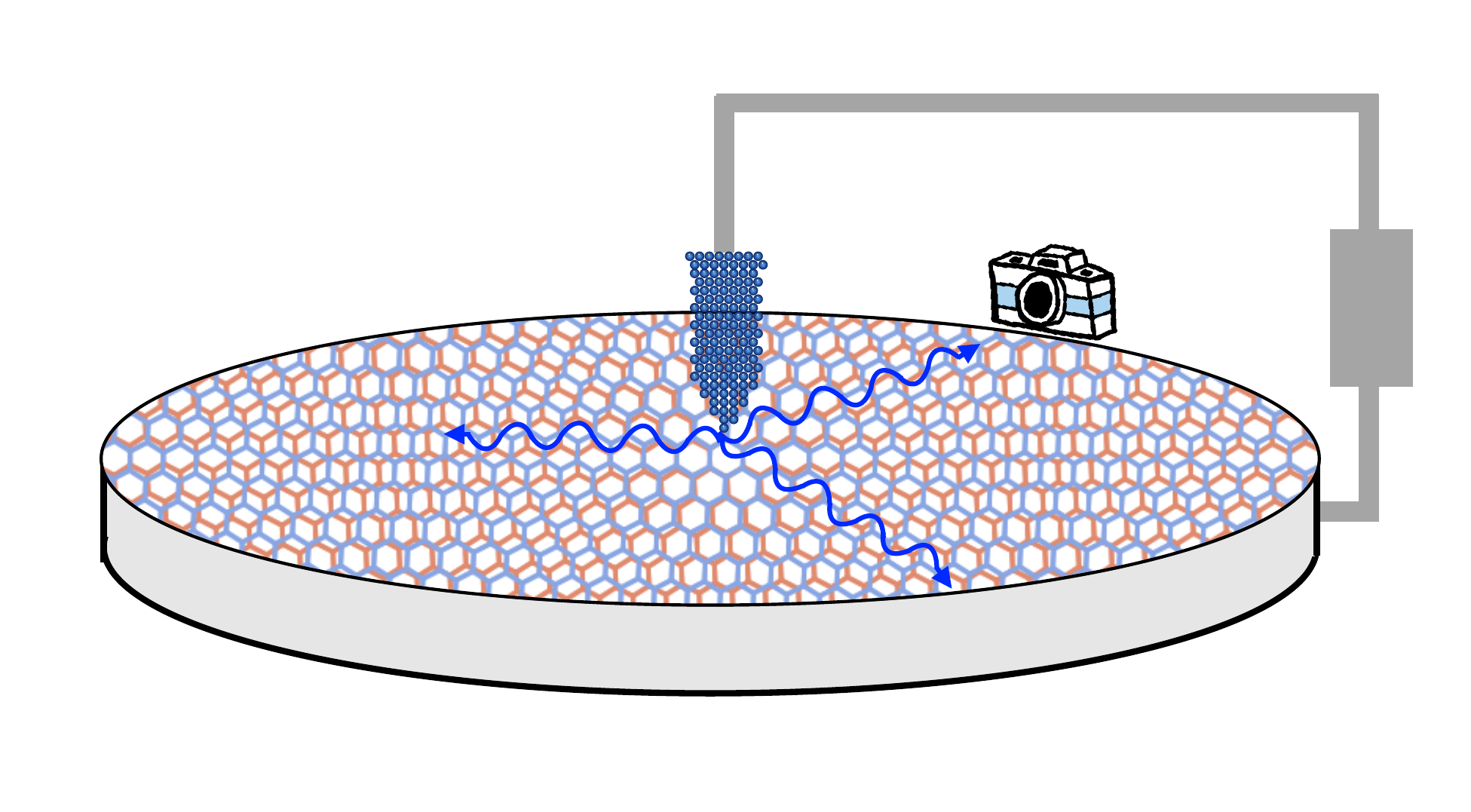}
\end{center}
\caption{a) To the left, analogous to the study of jet substructure at the Large Hadron Collider, ``jets" of slow neutrons can be produced in a variety of nuclear reactions. Here the production proceeds through a factorization channel of an ``un-nucleus,'' a state of definite quantum numbers in an NRCFT. b)  To the right, a schematic depiction of a tabletop experiment where nonrelativistic excitations are created at a point, propagate outwards, and are detected. }
\label{fig:tabletop}
\end{figure}

While measurements of asymptotic fluxes had long been considered in collider physics, an operator formulation of asymptotic measurements provides a direct connection between correlation functions of local operators and asymptotic measurements. This has had a significant impact in a number of distinct directions. First, it enabled an extension of operator product expansion (OPE) techniques to the study of asymptotic fluxes \cite{Hofman:2008ar,Kologlu:2019mfz, Chang:2020qpj}, which have provided new observables for the experimental collider physics program \cite{Moult:2025nhu}. Second, it provided a non-perturbative definition of collider physics, leading to the first non-perturbative calculation of the energy correlator in planar $\mathcal{N}=4$ super Yang-Mills \cite{Dempsey:2025yiv}. Finally, conformal collider physics observables have been used to place interesting bounds on the space of consistent CFTs \cite{Hofman:2008ar,Cordova:2017dhq,Chowdhury:2017vel,Cordova:2017zej,Meltzer:2017rtf,Meltzer:2018tnm,Afkhami-Jeddi:2018own,Delacretaz:2018cfk, Manenti:2019kbl,Belin:2019mnx,Mecaj:2025ecl, Mecaj:2026kji, Belin:2026wkc}, complementing the standard Euclidean conformal bootstrap. Additionally, beyond the specific use in ``conformal collider physics,'' light-ray operators have also played an important role in a variety of other contexts, including providing the analytic continuation in spin of CFT data \cite{Caron-Huot:2017vep,Kravchuk:2018htv}.

The specific relation between detector operators and light-ray operators is highly tied to the causal structure of CFTs. However, the general idea of collider physics, and the measurement of asymptotic fluxes is much more general. As a concrete physical example, consider the production of low energy neutrons in a nuclear collision, as shown in figure \ref{fig:tabletop}a. In ref.\ \cite{Hammer:2021zxb} it was shown that, since low-energy neutrons are described by a nonrelativistic conformal field theory (NRCFT) \cite{Nishida:2007pj,Boisvert:2025hex}, the cross section in the threshold region has a scaling behavior computable in terms of the scaling dimension of particular operators in the NRCFT.  
The idea is that the reaction can be factorized into a primary reaction in which an ``unnucleus" (described by a primary operator of the NRCFT) is created, and the decay of the unnucleus into neutrons.

Beyond the scaling of the total cross section, one might like to understand how the neutrons are distributed in terms of both angles and velocities. This is precisely equivalent to the study of jet substructure in relativistic particle colliders \cite{Larkoski:2017jix}. Indeed, calculations of energy correlators at the LHC rely on an exactly analogous factorization \cite{Dixon:2019uzg,Lee:2022ige}, however, instead of unnucleus states, energy flow operators are evaluated in boosted quark and gluon states. This has enabled the calculation, and measurement, of two, three and four-point correlators \cite{Komiske:2022enw,Chen:2019bpb,Chicherin:2024ifn}. Other 
experimentally realizable examples of NRCFTs include ultracold atoms at the BCS-BEC crossover, or, potentially, dispersive anyons of the fractional quantum Hall insulating states of moir\'e materials~\cite{Shi:2024cgb}.  A schematic experiment involving the latter is shown in figure \ref{fig:tabletop}b.

\begin{figure}
\includegraphics[width=0.5\linewidth]{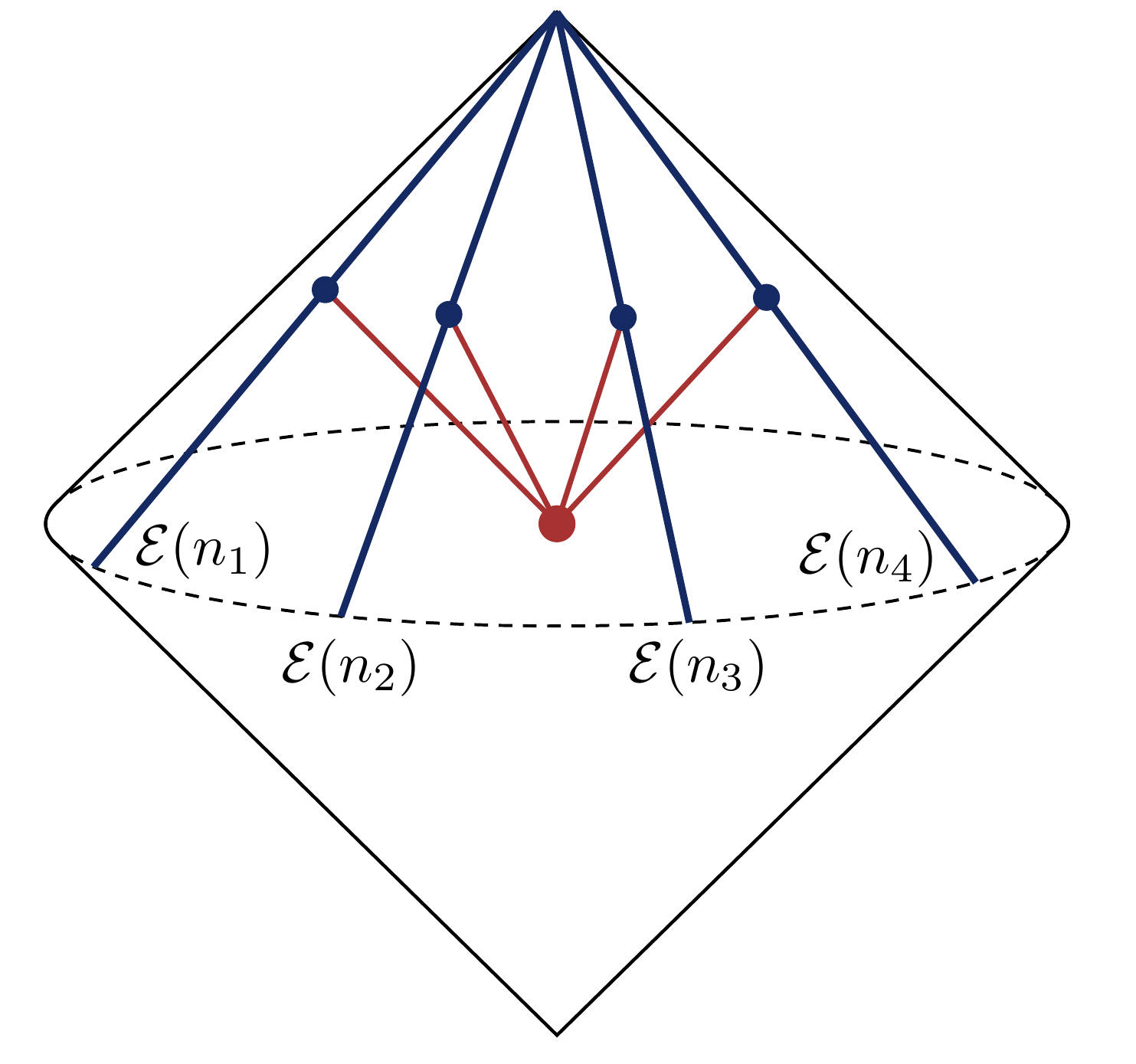}
\raisebox{1.5cm}{
\includegraphics[width=0.5\linewidth]{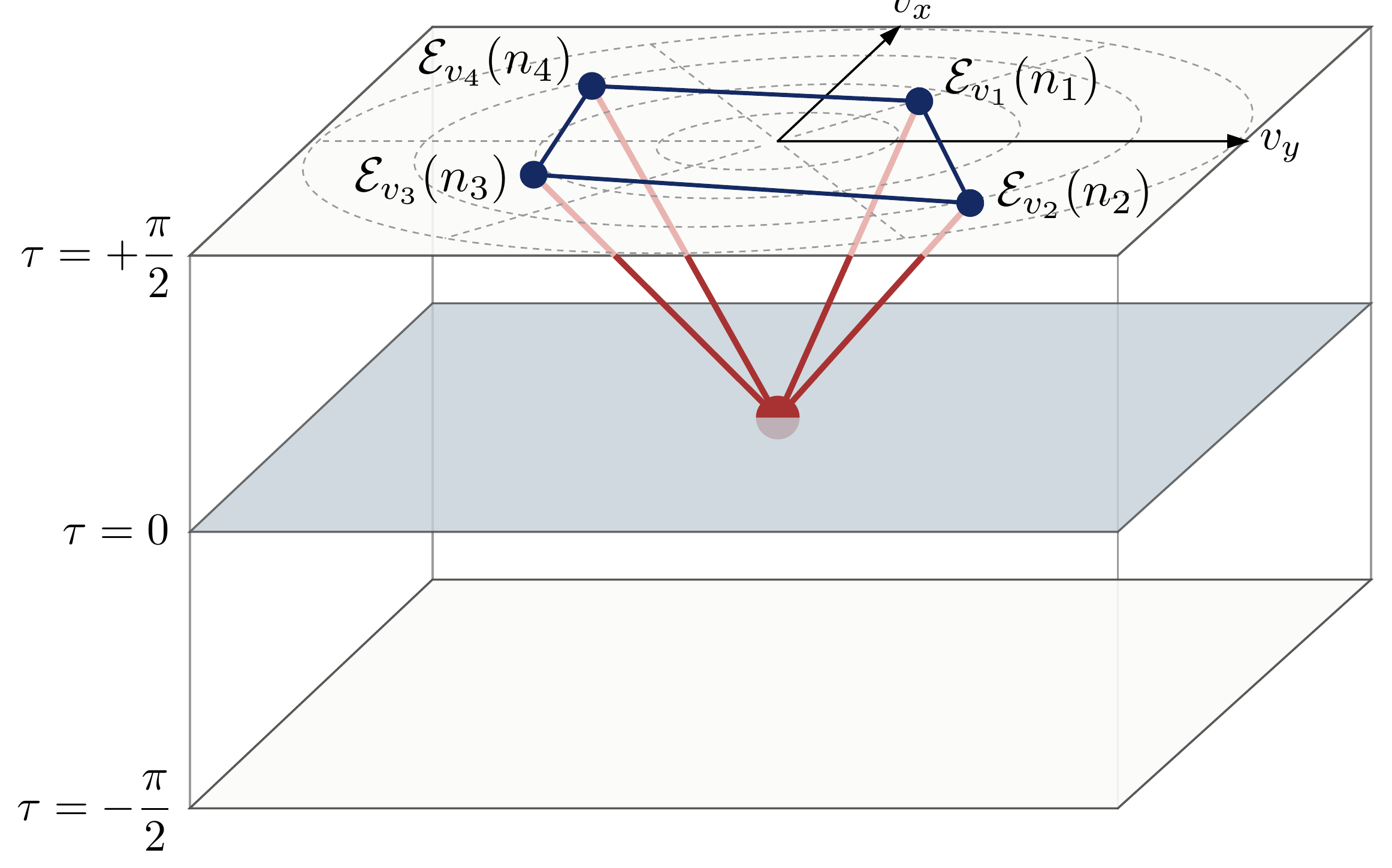}}
\caption{In a relativistic theory, radiation is detected by light-ray operators at future null infinity (Left). In a nonrelativistic theory, detector operators are local in the \emph{velocity space} at future infinity (Right).}
\label{fig:intro_detectors}
\end{figure}

Motivated by the many successes achieved by formalizing detectors in a field theoretic language in relativistic QFTs, in this paper we initiate an exploration of detector operators in nonrelativistic QFTs. For simplicity, we restrict ourselves to the case of nonrelativistic CFTs, and, following ref.\ \cite{Hofman:2008ar}, we call this formalism ``nonrelativistic conformal collider physics." However, we emphasize that the general paradigm is applicable to nonrelativistic QFTs more generally. The specific application of detector operators in the NRCFT of fermions at unitarity to point production of neutrons  was explored in ref.\ \cite{Chang:2026wrl}. In this paper, we provide a more general discussion of detector operators in NRCFTs.

We argue that the natural observables in NRCFTs are fluxes differential in both angle and velocity. We focus on the specific case of energy flux in NRCFTs in $3+1$ dimensions, and show that the energy flux in a particular direction $\hat n$ at asymptotic infinity and  graded by velocity, which we denote $\mathcal{E}_v({\hat n})$, can be expressed in terms of either the number density $n$ or current $\vec j$ as
\begin{align}\label{eq:Ev_definition}
\mathcal{E}_v({\hat n}) &=\frac{v}{2}\lim_{r \rightarrow \infty} r^3 n\left(\frac{r}{v}, r {\hat n} \right) = \frac{1}{2}\lim_{r \rightarrow \infty} r^3 {\hat n}\cdot \vec j\left(\frac{r}{v}, r {\hat n} \right). 
\end{align}
Similar velocity-resolved detectors were considered in relativistic theories to incorporate mass effects \cite{Mateu:2012nk}. 
The complete energy flux in a given direction can be obtained by integrating over the velocity
\begin{equation}
\cE(\hat n)=\int\! dv~ \cE_{v}(\hat n)\,. 
\end{equation}
We will see that, in NRCFT, velocity-resolved detectors, such as $\cE_{v}(\hat n)$, exhibit nicer symmetry properties than the velocity-unresolved ones.  An interesting feature of our detectors in NRCFTs is that, despite being local operators, they have precisely the properties of detector operators in relativistic theories. The fact that massless operators in NRCFTs have interesting analogies with light-ray operators was emphasized in ref.\ \cite{Boisvert:2025hex}. Here we emphasize that this is not only a mathematical analogy, but these are the genuine detectors in NRCFTs, measurable in experiments. 

To provide a non-perturbative definition of detector operators in NRCFTs without recourse to a limiting procedure, we study the geometry of asymptotic observables in NRCFTs. We show that this is most conveniently done by using the harmonic-trap geometry-the analog of the Lorentzian cylinder for relativistic CFTs. In the case of NRCFTs, timelike infinity corresponds to a spatial slice, and detector operators have a particularly elegant interpretation: they are local operators on this slice evaluated with a ``position" equal to their velocity
\begin{equation}
\cE_v(\hat n) = \frac{v^4}{2\omega^3}\tilde{n}\bigl(\tau=\tfrac{\pi}{2\omega},\vec y=\tfrac{v\hat n}{\omega}\bigr),
\end{equation}
where $\omega$ is the frequency of the harmonic trap. This is illustrated in figure \ref{fig:intro_detectors}, where it is compared with the case of a relativistic CFT.  This space exhibits the symmetries of a scale-, translation-, and rotation-invariant theory, but now in velocity space. We study the symmetries in this space. We believe that the realization of detector operators as local operators in velocity space provides a simplified setting to study a nonrelativistic analog of ``celestial holography."

We compute one-point functions of the detector operator, $\cE_v(\hat n) $ in a variety of two and three-body states, both in the free theory, and (non-perturbatively) in the interacting theory of fermions at unitarity. We find that  one-point functions of $\cE_v(\hat n) $ exhibit a remarkably rich structure, reflecting the complexity of the underlying wave-function. In particular, they exhibit singularities corresponding to the formation of ``jets" in kinematic configurations where two particles have almost the same velocity.

Additionally, we investigate a generalization of the Hofman-Maldacena ``conformal collider bounds" \cite{Hofman:2008ar} to NRCFTs. We use the positivity of our velocity dependent operator, $\langle \psi | \mathcal{E}_v(n)  | \psi \rangle \geq 0$, in spin 1 states to derive ``nonrelativistic conformal collider bounds,'' and we  show that they are non-trivially satisfied for three-body states in fermions at unitarity. These correspond to bounds on integrals of three-point functions in NRCFTs over the conformal cross-ratio.

An outline of this paper is as follows. In section \ref{sec:review} we provide a brief review of NRCFTs, emphasizing several aspects that will be important for understanding detector operators in these theories. In section \ref{sec:detectors} we introduce detector operators in NRCFTs and study their symmetries and algebra.  In section \ref{sec:calcs} we calculate one-point correlation functions of energy detectors in a variety of states within a variety of theories, including the two-, three-, and large-charge states for both free fermions and fermions at unitarity. We also propose a generalized Hofman-Maldacena collider bound in NRCFTs. In section \ref{sec:celestial}, we develop a ``celestial" approach to the calculation of detector operators in NRCFTs, showing how they can be obtained from a ``wave-function" living on future infinity. Finally, we conclude in section \ref{sec:conc} and discuss a number of directions for future study.

\section{Review of NRCFTs}\label{sec:review}

In this section, we provide a brief review of NRCFTs, highlighting only the aspects of NRCFTs that are important for understanding detector operators in these theories. For detailed discussion we refer the reader to ref.\ \cite{Nishida:2007pj}, to the recent discussion in ref.\ \cite{Boisvert:2025hex}, and to the review \cite{Baiguera:2023fus}. For simplicity, we will restrict our review to NRCFTs with only particles of the same mass, so for the rest of the discussion we set $m=1$.

\subsection{Symmetry Algebra}

We first review the symmetry algebra of NRCFTs, referred to as the Schrödinger algebra \cite{Hagen:1972pd,Niederer:1972zz}. It consists of rotation $M_{ij}$, spatial translation $P_i$, time translation $H$, Galilean boost $K_i$, dilatation $D$, special conformal transformation $C$, and particle number $N$ which is central. The non-vanishing commutators are
\begin{align}\label{eq:generators_algebra}
[M_{ij},\, M_{kl}] &= i(\delta_{ik}M_{jl}-\delta_{jk}M_{il}
+\delta_{il}M_{kj}-\delta_{jl}M_{ki}),& \nn \\
[M_{ij},\, P_k] &= i(\delta_{ik}P_j-\delta_{jk}P_i),&
[M_{ij},\, K_k] &= i(\delta_{ik}K_j-\delta_{jk}K_i), \nn \\
[K_i,\, P_j] &= i\delta_{ij}N,&
[C,\, P_i] &= i K_i, \nn \\
[D,\, P_i] &= iP_i,&
[D,\, K_i] &= -iK_i, \nn \\
[D,\, C] &= -2iC,&
[H,\, K_i] &= -iP_i,\nn \\
[H,\, C] &= -iD, & [D,\, H]&= 2 i H.
\end{align}

For the class of free and interacting nonrelativistic fermionic theories
considered in this work,  the generators of the Schrödinger algebra (except $H$) can be expressed as spatial integrals involving the particle number density $n(\vx)$ and momentum density $j_i(\vx)$, and the density operators form a closed algebra
\begin{align}
  & N = \int\!d^3\vx\, n(\vx), \qquad
  P_i = \int\!d^3\vx\, j_i(\vx), \qquad
  M_{ij} = \int\! d^3\vx\, \bigl(x_i j_j(\vx)-x_j j_i(\vx)\bigr),\nn \\
  & K_i = \int\!d^3\vx\, x_i n(\vx), \qquad
  C = \int\!d^3\vx\, \frac{x^2}{2} n(\vx), \qquad
  D = \int\!d^3\vx\, x_i j_i(\vx),\nn \\
   &[n(\vx),\, n(\vy)] = 0, \qquad
    [n(\vx),\, j_i(\vy)] = -i n(\vy) \partial^x_i \delta^{(3)}(\vx-\vy), \nn \\
  & [j_i(\vx),\, j_j(\vy)] = -i \left( j_j(\vx)\partial^x_i + j_i(\vy)\partial^x_j \right)
     \delta^{(3)}(\vx-\vy). \label{eq:nj_algebra}
\end{align}
The algebra formed by the momentum density operator $j_i(\vx)$ is that of spatial diffeomorphism \cite{Dzyaloshinskii:1980PoissonBrackets}.
A major difference between the ``nonrelativistic conformal collider physics" and the conventional (relativistic) ``conformal collider physics" is the presence, in the former, of a conserved particle number. For states produced from local operators, the particle number is conserved as the state evolves towards the detector at infinity. This is quite distinct from, e.g., QCD, where collisions produce high multiplicity final states. This feature will greatly simplify the calculation of detector correlators and allow exact results in interacting theories.

In NRCFTs, a local operator $\cO$ is characterized by its ``particle number" $N_{\cO}$, scaling dimension $\De_{\cO}$, and angular momentum $l$
\begin{equation}\label{eq:generators_origin_1}
[N,\, \cO(0)] = N_{\cO} \cO(0),\quad [D,\, \cO(0)]=i\De_{\cO}\cO(0),\quad [M_{ij},\, \cO(0)] = S^{l}_{ij}\.\cO(0), 
\end{equation}
where $S^l_{ij}$ are the matrices representing $M_{ij}$ in the spin-$l$ representation of the $\mathfrak{so}(3)$ algebra. In eq.\ (\ref{eq:generators_origin_1}) we adopt the convention $N_\psi=-1$, $N_{\psi^{\dagger}}=1$: $\psi$ annihilates a particle, while $\psi^\dagger$ creates one. Note that both $n$ and $\vec j$ have particle number zero. 

Let us also review the notion of primary operators. By definition, they are annihilated by the Galilean boost generators
$K_i$ and the special conformal generator $C$ at the origin,
\begin{equation}\label{eq:generators_origin_2}
[K_i,\, \cO(0)]=[C,\, \cO(0)]=0.
\end{equation}

The transformation laws at an arbitrary spacetime point follow from translation in time and space generated by $H$ and $P_i$,
\begin{equation}\label{eq:generators_3}
[H,\, \cO(t,\vec x)] = -i\ptl_t \cO(t,\vec x),\quad [P_i,\, \cO(t,\vec x)] = i\ptl_i \cO(t,\vec x),
\end{equation}
which gives $\cO(t,\vx) =e^{iHt-i\vec P\cdot\vx}\cO(0)e^{-iHt+i\vec P\cdot\vx}$. Together with the commutation relations of the Schrödinger algebra, we find
\be\label{eq:generators_general}
[D,\, \cO(t,\vx)] &=i\left(2t\partial_t+x_i\partial_{i}+\Delta_{\cO}\right)\cO(t,\vx),\nonumber\\
[M_{ij},\, \cO(t,\vx)]
   &=i(x_i\ptl_j-x_j\ptl_i)\cO(t,\vec x)+ S^l_{ij}\.\cO(t,\vec x),\nonumber\\
[K_i,\, \cO(t,\vx)]
   &=\left(-it\,\partial_i+x_iN_{\cO}\right)\cO(t,\vx),\nonumber\\
[C,\, \cO(t,\vx)]
   &=-i\left(t^2\partial_t+t\,x_i\partial_{i}+t\Delta_{\cO}\right)\cO(t,\vx)
   +\frac{x^2}{2}N_{\cO}\,\cO(t,\vx).
\ee

We note that while the density operator $n$ is a primary operator, the current operator $\vec j$ is not. It is an example of ``alien operators" which are neither primaries nor descendants \cite{Bekaert:2011qd, Golkar:2014mwa}. For a recent detailed discussion of alien operators see ref.\ \cite{Boisvert:2026mgw}. In particular, the transformation laws of $\vec j$ under Galilean boosts and special conformal transformations are different from the ones in eq.\ \eqref{eq:generators_general}, and they are given by
\begin{align}\label{K_C_comm_ji}
[K_i,\, j_j (t,\vx)]&= - it \partial_i j_j(t,\vx)+ i \delta_{ij} n (t,\vx), \nonumber\\
[C,\, \vec j(t,\vx)]&=-i \left( t^2 \partial_t + t x_i \partial_i+ t \Delta_j\right) \vec j(t,\vx) + i \vx n(t,\vx),
\end{align}
where $\De_j = \De_n + 1 =4$. This can be obtained using the current algebra \eqref{eq:nj_algebra}.

\subsection{Correlators}

Much like in CFTs, the structure of low point correlators in NRCFT is largely fixed by symmetry \cite{Nishida:2007pj,Goldberger:2014hca}, albeit with several important differences that will play a key role in our calculation of detector correlators, so we review them here. 

Two-point functions are completely fixed up to normalization, and are given by
\begin{equation}\label{eq:2pt_convention}
\langle \Omega | \mathcal{O}_1(x_1) \mathcal{O}_2 (x_2) | \Omega \rangle = \delta_{\Delta_{\mathcal{O}_1} \Delta_{\mathcal{O}_2}} \frac{C}{(t_{12}-i \epsilon_{12})^{\Delta_1}} e^{-\frac{i N_{\mathcal{O}_1}\vec x^2_{12}}{2t_{12}}}\,,
\end{equation}
where particle number conservation requires $N_{\cO_1}+N_{\cO_2}=0$.

However, unlike relativistic CFTs, where three-point functions are fixed up to constants, the NRCFT three-point functions  \cite{Volovich:2009yh}
\begin{equation}\label{eq:3ptfunc_general}
\langle \Omega | \mathcal{O}_1(x_1) \mathcal{O}_2(x_2)  \mathcal{O}_3 (x_3) | \Omega \rangle =F(v_{123}) \frac{e^{-\frac{iN_{\mathcal{O}_1}\vec x^2_{13}}{2t_{13}}-\frac{iN_{\mathcal{O}_2}\vec x^2_{23}}{2t_{23}}}}{t_{12}^{(\Delta_1+\Delta_2-\Delta_3)/2}  t_{23}^{(\Delta_2+\Delta_3-\Delta_1)/2}  t_{13}^{(\Delta_1+\Delta_3-\Delta_2)/2}   }\,,
\end{equation}
depend on a function of the cross ratio
\begin{align}
v_{ijk}=\frac{1}{2}\left(\frac{\vec x_{jk}^2}{t_{jk}}+\frac{\vec x_{ij}^2}{t_{ij}}-\frac{\vec x_{ik}^2}{t_{ik}}\right).
\end{align}
Particle number conservation requires $N_{\cO_1}+N_{\cO_2}+N_{\cO_3}=0$.
This extra dependence will play an important role in our study of one-point functions of detector operators. As famously shown in ref.\ \cite{Hofman:2008ar}, one-point functions of detector operators in CFTs are related to three-point functions, and in CFTs are therefore fixed up to several constants for different tensor structures. The conformal collider bounds \cite{Hofman:2008ar} then place constraints on these constants. In an NRCFT, we will see that the one-point function of $\mathcal{E}_v$ is again related to a three-point function. Its non-trivial dependence on velocity arises precisely from the presence of a cross-ratio in the three-point correlator. Our nonrelativistic generalizations of the Hofman-Maldacena bounds discussed in section \ref{sec:NR_bounds} will place bounds on $F(v)$.

\subsection{Harmonic-Trap Geometry}\label{sec:trap_review}

Detector operators are designed to implement measurements at ``asymptotic infinity," and as such rely on an understanding of the structure of asymptotic infinity. To understand the structure of detector operators in NRCFT, it will be convenient to work in coordinates that avoid a limiting procedure. This will also allow us to review the state-operator correspondence in NRCFTs.

The nonrelativistic state-operator correspondence \cite{Nishida:2007pj, Goldberger:2014hca, Boisvert:2025hex} states that operators of the NRCFT are in one-to-one map with states in the harmonic-trap geometry. The geometry can be thought of as the nonrelativistic version of the Lorentzian cylinder \cite{Luscher:1974ez}. For a detailed discussion from a modern perspective, see ref.\ \cite{Boisvert:2025hex}. It is obtained by the coordinate transformation\footnote{Conventionally, the harmonic-trap geometry comes with a parameter $\omega$-the frequency of the harmonic trap-so that $\omega t = \tan(\omega \tau), \vec x = \vec y \sec(\omega \tau)$. In this paper we always set the trap frequency to unity, preserving the letter
$\omega$ for the energy of the state.}
\begin{equation}\label{coord_ht_flat}
  \begin{cases}  t = \tan\tau \\  \vx  = \vy \sec \tau \end{cases}  \leftrightarrow \quad
  \begin{cases} \tau = \arctan t \\ \vy = \displaystyle{\frac{\vx}{\sqrt{1+t^2}}}\end{cases}
\end{equation}
and depicted in figure \ref{fig:intro_detectors}.

The harmonic-trap geometry makes clear the structure of asymptotic infinity, similar to a Penrose diagram in the relativistic case. Future infinity is a spacelike slice of dimension $d$ at $\tau=\pi/2$. Our detector operators will be defined to live on this spacelike slice.

In the harmonic-trap geometry, the time evolution in $\tau$ is generated by $H+C$. Under this mapping, a primary operator $\cO$ transforms as \cite{Goldberger:2014hca}
\begin{align}
\cO(t,\vx)
&= (\cos \tau)^{\Delta_\cO}
\exp\!\left(-\frac{i}{2} N_\cO\, \vy^{\,2} \tan \tau \right)
\tilde{\cO}(\tau,\vy),
\end{align}
where $\tilde{\cO}$ denotes the operator expressed in harmonic-trap coordinates
\be\label{eq:O_HT_def}
\tilde{\cO}(\tau, \vy) = e^{i (H+C)\tau} \cO(0,\vy) e^{-i (H+C)\tau},
\ee
and $\cO(0,\vy)= \tilde{\cO}(0,\vy)$. Combining this with eq.\ \eqref{K_C_comm_ji}, we find that the number density $n$ and current $\vec j$ transform as (assuming $d=3$)
\begin{align}\label{En_osc}
j_i(t,\vx)
&= (\cos\tau)^{4}
\left[
\tilde{j}_i(\tau,\vy)
+ \tan\tau \, y_i \tilde{n}(\tau,\vy)
\right], \qquad
n(t,\vx)
= (\cos\tau)^{3}\,\tilde{n}(\tau,\vy).
\end{align}

\subsection{Wave-functions}\label{sec:wave-function}

Due to the conservation of particle number in NRCFTs, we can decompose the Hilbert space into sectors characterized by particle number. In this paper we will primarily focus on NRCFTs of fermions. In this case, we can decompose into sectors characterized by a fixed number of spin-up and spin-down particles \cite{Chowdhury:2023oas},
\begin{equation}\label{sub_hilbert_space}
  \mathcal H = \mathcal H_0 \oplus \mathcal H_{1,0} \oplus \mathcal H_{0,1} \oplus
  \mathcal H_{1,1} \oplus \mathcal H_{2,1} \oplus \cdots,
\end{equation}
where $\mathcal H_{m,n}$ is a sub-Hilbert space of states with $m$ spin-up and $n$ spin-down fermions. We can characterize the states in each Hilbert space by a wave-function
\begin{equation}\label{wfnsdef}
  \Psi(\{\vx_i; \vy_j\}) \equiv \Psi(\vx_1, \vx_2, \ldots, \vx_m; \vy_1, \vy_2, \ldots, \vy_n)\,.
\end{equation}
Fermi-Dirac statistics requires that the many-body wave-function $\Psi$ be totally antisymmetric under the exchange of any pair of spatial coordinates among the spin-up ($x$'s) or spin-down ($y$'s) particles. This antisymmetry is a direct consequence of the Pauli exclusion principle, which forbids two identical fermions from occupying the same quantum state simultaneously.

For the NRCFT of free-fermions, the wave-function describing $m$ spin-up and $n$ spin-down fermions factorizes and is given by the Slater determinant of the individual single-particle wave-functions,
\begin{align}\label{free_wvfn_p_q}
    \Psi(\vx_1,\ldots,\vx_m;\,\vy_1,\ldots,\vy_n) = \frac{1}{m!\,n!}
    \begin{vmatrix}
        \phi_1(\vx_1) & \cdots & \phi_1(\vx_m) \\
        \vdots & \ddots & \vdots \\
        \phi_m(\vx_1) & \cdots & \phi_m(\vx_m)
    \end{vmatrix}
    \times
    \begin{vmatrix}
        \chi_1(\vy_1) & \cdots & \chi_1(\vy_n) \\
        \vdots & \ddots & \vdots \\
        \chi_n(\vy_1) & \cdots & \chi_n(\vy_n)
    \end{vmatrix},
\end{align}
where $\phi, \chi$ are the single particle wave-functions and the determinant structure automatically encodes the required antisymmetry.

In this paper we will also consider the case of fermions at unitarity. This can be formulated as quantum mechanics with the Hamiltonian
\begin{align}
H=-\frac{1}{2}\sum\limits_{i=1}^{m}\frac{\partial^2}{\partial \vec x_i^{\,2}}-\frac{1}{2}\sum\limits_{i=1}^{n}\frac{\partial^2}{\partial \vec y_i^{\,2}}\,,
\end{align}
along with the Bethe-Peierls boundary condition
\begin{align}\label{eq:Bethe_Peierls}
\Psi(\vec x_i, \vec y_j) \to \frac{C_0\left( \frac{\vec x_i +\vec y_j}{2} \right) }{|\vec x_i -\vec y_j|} + |\vec x_i -\vec y_j|  C_1\left( \frac{\vec x_i +\vec y_j}{2} \right)+\cdots\,,
\end{align}
namely that the $s$-wave component of wave-function is expandable in odd powers of $|\vec x_i -\vec y_j|$ starting from $1/|\vec x_i -\vec y_j|$. This defines a consistent quantum mechanics.

Using wave-functions, we can define operators via their corresponding states. In the free-theory setting, the local operators $\mathcal{O}_n(x)$ annihilating $n$-particle states take a particularly transparent form: they are constructed as degree-$n$ monomials in the fermionic field $\psi_\sigma(x)$, where $\sigma = \uparrow, \downarrow$ denotes the spin projection, supplemented by appropriate spatial and time derivatives, essential to circumvent the Pauli exclusion principle. However, this construction also applies to the case of interacting theories, such as fermions at unitarity.  For example, for two-body operators in the theory of fermions at unitarity, we can write 
\begin{align}
|\Psi \rangle =\int d^3\vec x d^3\vec y\, \Psi(\vec x, \vec y) \psi_\downarrow^\dagger (\vec y)   \psi_\uparrow^\dagger (\vec x) |0\rangle\,.
\end{align}
We can then define a charge-2 operator by its matrix elements
\begin{align}
\langle 0| \cO_2(\vec X) |\Psi \rangle = \lim_{\vec x \to \vec X, \vec y \to \vec X} | \vec x-\vec y| \Psi(\vec x, \vec y)\,.
\end{align}
This allows us to compute the spectrum of two-body operators by solving two-body Schrödinger equations. Similarly, the spectrum of three-body operators can be obtained by solving the three-body Schrödinger equation with Bethe-Peierls boundary conditions \cite{Efimov:1973awb,Efimov:1971zz,Efimov:1970zz}. The explicit form of the wave-functions can be found in ref.\ \cite{Chowdhury:2023oas}. We will review the explicit form of the two and three body wave-functions in sections \ref{sec:unitarity_2body} and \ref{sec:unitarity_3body}.  These  wave-functions will allow us to compute detector correlators exactly in two and three-body states in the theory of fermions at unitarity.

\subsection{NRCFTs at Large Charge}\label{sec:large_charge}

Beyond the three-body sector, there do not exist analytic results for wave-functions in interacting theories (see, e.g., refs.\ \cite{Platter:2004he,Hammer:2006ct} for explorations of the four-body sector). One way of accessing higher body states is to use a large-charge expansion. This has been explored extensively in the case of relativistic theories \cite{Hellerman:2015nra,Alvarez-Gaume:2016vff,Monin:2016jmo,Loukas:2016ckj,Hellerman:2017efx,Hellerman:2017veg,Banerjee:2017fcx,Loukas:2017lof,Hellerman:2017sur}. Using a nonrelativistic superfluid effective field theory (EFT) \cite{Son:2005rv}, it was generalized to NRCFTs to perform a number of calculations at large charge
\cite{Kravec:2018qnu,Kravec:2019djc,Hellerman:2023myh,Hellerman:2021qzz,Hellerman:2020eff,Beane:2025tum,Beane:2024kld}. Here we briefly review this approach to the study of large-charge operators in NRCFTs.

We focus on the specific case of fermions at unitarity, which is described by a superfluid EFT in the large-charge limit \cite{Beane:2024kld, Beane:2025tum}. The EFT has a Nambu-Goldstone boson $\th(x)$ for the broken $U(1)$ particle number symmetry. In Euclidean signature, the leading order Lagrangian of the EFT is given by \cite{Son:2005rv} (setting $M=1$)
\be
\cL =  -\frac{2^{\frac{5}{2}}}{15\pi^2\xi^{\frac{3}{2}}} X^{\frac{5}{2}}, \qquad X=\p{i\ptl_{t_E} \th-\frac{(\ptl_i\th)^2}{2}},
\ee
where $\xi$ is the Bertsch parameter, whose value is not relevant for our discussion here. The number density $n$, to the leading semi-classical order, in the EFT is given by
\be\label{eq:EFT_nOp}
n(t_E,\vec x) \equiv-\frac{\partial \cL }{\partial X}=\frac{2^{\frac{3}{2}}}{3\pi^2\xi^{\frac{3}{2}}} X^{\frac{3}{2}}.
\ee
Additionally, the scaling dimension of the leading charge-$Q$ operator $\cO_Q$ to leading order at large $Q$ is \cite{Kravec:2018qnu}
\be
\De_Q = \frac{3^{\frac{4}{3}}}{4}\xi^{\frac{1}{2}}Q^{\frac{4}{3}}.
\ee

The large-charge EFT also provides a systematic framework for computing correlation functions. For example, at leading order in the EFT, the Euclidean two-point function $\langle \cO_Q \cO^\dagger_Q \rangle$ of the primary $\cO_Q = \mathcal{ N } \left(\frac{2}{\gamma} \right)^{\frac{\Delta_Q}{2}} X^{\frac{\Delta_Q}{2}} e^{-i Q \theta}$ admits the path-integral representation \footnote{Note that $\cO^{\rm here}_Q\equiv \cO_{\Delta, -Q}$ of refs.\ \cite{Beane:2024kld, Beane:2025tum}.},
\be\label{eq}
G_Q(t_{E1},\vec x_1; t_{E2},\vec x_2)\equiv \langle\cO_Q \cO^\dagger_Q \rangle=\int \cD \th\, \cO_Q(t_{E1},\vec x_1)\cO_Q^{\dag}(t_{E2},\vec x_2)e^{-\int dt_E d^3 \vec x \cL},
\ee
where $\g=3^{\frac{1}{3}}\xi^{\frac{1}{2}}Q^{\frac{1}{3}}$ and $\mathcal N$ is a normalization constant. In ref.\ \cite{Beane:2024kld}, this path integral was evaluated semiclassically by exponentiating the operator insertions and solving the resulting equations of motion in the presence of localized sources. The saddle-point solution is
\be\label{eq:saddle_theta}
\th_s(t_E ,\vec x) = \frac{i}{2}\g \log\p{\frac{t_{E1}-t_E}{t_E-t_{E2}}}-\frac{i }{4}\left[\frac{(\vec x-\vec x_2)^2}{(t_E-t_{E2})}-\frac{(\vec x-\vec x_1)^2}{(t_{E1}-t_E)}\right],
\ee
where  $t_{E2}<t_E<t_{E1}$. Evaluating both the action and the operator insertions on this saddle yields the Schrödinger-invariant two-point function at leading semiclassical order,
\begin{align}
G_Q(t_{E1},\vec x_1; t_{E2},\vec x_2)&=\mathcal{N}^2 t_{E12}^{-\Delta_Q} e^{\frac{-Q \vec x^2_{12}}{2 t_{E12}}}.
\end{align}
Similar approaches can be used to compute higher point functions \cite{Beane:2024kld}, as will be required for the calculation of detector correlators.

\section{Detector Operators in NRCFTs}\label{sec:detectors}

In this section we define detector operators in NRCFTs.  A key aspect that we want to emphasize in this paper is that not only can one define detector operators in NRCFTs, but that they satisfy nearly identical properties to their relativistic analogs.  We therefore begin by reviewing the properties of detector operators in relativistic CFTs, and then stating the results for how they generalize to the case of NRCFTs. The remainder of this section will then focus on showing that these properties indeed hold.

Detector operators in relativistic theories satisfy a number of properties. While these properties are physically expected of detectors in real world experiments, they can also be rigorously proven due to the relationship between detector operators and the ANE operator.  In particular, the ANE operator satisfies the following conditions:
\begin{itemize}
\item Translation invariance, $[P^\mu, \mathcal{E}]=0$, which makes it behave as a ``passive" detector.\footnote{Throughout this paper, we will use the terminology of ``active" and ``passive" detectors. An ``active" detector modifies the quantum numbers of the state it is measuring, while a ``passive" detector does not. Examples of ``active" detectors are the generalized detectors of refs.\ \cite{Korchemsky:2021htm,Korchemsky:2021okt}, which inject/absorb energy, or memory/BMS detectors \cite{Strominger:2014pwa,Himwich:2026exq,Himwich:2025ekg}, which modify the particle number.}
\item Ward identities: $\int D^{d-2}z\, z^\mu \mathcal{E}(z)=P^\mu$.\footnote{Here, $z=(1,\hat n)$ and $D^{d-2}z=\frac{2d^d z \delta(z^2)\theta(z^0)}{\vol{\SO(d-1,1)}}$ is the standard measure of the projective null cone \cite{Simmons-Duffin:2012juh}.} Integration over the transverse coordinate relates the ANE to the  charges of the Lorentz group, which are topological operators.
\item The ANE annihilates the vacuum, $\mathcal{E}(\hat n) |0 \rangle =0$.
\item The ANE is a positive operator, $\mathcal{E}\succeq 0$. Namely, $\langle \psi | \mathcal{E}(\hat n)  | \psi \rangle \geq 0$ for any state $| \psi \rangle$ \cite{Faulkner:2016mzt,Hartman:2016lgu}. This is also referred to as the Average Null Energy Condition (ANEC).
\item Commutativity:  $[\mathcal{E}(\hat n_1), \mathcal{E}(\hat n_2) ]=0$ \cite{Kologlu:2019bco,Cordova:2018ygx}.
\item Existence of an operator product expansion \cite{Hofman:2008ar,Kologlu:2019mfz, Chang:2020qpj}
\begin{equation}
    \mathcal{E}(\hat n_1)\mathcal{E}(\hat n_2) \sim \sum_i C_i \, (1-\hat n_1\cdot \hat n_2)^{\frac{\tau_i-4}{2}} \mathbb{O}_{i}^{[J=3]} (\hat n_2) + \text{transverse derivatives}\,.
\end{equation}
\end{itemize}
Note that any positive operator with vanishing vacuum expectation values must annihilate the vacuum \cite{Epstein:1965zza}, and therefore these properties are not independent. While intuitive for detector operators, these properties cannot be obeyed by local operators in relativistic QFTs, highlighting the non-local nature of detector operators. Indeed, the Reeh-Schlieder theorem states that the vacuum is cyclic and separating \cite{Reeh:1961ujh}, in particular, implying that in relativistic theories, there do not exist positive local operators other than the identity.

In this section, we will show that the natural generalization of energy flux detectors to the case of NRCFTs in $3+1$ dimensions is given by
\begin{align}\label{eq:Ev_definition_2}
\mathcal{E}_v({\hat n}) =\frac{v}{2}\lim_{r \rightarrow \infty} r^3 n\left(\frac{r}{v}, r {\hat n} \right) = \frac{1}{2}\lim_{r \rightarrow \infty} r^3 {\hat n}\cdot \vec j\left(\frac{r}{v}, r {\hat n} \right)\,. 
\end{align}
As compared to the case of relativistic detector operators, eq.\ (\ref{eq:Ev_definition_2}) does not involve integration, one might therefore expect that they exhibit quite different properties from those in the relativistic case. However, quite interestingly, the distinct properties of NRCFTs enable them to have precisely the properties of detector operators in relativistic theories, namely
\begin{itemize}
\item Commutativity with energy $[H,\, \mathcal{E}_v]=0$, momentum $[\vec P,\, \mathcal{E}_v]=0$, and particle number $[N,\, \mathcal{E}_v]=0$, making them ``passive" detectors.
\item Ward identities: Integration over both velocity and transverse position reduces them to the topological charges for energy  $\int\! d\Omega \int\! dv\,  \mathcal{E}_v(\hat n) =H$, momentum $\int\! d\Omega \int\! dv\,  \mathcal{E}_v(\hat n) \frac{2\vec v}{v^2}=\vec P$, and particle number  $\int\! d\Omega \int\! dv\,  \mathcal{E}_v(\hat n) \frac{2}{v^2}=N$.
\item Annihilation of the vacuum $\mathcal{E}_v|\Omega \rangle=0$.
\item Positive definiteness $\mathcal{E}_v\succeq 0$.
\item Commutativity $[\mathcal{E}_{v_1}(\vec n_1),\, \mathcal{E}_{v_2}(\vec n_2) ]=0$.
\item Existence of an OPE in the limit $v_1 \hat n_1 \to v_2 \hat n_2$ \cite{Boisvert:2025hex}. Schematically,
\begin{align}
    \mathcal{E}_{v_1}(\hat n_1)\mathcal{E}_{v_2}(\hat n_2) &\sim \sum_i C_i \, (\vec v_{12}^2)^{\frac{\Delta_i-6}{2}} \mathbb{O}^{i}_{v_2} (\hat n_2)\,,
\end{align}
where $\vec v_{12} = v_1\hat n_1 -v_2 \hat n_2$.
\end{itemize}
We note that detector operators in NRCFTs have particle number zero, and are what are referred to as ``massless operators." Much like for light-ray operators in relativistic theories, the OPE of massless operators is more subtle than for operators with non-zero particle number. The OPE for massless operators in NRCFTs and its convergence properties are discussed in ref.\ \cite{Boisvert:2025hex}. The harmonic-trap representation relates the detector OPE in the coincident-velocity limit to the equal-time spatial OPE of density operators. Reference \cite{Boisvert:2025hex} also emphasizes the analogy between the properties of massless operators in NRCFTs and light-ray operators in CFTs.

In the remainder of this section we develop the definition of detector operators in NRCFTs, and show that they indeed satisfy these properties. In the case of CFTs, there are two approaches to defining detector operators. In the case that one has asymptotic states, one can define detector operators via their action on these asymptotic states, as was originally done by Sterman \cite{Sterman:1975xv}. A more general approach, which applies also in interacting CFTs, is to define them as integrals of the stress tensor, or other currents in the theory \cite{Korchemsky:1997sy,Hofman:2008ar}. In this section, we consider both approaches. We first begin with the simplified case of free theories, and define detector operators. We then extend this to the case of interacting theories, and provide an operator definition. In the following sections, we evaluate the one-point function of the $\cE_v(\hat n)$ detector in a variety of states relevant to final-state neutron production for vanishing momentum $\vk=0$. These examples provide explicit checks of the Ward identities \eqref{eq:Ward_E_generalk}, \eqref{eq:Ward_P_N_generalk} and allow us to verify the positivity constraints implied by eq.~\eqref{non_rel_HM_bound}.

\subsection{Warm Up: Energy Detectors in Free NRCFTs}

As a  starting point, we first consider the simple case of the NRCFT of free fermions in $3+1$ spacetime dimensions. The Lagrangian for free fermions of mass $m$ is given by
\begin{equation}\label{eq:free_fermions}
  \mathcal L = \sum_{\sigma=\uparrow\!\downarrow} \psi^\dagger_\sigma 
   \left( i \partial_t + \frac{\nabla^2}{2m} \right) \psi_\sigma.
\end{equation}
We would like to 
construct an operator representing the total energy deposited in the calorimeter 
at spatial infinity. For the free theory, in which the particle interpretation 
is unambiguous, this operator takes a particularly transparent form: it is 
simply the sum of the individual kinetic energies of all outgoing fermions observed along the detector direction ${\hat n}$. The on-shell free fermion states of momentum $\vk$ and mass $m$ can be written as\footnote{The convention for Fourier transform in this paper is
\begin{align}
f(t,x)=\int\! \frac{d^3\vk\,d\omega}{(2\pi)^4} e^{- i \omega t + i \vk\cdot \vx} \tilde{f}(\omega, \vk).
\end{align}
}
\begin{align}
\psi_\sigma(t, \vec x)= \int_{\vec k} e^{i \left( \vec k\cdot \vec x -\frac{\vec k^2}{2m} t\right)} a_\sigma(\vec k),\quad \int_\vk\equiv \int\! \frac{d^3 \vk}{(2\pi)^3}\,.
\end{align}
The energy flux operator at spatial infinity along the direction ${\hat n}$ in the free 
theory is then given by the following operator on the general on-shell particle states
\begin{equation}\label{En_def}
\mathcal{E}({\hat n})=\sum_{\sigma=\uparrow, \downarrow}\int_{\vec k} \frac{k^2}{2m} a_\sigma^\dagger(\vec k)a_\sigma(\vec k) \delta^{(2)}(\hat{k}-{\hat n})\,.
\end{equation}

While the above expression is perfectly well-defined for the free theory, it  suffers from a fundamental conceptual drawback when one attempts to generalize 
it to the interacting case. Specifically, in the presence of interactions, the total energy of the system is no longer simply the sum of the individual single-particle kinetic energies; interaction terms contribute additional, momentum-correlated contributions to the Hamiltonian that cannot be decomposed particle-by-particle. A definition of the calorimeter energy operator that relies explicitly on the particle content of the state is therefore not suitable as a starting point for the interacting theory. It is thus desirable to reformulate the energy operator in a manner that makes  no reference to the individual particle momenta, and which remains well-defined even when interactions are present. In the relativistic case \eqref{eq:ANEC_op}, the energy detector is defined using the stress tensor $T^{\mu\nu}$. We will argue that in NRCFTs, the correct definition of the analogous detector involves the number density $n$ or the particle-number current $\vec j$. In free theory, these are given by
\be\label{ji_def}
n(t,\vec x)&=\sum_{\sigma=\uparrow,\downarrow}\psi^\dagger_{\sigma}(t,\vec x)\psi_\s(t,\vec x), \nn \\
j^i(t,\vx)&= \frac{-i}{2m}\sum_{\sigma=\uparrow,\downarrow}\left( \psi^\dagger_\sigma(t,\vx)\partial^i_\vx \psi_\sigma(t,\vx)-\partial^i_\vx\psi^\dagger_\sigma(t,\vx) \psi_\sigma(t,\vx) \right).
\ee
We can re-express our energy detector \eqref{En_def} in terms of a limit of $n$ and $\vec j$ as follows. We look at the mode decomposition of our fermionic degrees of freedom in the following limit, which localizes due to a saddle-point approximation  
\begin{align}\label{saddle_psi}
\lim_{r \rightarrow \infty}\psi\left(\frac{m}{k_0} r, r {\hat n} \right)&= \lim_{r \rightarrow \infty}\int_{\vec k} e^{i \left( k_n r -\frac{\left(k^2_n+k_\perp^2\right)}{2m}\frac{m}{k_0} r \right)} a_\sigma(\vec k)\nonumber\\
&\simeq e^{\frac{ik_0 r}{2}}\left( \frac{-i k_0}{2\pi r}\right)^\frac{3}{2} a_\sigma(k_0 {\hat n})+ O\left( \frac{1}{r^\frac{5}{2}}\right),
\end{align}
where $k_n= {\hat n}\cdot \vk , \vk_\perp= \vk- k_n\hat n$. Plugging this into eq.\ \eqref{ji_def}, we find that the energy detector $\cE(\hat n)$ given in eq.\ \eqref{En_def} can be written as two equivalent expressions,
\begin{align}\label{En_def_ji}
\mathcal{E}({\hat n})&=\frac{1}{2m}\int dk_0\, k_0 \lim_{r \rightarrow \infty} r^3 n\left(\frac{m}{k_0} r, r {\hat n} \right)=\frac{1}{2}\int dk_0\, \lim_{r \rightarrow \infty} r^3 {\hat n}\cdot \vec j\left(\frac{m}{k_0} r, r {\hat n} \right).
\end{align}
This reformulation provides a natural and well-motivated definition of the energy observable in the collider setting. We will show in a later subsection that this operator definition of the energy detector holds true even in interacting theories invariant under Schr\"odinger symmetries.

We gain further physical intuition by re-expressing the integral of our energy operator \eqref{En_def_ji} in terms of the ``velocity" $v=\frac{k_0}{m}$,
\begin{align}
\mathcal{E}({\hat n})=\int_0^\infty dv\,  \mathcal{E}_v({\hat n})\,,
\end{align}
where we have defined a detector
\begin{align}\label{En_def_ji_v}
\mathcal{E}_v({\hat n}) =\frac{mv}{2}\lim_{r \rightarrow \infty} r^3 n\left(\frac{r}{v}, r {\hat n} \right) = \frac{m}{2}\lim_{r \rightarrow \infty} r^3 {\hat n}\cdot \vec j\left(\frac{r}{v}, r {\hat n} \right)\,.
\end{align}
This operator counts particles reaching the detector located at distance $r$ along the direction $\hat n$ at time $r/v$, and thus measures the energy density carried by particles moving with velocity $v\hat n$ (i.e., energy per unit $v$ and per unit solid angle)\footnote{The setup here is exactly that of a time-of-flight measurement. We thank Antoine Georges for this comment.}, and thus the total energy deposited in the calorimeter along ${\hat n}$ is a sum over all the possible velocities measured by the calorimeter. Integrating over ${\hat n}$ then gives us the total energy of the initial states. We will see that in NRCFTs, it is in fact $\mathcal{E}_v({\hat n})$ that will play the central role in the study of detector operators. In the rest of the paper we set $m=1$ for convenience.

Let us make some comments on the distinction between the nonrelativistic and relativistic detector observables. In massless theories and CFTs, the energy flow operator $\mathcal{E}(\hat{n})$ is defined as a limit to \textit{null infinity} $\mathcal{I}^+$ - following null rays $u= c t - r = \text{const}$ as $r \to \infty$. This is natural for massless particles, since they travel at $c$ and precisely reach $\mathcal{I}^+$. In the nonrelativistic setting, however, all particles are massive and travel at speeds $v < \oo$. Their worldlines satisfy $\vx/t = v\hat n= \text{const}$ as $t \to \infty$. The appropriate detector observable is therefore defined by the limit
\begin{equation}
    t = \frac{r}{v}, \qquad \vx = r\hat{n}, \qquad r \to \infty\,,
\end{equation}
at fixed velocity $\vec v = v\hat{n}$, which precisely follows the trajectories of these massive particles to infinity. This naturally produces a distribution over velocity space $\vec v \in \mathbb{R}^3$, and is the quantum field theoretic formalization of what is measured in time-of-flight experiments in cold atomic systems. Indeed, it is known that expanding unitary fermions released from an isotropic harmonic trap achieved an asymptotic momentum distribution function of the same shape as the coordinate-space distribution before release (eq.~(123) of ref.~\cite{2012LNP...836..127C}).  For mass-dependent corrections to relativistic detector observables, a similar approach has been employed in ref.\ \cite{Mateu:2012nk}. One can also further integrate over the velocity to get a massive detector with definite Lorentz spin \cite{Chang:2025zib}.  It would be interesting to better understand the connection between the $\cE_v(\hat n)$ detector and the relativistic massive detectors.

The quantity we will be interested in is the $N$-point function of $\cE_v(\hat n)$ in a $Q$-particle state created by an operator $\cO^\dagger_{Q,l}(\omega,\vec k)$. We will argue in the next subsection that $\cE_v(\hat n)$ preserves energy and momentum, and thus its $N$-point function can be written as
\be\label{eq:Ev_Npt_general}
&\<0|\cO_{Q,l}(\omega',\vec k')\cE_{v_1}(\hat n_1)\cE_{v_2}(\hat n_2)\cdots \cE_{v_N}(\hat n_N)\cO^\dagger_{Q,l}(\omega,\vec k)|0\> \nn \\
&=(2\pi)^{4}\de(\omega-\omega')\de^{(3)}(\vec k-\vec k') F(v_i\hat n_i,\omega,\vec k).
\ee
We will often abuse notation and write
\be
\<0|\cO_{Q,l}(\omega,\vec k)\cE_{v_1}(\hat n_1)\cE_{v_2}(\hat n_2)\cdots \cE_{v_N}(\hat n_N)\cO^\dagger_{Q,l}(\omega,\vec k)|0\> = F(v_i\hat n_i,\omega,\vec k).
\ee
Here, $\ell$ denotes some other quantum numbers (such as angular momentum) of the state. We also keep the dependence of $F$ on $Q$ and $\ell$ implicit. The correlators defined above are Wightman correlators: expectation values of products of operators in a fixed ordering, with no time-ordering prescription imposed. For the rest of the paper we restrict ourselves to the one point function of the $\cE_v(\hat n)$ detector
\be\label{eq:Ev_onept_normalized_def}
\<\cE_v(\hat n)\>_{\cO(\omega,\vec k)} \equiv \frac{\frac{v}{2}\lim_{r\to \oo}r^3 \<0|\cO(\omega,\vec k)n(\tfrac{r}{v},r\hat n)\cO^\dagger(\omega,\vec k)|0\>}{\<0|\cO(\omega,\vec k)\cO^\dagger(\omega,\vec k)|0\>}.
\ee

\subsection{Energy Detectors in General NRCFTs}\label{DigNRCFT}

Our previous definitions of detector operators in NRCFTs were formulated using a limiting procedure. In the case of relativistic CFTs, ref.\ \cite{Hofman:2008ar} showed that, in the case of CFTs, detector operators could be defined without the limiting procedure, and hence are non-perturbatively defined. In this section, we use the harmonic-trap coordinates, analogous to the Lorentzian cylinder for relativistic CFTs, to show that we can provide a general non-perturbative definition of detector operators in NRCFTs that does not make use of a limiting procedure. We will then be able to show, purely from symmetry arguments, that our detector operators satisfy all the desired Ward identities.

\begin{figure}
\includegraphics[width=0.5\linewidth]{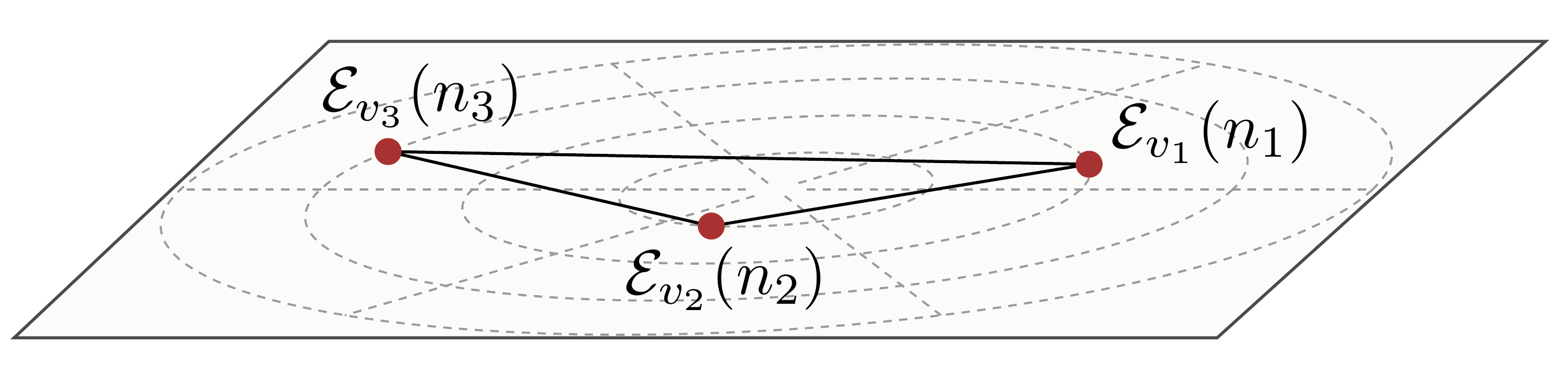}
\includegraphics[width=0.5\linewidth]{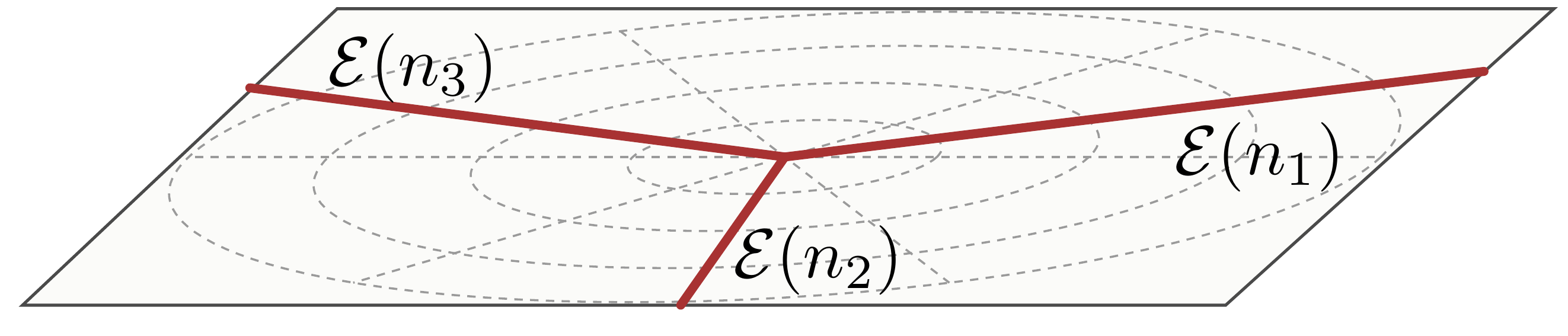} 
\caption{An illustration of the different detector operators on future infinity. a) To the left, Resolved detectors, $\mathcal{E}_v (n)$, are local on future infinity. b) To the right, the integrated energy detector $\cE(n)$ is a non-local line operator at future infinity. }
\label{fig:integrated_v_local}
\end{figure}

Recall from section \ref{sec:trap_review} that in the harmonic-trap geometry, future infinity corresponds to the spacelike slice of dimension $d$ at $\tau=\pi/2$. Our $\cE_v(\hat n)$ detector is defined in the limit $r\to \oo$ of the coordinates which scale as $t=\frac{r}{v}, \vec x=r\hat n$. In the harmonic-trap coordinates eq.\ \eqref{coord_ht_flat}, this corresponds to $\tau=\frac{\pi}{2},\vec y=v\hat n$. Furthermore, using eq.\ \eqref{En_osc}, we find that the two definitions in eq.\ \eqref{En_def_ji_v} using $n$ and $\vec j$ are equivalent, and can both be written as
\begin{align}\label{En_def_ji_ht}
\cE_v(\hat n) = \frac{v^4}{2}\tilde{n}\p{\frac{\pi}{2},v \hat n}.
\end{align}
Similarly, we can define the integrated energy detector as
\begin{align}\label{E_def}
\cE(\hat n) = \int dv \frac{v^4}{2}\tilde{n}\p{\frac{\pi}{2},v \hat n}\,.
\end{align}
The harmonic-trap geometry therefore provides a clear geometric picture of the detector operators. They are localized at $\tau=\frac{\pi}{2}$. The detectors $\cE_v(\hat n)$ are local on this spacelike slice, and are labeled by their velocity $v \hat n$. The integrated detectors $\cE(\hat n)$ are non-local line operators on this spacelike slice, and are labeled by $\hat n$. This is illustrated in figure \ref{fig:integrated_v_local}. The harmonic-trap geometry also makes clear one of the key properties of detector operators, namely that due to their spacelike separation
\begin{align}
[\cE_{v_1}(\hat n_1),\cE_{v_2}(\hat n_2)]=0\,.
\end{align}

We now explain how the matrix elements of our detector operators are constrained by the Schrödinger symmetry. Motivated by the harmonic-trap picture, we find that it would be more convenient to study the harmonic-trap density operator $\cD(\vec v)$
\be\label{eq:Dv_nHT_def}
\cD(\vec v)=\tilde{n}\!\left(\frac{\pi}{2}, \vec v\right), \quad  \mathcal{E}_v({\hat n})=\frac{v^4}{2} \cD(v\hat n).
\ee
For simplicity, we first consider the one-point function of $\cD(\vec v)$. Combining eqs.\ \eqref{eq:generators_3}, \eqref{eq:generators_general}, \eqref{coord_ht_flat}, and \eqref{En_osc}, the actions of conformal generators on the harmonic-trap density operator $\tn$ can be written as (see also ref.\ \cite{Boisvert:2025hex})
\be\label{eq:Dop_conformalgenerators}
[D,\tilde{n}(\tau, \vy)] &= i\left(\sin (2 \tau)\partial_\tau  +  \cos(2\tau) \left(y_i \partial_i +\Delta_n\right)\right)\tn(\tau,\vec y),\qquad [K_i,\tilde{n}(\tau, \vy)] =-i \sin \tau \partial_i \tn(\tau,\vec y), \nonumber\\
[C,\tilde{n}(\tau, \vy)] &= -i \left( \sin^2\tau \partial_ \tau+\sin\tau \cos \tau \left( y_i \partial_i+\Delta_n\right)\right)\tn(\tau,\vec y),\qquad  [P_i,\tilde{n}(\tau,\vy)] = i\cos \tau \ptl_i \tn(\tau,\vec y), \nonumber \\
[H,\tn(\tau,\vy)] &= -i(\cos^2\tau \ptl_\tau - \sin\tau \cos\tau \left( y_i \partial_i+\Delta_n\right))\tn(\tau,\vy).
\ee
Setting $\tau=\frac{\pi}{2}$, we see that the $\cD(\vec v)$ detector preserves energy and momentum
\be\label{eq:H_C_Dv_comm}
[H,\cD(\vec v)] = [P_i,\cD(\vec v)] = 0.
\ee
As a result, the one-point function $\<0|\cO(\omega,\vec k)\cD(\vec v)\cO^\dagger(\omega',\vec k')|0\>$ is proportional to delta functions $\de(\omega-\omega')\de^{(3)}(\vec k-\vec k')$, which then implies eq.\ \eqref{eq:Ev_Npt_general}.

Additionally, $K_i$ acts as translation and $D$ acts as rescaling in $\vec v$ at $\tau=\frac{\pi}{2}$. Inserting this into the one-point function, we obtain that the constraints from Galilean boost invariance and scale invariance are given by
\be
&\left(- \vec k \partial_\omega + N_\cO \partial_\vk - \partial_{\vec{v}}\right) \<0|\cO(\omega,\vec k)\cD(\vec v)\cO^\dagger(\omega,\vec k)|0\>=0, \nonumber\\
&\left(2 \omega \partial_\omega + \vk\. \partial_{\vk}+ \vec v\.\partial_{\vec v}+ 2\left(4-\Delta_\cO\right) \right)\<0|\cO(\omega,\vec k)\cD(\vec v)\cO^\dagger(\omega,\vec k)|0\>=0.
\ee
These equations can be solved by the method of characteristics, yielding the general form
\begin{align}\label{constraints_symm}
\<0|\cO(\omega,\vec k)\cD(\vec v)\cO^\dagger(\omega,\vec k)|0\>&=\p{\omega+\frac{k^2}{2N_\cO}}^{\De_\cO-4}f\left( \frac{\vec v+ \frac{\vk}{N_O}}{\sqrt{\omega+\frac{k^2}{2N_O}}}\right).
\end{align}

This expression makes the consequences of Schrödinger symmetry manifest. Galilean invariance restricts the dependence of the correlator to the invariant combinations
\begin{align}
\vec v + \frac{\vk}{N_O},
\qquad
\omega+\frac{k^2}{2N_O},
\end{align}
which are unchanged under boosts. Scale invariance then fixes the overall dependence on the second invariant, leaving only an arbitrary dimensionless function $f$ of the ratio appearing above. This result can also be generalized to $N$-point function straightforwardly, which gives
\be\label{eq:npoint_symmetry}
&\<0|\cO(\omega,\vec k)\cD(\vec v_1)\cdots \cD(\vec v_N)\cO^\dagger(\omega,\vec k)|0\>\nn \\
&=\p{\omega+\frac{k^2}{2N_\cO}}^{\frac{2\De_\cO-5-3N}{2}}f\left( \frac{\vec v_1+ \frac{\vk}{N_O}}{\sqrt{\omega+\frac{k^2}{2N_O}}},\ldots,\frac{\vec v_N+ \frac{\vk}{N_O}}{\sqrt{\omega+\frac{k^2}{2N_O}}}\right).
\ee

Finally, imposing the special conformal generator $C$ does not lead to any additional constraints on $f$. This is consistent with the general structure of Schrödinger-invariant theories: unlike relativistic conformal symmetry, Schrödinger symmetry is not sufficient to completely determine three-point functions of primary operators, and an undetermined function of invariant cross-ratios remains. For all practical purposes, eq.~\eqref{constraints_symm} implies that one may set $\vk=0$ when computing the one-point function, with the full momentum dependence subsequently restored via Galilean invariance.

\subsection{Ward Identities}

In relativistic theories, expectation values of the ANE operator (eq. \eqref{eq:ANEC_op}) are constrained by Ward identities. This follows from the fact that integrating the ANE operator over all directions with appropriate weight can produce the translation generator $P^\mu$ \cite{Hofman:2008ar}. We now show that the nonrelativistic detector also satisfies Ward identities. The idea is similar to the relativistic case, and we will argue that certain integrals of the detector operator can be expressed in terms of generators of the Schrödinger symmetry. In fact, it turns out that there are three different Ward identities measuring energy, momentum, and particle number.

To derive the Ward identities, it is helpful to consider the conformal generators evolved by the harmonic-trap Hamiltonian $H+C$:
\be\label{eq:Atau}
A(\tau) \equiv e^{i(H+C)\tau} A e^{-i(H+C)\tau},
\ee
where $A$ is a generator of the Schrödinger symmetry. Importantly, we note that by combining eqs.\ \eqref{eq:nj_algebra}, \eqref{eq:O_HT_def}, and \eqref{eq:Dv_nHT_def}, $N(\tau), K_i(\tau), C(\tau)$ at $\tau=\frac{\pi}{2}$ can be written as integrals of the $\cD(\vec v)$ detector. Using the Schrödinger algebra given in eq.\ \eqref{eq:generators_algebra}, we obtain
\be\label{eq:Atau_CKN}
C(\tau)=C \cos^2 \tau+ \frac12 D \sin 2\tau + H \sin^2 \tau, \nn \\
K_i(\tau) = K_i \cos\tau + P_i \sin\tau,\quad N(\tau) = N.
\ee
Setting $\tau=\frac{\pi}{2}$, we get the complete set of Ward identities obeyed by the $\cD(\vec v)$ detector
\be\label{eq:Ward_forDv}
\int d^3 \vec v\, \frac{\vec v^2}{2} \cD(\vec v) = H,\quad \int d^3\vec v\, v_i \cD(\vec v) = P_i,\quad \int d^3 \vec v\, \cD(\vec v) = N.
\ee

We can rewrite the above identities as constraints on the one-point function $\<\cE_v(\hat n)\>_{\cO(\omega,\vec k)}$ defined by eq.\ \eqref{eq:Ev_onept_normalized_def}. For example, for the integral giving the $H$ generator, we get
\be\label{eq:Ward_E_generalk}
\int d\O_{\hat n}\int dv\, \<\cE_v(\hat n)\>_{\cO(\omega,\vec k)} = \omega.
\ee
Thus, we have shown on general grounds that in generic NRCFTs obeying Schr\"odinger symmetry furnished by the current algebra in eq.\ \eqref{eq:nj_algebra}, the detector observable defined in eq.\ \eqref{En_def_ji_v} for free theories continues to be a good definition of an observable detecting energies. The other two identities from $P_i$ and $N$ give
\be\label{eq:Ward_P_N_generalk}
\int d\O_{\hat n}\int dv\, \frac{2\hat n}{v} \<\cE_v(\hat n)\>_{\cO(\omega,\vec k)} =\vec k,\quad \int d\O_{\hat n}\int dv\, \frac{2}{v^2} \<\cE_v(\hat n)\>_{\cO(\omega,\vec k)}  = N_{\cO^\dagger} = -N_\cO.
\ee
These relations provide a non-trivial set of constraints on any admissible nonrelativistic detector distribution. The first identity fixes the total energy, the second measures the total momentum, and the third determines the total particle number carried by the state. Together, they constitute the nonrelativistic analog of the familiar event-shape sum rules encountered in relativistic CFTs. 

Another way of explicitly deriving the Ward identities is by taking the detector limit eq.\ \eqref{En_def_ji_v} of the flat space three-point functions $\<\cO n \cO^\dagger\>, \<\cO \vec j \cO^\dagger\>$, and one can show that the Ward identities for $\<\cE_v(\hat n)\>_{\cO(\omega,\vec k)}$ follow from the constraints on $\<\cO n \cO^\dagger\>, \<\cO \vec j \cO^\dagger\>$. This is similar to how the Ward identities for the relativistic energy detector follow from the Ward identities of the $\l_{\cO T \cO}$ OPE coefficients. We do this calculation for scalar $\cO$ in appendix \ref{app:ward_id_flat}.

Note that $\<\cE_v(\hat n)\>_{\cO(\omega,\vec k)}$ also obeys symmetry constraints following from eq.\ \eqref{constraints_symm}. These constraints imply that it is sufficient to verify the Ward identities for $\vk=0$, and the identities for general $\vec k$ will be automatically satisfied. The crucial observation is that $\<\cE_v(\hat n)\>_{\cO(\omega,\vec k)}$ depends on $\vk$ and $\omega$ only through the Galilean-invariant combinations identified in eq.\ \eqref{constraints_symm} (up to the $v^4$ factor in eq.\ \eqref{eq:Dv_nHT_def}). Let us consider the energy Ward identity as an example. Assuming the $\vec k=0$ Ward identities are true, then for general $\vec k$ we have
\be
\int_{\Omega_{{\hat n}}} \int dv\, \<\cE_v(\hat n)\>_{\cO(\omega,\vec k)}
&=\int d^3 \vec v\, \frac{\vec{v}^2}{2} \frac{\<0|\cO(\omega+\tfrac{k^2}{2N_\cO},0)\cD(\vec v+\tfrac{\vec k}{N_\cO})\cO^\dagger(\omega+\tfrac{k^2}{2N_\cO},0)|0\>}{\<0|\cO(\omega+\tfrac{k^2}{2N_\cO},0)\cO^\dagger(\omega+\tfrac{k^2}{2N_\cO},0)|0\>} \nonumber\\
&=\int d^3 \vec v' \, \frac{(\vec{v}'-\tfrac{\vec k}{N_\cO})^2}{2} \frac{\<0|\cO(\omega+\tfrac{k^2}{2N_\cO},0)\cD(\vec v')\cO^\dagger(\omega+\tfrac{k^2}{2N_\cO},0)|0\>}{\<0|\cO(\omega+\tfrac{k^2}{2N_\cO},0)\cO^\dagger(\omega+\tfrac{k^2}{2N_\cO},0)|0\>} \nonumber\\
&= \left(\omega+\frac{k^2}{2N_\cO}\right)+\frac{k^2}{2N_\cO^2}(-N_\cO) = \omega.
\ee
In the first line, we have used the relation between $\cE_v(\hat n)$ and $\cD(\vec v)$ and Galilean invariance to express the integrand solely in terms of the invariant combinations
$\vec v+\frac{\vk}{N_O}, \omega+\frac{k^2}{2N_O}.$ We then perform the change of variables $\vec{v}'=\vec v+\frac{\vk}{N_O}$,
which leads to the second line. The third line follows from the $\vec k=0$ Ward identities. We therefore recover the energy Ward identity eq.\ \eqref{eq:Ward_E_generalk} for arbitrary momentum. An identical argument applies to the other two identities, showing that all Ward identities at non-zero $\vk$ follow directly from their $\vk=0$ counterparts together with Galilean invariance.

\subsection{Positivity}\label{sec:positivity}
In the harmonic-trap geometry, the velocity detector operator defined in eq.\ \eqref{En_def_ji_v} becomes
\begin{align}
\mathcal{E}_v({\hat n}) = \frac{v^4}{2}\,\cD(  v {\hat n}).
\end{align}
 From a physical standpoint, one expects this quantity to be non-negative. Unlike in relativistic quantum field theories, where local energy densities can exhibit negative contributions, such effects are not expected to arise in the nonrelativistic regime considered here. The positivity is also explicitly manifest in the oscillator coordinate expression, due to the fact that it is the conformally transformed $U(1)$ particle density in flat coordinates with a positive measure (see eq.\ \eqref{En_osc}). This motivates the proposal of a nonrelativistic analog of the Hofman–Maldacena bound \cite{Hofman:2008ar}:
\begin{align} \label{non_rel_HM_bound}
\mathcal{E}_v({\hat n}) \geq 0.
\end{align}

It is worth noting that the proposal implies the existence of positive local operators in NRCFTs. This is in contrast to relativistic QFTs, where nonzero positive local operators with vanishing vacuum expectation values do not exist \cite{Epstein:1965zza}. Let us briefly review the argument for the relativistic case. Suppose there exists an operator $\cO$ that is positive and has vanishing one-point function. First, by considering the expectation value of $\cO$ in a linear combination of the vacuum and an arbitrary state, one can show that $\cO$ must annihilate the vacuum. Since $\cO$ is a local operator, we can consider an open set in spacetime $\cR$ that is spacelike separated from $\cO$. Then, for any operators $\cO_1, \cO_2$ smeared in $\cR$, we have $\<\O|\cO_1 \cO \cO_2|\O\> = \<\O|\cO_1  \cO_2 \cO|\O\> = 0 $. Finally, we can apply the Reeh-Schlieder theorem \cite{Reeh:1961ujh}, which states that any state can be approximated to arbitrary precision by states created by (smeared) local operators in an open set in spacetime. Since $\cR$ is open, we conclude that $\cO$ has vanishing matrix elements in all states, and therefore is identically zero.

One important step of this argument is the existence of an open set in spacetime that is spacelike separated from the operator. This is why it does not apply to the relativistic average null energy operator, since the region that is spacelike separated from a null line is not open. For local operators in nonrelativistic spacetime, the argument breaks down for the exact same reason \cite{Boisvert:2025hex}---the region spacelike separated from a local operator $\cO(t,\vec x)$ is simply the spatial slice at time $t$, which is not an open set in spacetime. Hence, there is no contradiction in having positive local operators in NRCFTs.

We can also understand the positivity from the perspective of the non-perturbative ``non-renormalization theorems" and OPE recently discussed in ref.\ \cite{Boisvert:2025hex}. We assume that the NRCFT has no nontrivial genuine massless sector and that its harmonic-trap spectrum is discrete, with no finite accumulation points. Within the OPE framework of ref.\ \cite{Boisvert:2025hex}, these assumptions lead to the identification of a canonical non-genuine primary $(\cO_1^\dagger\cO_2)$, with scaling dimension $\Delta_1+\Delta_2$, appearing schematically as 
\begin{align}\label{eq:OPE_OO}
\mathcal{O}_1^\dagger(x) \mathcal{O}_2(0) \sim (\mathcal{O}_1^\dagger \mathcal{O}_2)(0)+\cdots
\end{align}

We use this framework to prove positivity of a wide class of operators in NRCFTs, namely the canonical normal ordered operators of the form $(\mathcal{O}^\dagger \mathcal{O})$. We further reasonably assume that every other non-genuine primary appearing in this OPE has $\Delta>2\Delta_{\cO}$ and the leading structure function $c_0\left(\xi \equiv \frac{x^2}{t}\right)$, associated with
$\Op_0=\NO{\Op^\dagger\Op}$ and $\Delta_0=2\Delta_{\cO}$, is continuous at $\xi=0$. Then, we consider the smeared norm involving the two-point function
\begin{equation}\label{eq:gram}
 \int\! dw_1\,dw_2\;\overline{g_{\varepsilon, z}(w_1)}\,g_{\varepsilon, z}(w_2)\;\langle\Psi\rvert\,\Op^\dagger(w_1)\Op(w_2)\,\lvert\Psi\rangle\,
  \;\ge\;0 ,
\end{equation}
where $\Psi$ denotes an arbitrary state and $g_{\varepsilon, z}$ is a smearing function localized near $(t_z, \vec x_z)$. Due to the nonrelativistic scaling, it is convenient to consider a family of smearing functions with anisotropic scaling. We can define, for $\varepsilon>0$ and a fixed
$\delta>0$
\begin{equation}\label{eq:reg}
  g_{\varepsilon,z}(w)\;=\;\varepsilon^{-(3(1+\delta)+2)}\,
  g\!\left(\frac{t-t_z}{\varepsilon^{2}},\,\frac{\xx-\xx_z}{\varepsilon^{1+\delta}}\right). 
  \end{equation}
Assuming the OPE, and the absence of singular terms, as stated in the assumptions above\footnote{Here we assume that the OPE is sufficiently well-behaved which allows for exchanging the $\varepsilon \to 0$ limit with the integrals.}, it then follows that one can smoothly take the $\varepsilon \to 0$ limit, and prove that $\mathcal{O}^\dagger \mathcal{O}$ is a positive operator. Corresponding evolution in the harmonic-trap frame only changes this by a positive measure and hence, the detector operator defined in terms of the corresponding harmonic-trap primary at $\tau=\frac{\pi}{2}$ is also positive. Therefore in NRCFTs without genuine massless sectors we expect a variety of positive operators, and our detector operators are specific examples of them.

More interestingly, we can use positivity of the detector operator $\mathcal{E}_v({\hat n})$ as a test for the presence of genuine massless operators, or non-genuine primaries with $\De<2\De_\cO$. In this paper the only interacting theory we will consider is the case of fermions at unitarity, and we will find that it is positive.

In relativistic QFTs, the ANEC can be rigorously proven, and is related to either causality \cite{Hartman:2016lgu}, or monotonicity of entanglement entropy \cite{Faulkner:2016mzt}. It would be extremely interesting to explore whether it is possible to understand positivity of $\mathcal{E}_v({\hat n})$, and related operators in NRCFTs from a more algebraic perspective. See, e.g., ref.\ \cite{Sorce:2024pte} for a review.

\subsection{Schrödinger Charge Detectors}\label{sec:general}

Beyond the case of energy, momentum and particle number  (given by eq.\ \eqref{eq:Ward_forDv}),  it is interesting to ask whether there are detectors whose Ward identities give the remaining generators of the Schrödinger group. In the relativistic case, the algebra of light-ray operators on a null plane whose integrals give rise to different conformal generators has been explored in refs.\ \cite{Cordova:2018ygx,Gonzo:2020xza}, and more general detectors have been systematically studied in refs.\ \cite{Himwich:2026exq,Himwich:2025ekg,Strominger:2026yrh}.  In this subsection, we briefly investigate the construction of other detector operators which generate the remaining Schrödinger charges of Galilean transformations, dilatations, rotations, and special conformal transformation.

One such detector can be built from the probability current $\vec j$ in the harmonic-trap frame. More precisely, let us define
\be
\cD_{1i}(\vec v) &\equiv \tilde{j}_i (\tau=\tfrac{\pi}{2},\vec v).
\ee
Here, $\tilde{j}(\tau)$ is the probability current in the harmonic-trap frame defined by eq.\ \eqref{eq:O_HT_def}. Using eq.\ \eqref{En_osc}, we can also write the $\cD_{1i}(\vec v)$ detector in terms of the operators in flat space
\be
\cD_{1i}(v\hat n) &= \lim_{r\to \oo}\frac{r^4}{v^4}\p{j_i(t=\tfrac{r}{v},r \hat n)-v_i n(t=\tfrac{r}{v},r \hat n)}.
\ee
Note that both $j_i(t=\tfrac{r}{v},r \hat n)$ and $v_i n(t=\tfrac{r}{v},r \hat n)$ have the same leading term in the $r\to \oo$ limit, and therefore their difference goes like $1/r^4$ instead of $1/r^3$. As in the case of $\cD(\vec v)$, we can obtain the symmetry constraints of this new detector operator by studying the action of Schr\"odinger generators on $\tilde{j}_i(\tau, \vy)$
\be\label{eq:Dop_conformalgenerators_2}
[D,\tj_i(\tau, \vy)] &= i\left(\sin 2 \tau \ptl_\tau  +  \cos 2\tau \left(y_a \partial_a + \Delta_j\right)\right)\tj_i(\tau,\vec y)+ 2 i  y_i \sin 2 \tau \, \tn(\tau, \vy),\nonumber\\
[K_j,\tj_i(\tau, \vy)] &=-i \sin \tau \partial_j  \tj_i(\tau,\vec y)+ i \delta_{ij} \cos \tau \tn(\tau, \vy), \qquad  [P_j ,\tj_i(\tau,\vy)] = i\cos \tau \partial_j \tj_i(\tau, \vy)+ i \delta_{ij} \sin \tau \tn(\tau,\vec y),\nonumber\\
[C,\tj_i(\tau, \vy)] &= -i \left(\sin^2 \tau \ptl_\tau +\sin\tau \cos \tau \left( y_a \partial_a+\Delta_j \right)\right)\tj_i(\tau,\vec y) + i y_i \cos 2\tau \, \tn(\tau, \vy), \nonumber \\
[H,\tj_i(\tau,\vy)] &= -i(\cos^2\tau \ptl_\tau - \sin\tau \cos\tau \left( y_a \partial_a+\Delta_j\right))\tj_i(\tau,\vy)- i y_i \cos 2\tau \tn(\tau, \vy),
\ee
where $\De_j=\De_n+1=4$. We see that the detector is an ``active" detector which does not preserve energy and momentum\footnote{Another example of ``active" detectors that has been studied in the relativistic case are time resolved detectors \cite{Korchemsky:2021okt}.}. More precisely, we have
\be
[H,\cD_{1i}(\vec v)] = i v_i \cD(\vec v),\quad [P_j,\cD_{1i}(\vec v)] = i\de_{ij}\cD(\vec v),
\ee
where $\cD(\vec v)=\tilde n(\tfrac{\pi}{2},\vec v)$. Consequently, whereas the actions of $H$ and $P$ constrain correlators of $\cD(\vec v)$ to take the forms given in eq.\ \eqref{eq:Ev_Npt_general}, the correlators of $\cD_{1i}(\vec v)$ instead satisfy
\begin{align}
&\<0|\cO(\omega',\vec k')\cD_{1i}(\vec v)\cO^\dagger(\omega,\vec k)|0\> \\
&= (2\pi)^4\delta^{(3)}(\boldsymbol{k})\delta(\boldsymbol{\omega}) g_i(\vec v, \bar{\omega}, \bar{k}) + i(2\pi)^4 \left( v_i\partial_{\boldsymbol{\omega}} + \partial_{\boldsymbol{k}_i}\right)\delta^{(3)}(\boldsymbol{k})\delta(\boldsymbol{\omega}) \<0|\cO(\bar \omega,\bar k) \cD(\vec v) \cO^\dagger(\bar \omega,\bar k)|0\>,\nn 
\end{align}
where we use $\bar z= \frac{z+z'}{2},\boldsymbol{z} = z-z'$, and $g_i(\vec v, \bar{\omega}, \bar{k})$ is an unfixed function. 

Now, let us write down the Ward identities for $\cD_{1i}(\vec v)$, in other words the conserved charges that can be obtained from integrals of this detector operator. Similar to the derivation of the Ward identities for $\cD(\vec v)$, we consider the conformal generator evolved by the harmonic-trap Hamiltonian eq.\ \eqref{eq:Atau} and use eq.\ \eqref{eq:generators_algebra},\footnote{Comparing eqs.\ \eqref{eq:Atau_CKN} and \eqref{eq:Atau_MDP}, we see that $K_i'(\tau)=P_i(\tau)$ and $C'(\tau)=D(\tau)$. This is due to the continuity equation $\partial_t n + \partial_{x_i}j_i=0$. In the harmonic-trap frame, it becomes $\partial_\tau \tilde{n} + \partial_{y_i}\tilde{j}_i =0 $, and thus we have $K_i'(\tau) = \int d^3 \vec y\, y_i \ptl_\tau \tilde{n}(\tau,\vec y) = \int d^3 \vec y\, \tilde{j}_i(\tau,\vec y) = P_i(\tau)$. A similar argument also gives $C'(\tau)=D(\tau)$.}
\be\label{eq:Atau_MDP}
P_i(\tau) = -K_i \sin \tau + P_i \cos \tau,\quad D(\tau) = D\cos 2\tau + (H-C)\sin 2\tau,\quad M_{ij}(\tau) = M_{ij}.
\ee
Setting $\tau=\frac{\pi}{2}$, we obtain
\be
\int d^3 \vec v\, \cD_{1i}(\vec v) = -K_i,\quad \int d^3 \vec v\, v_i \cD_{1i}(\vec v) = -D,\quad \int d^3\vec v\, (v_i\cD_{1j}(\vec v) - v_j \cD_{1i}(\vec v)) = M_{ij}.
\ee

The Ward identities of $\cD_{1i}(\vec v)$ give the generators $K_i, D, M_{ij}$. From eq.\ \eqref{eq:generators_general}, we can see that the actions of these generators on a local operator in Fourier space $\cO^\dagger(\omega,\vec k)$ all involve first-order derivatives with respect to $\omega$ or $\vec k$. This is consistent with the fact that the detector comes from the subleading terms $\sim 1/r^4$ of the flat space operators $n$ and $\vec j$ in the $r \to \oo$ limit. In fact, it seems that studying these detectors could also be useful for understanding how to construct similar ``active" detectors at infinity in the relativistic case.

There is still one remaining generator that is not related to the Ward identities of $\cD(\vec v)$ and $\cD_{1i}(\vec v)$, the special conformal generator $C$. The action of $C$ on $\cO^\dagger(\omega,\vec k)$ involves second-order derivatives $\ptl_\omega^2,\ptl_{\vec k}^2$, and hence one should expect that it corresponds to some combination of flat space operators which goes like $1/r^5$ in the detector limit. One way to construct this detector is to consider the stress tensor $\Pi_{ij}$, which is another alien operator. Let us review some properties of this operator. First, $j_i$ and $\Pi_{ij}$ obey a momentum conservation equation
\be\label{eq:Pi_momentum_conservation}
\ptl_t j_i + \ptl_j \Pi_{ij} = 0.
\ee
Reference \cite{Nishida:2007pj} (see eq.\ (B7) therein) constructed an expression for $\Pi_{ij}$ in terms of $\psi,\psi^\dagger$ that satisfies the momentum conservation equation. Using this expression, we obtain the commutators of $\Pi_{ij}$ with the Galilean boost generator $K_k$ and the special conformal generator $C$,
\be\label{eq:Pi_comm_explicit}
[K_k,\Pi_{ij}(t,\vec x)] &= -it\ptl_{k}\Pi_{ij}(t,\vec x) + i(\de_{ki}j_j(t,\vec x)+\de_{kj}j_i(t,\vec x)), \nn \\
[C,\Pi_{ij}(t,\vec x)] &= -i(t^2\ptl_t+t x_a\ptl_a+t\De_\Pi)\Pi_{ij}(t,\vec x) + i(x_i j_j(t,\vec x) + x_j j_i(t,\vec x)).
\ee
The stress tensor in the harmonic-trap frame $\tilde \Pi_{ij}(\tau ,\vec y)$, defined by eq.\ \eqref{eq:O_HT_def}, is therefore related to its flat-space counterpart by
\be
\tilde{\Pi}_{ij}(\tau ,\vec y) = \sec^5\! \tau\, \Pi_{ij}(t,\vec x) -  \sec^4\! \tau \tan \tau (y_i j_j(t,\vec x) + y_j j_i(t,\vec x)) +  \sec^3\! \tau \tan^2\! \tau\, y_i y_j n(t,\vec x).
\ee
Using the corresponding transformations of the current and number density, the inverse relation takes the form
\be
\Pi_{ij}(t,\vec x) = \cos^5\! \tau (\tilde{\Pi}_{ij}(\tau,\vec y) + \tan \tau  (y_i \tilde{j}_j(\tau,\vec y) + y_j \tilde{j}_i(\tau,\vec y))+\tan^2\! \tau \, y_i y_j \tilde{n}(\tau,\vec y)).
\ee

The detector we will study is simply given by $\tilde{\Pi}_{ij}(\tau,\vec y)$ evaluated at $\tau=\frac{\pi}{2}$,
\be
\cD_{2ij}(\vec v)\equiv \tilde{\Pi}_{ij}(\tfrac{\pi}{2},\vec v).
\ee
In terms of the flat space operators, we have
\be
&\cD_{2ij}(v\hat n)= \lim_{r\to \oo}\frac{r^5}{v^5}\p{\Pi_{ij}(t=\tfrac{r}{v},r\hat n)-(v_i j_j(t=\tfrac{r}{v},r\hat n) + v_j j_i(t=\tfrac{r}{v},r\hat n)) + v_i v_j n(t=\tfrac{r}{v},r\hat n))}.
\ee
As expected, the detector comes from the subsubleading terms with $1/r^5$ behavior in the $r\to \oo$ limit. Finally, to obtain the Ward identity, we can consider the generator $H(\tau)$\footnote{Note that we have $H(\tau) = C(\tau) + \frac{1}{2}D'(\tau)$. This can be shown using the momentum conservation equation \eqref{eq:Pi_momentum_conservation}, which becomes $\ptl_\tau \tilde{j}_i(\tau,\vec y) + \ptl_{y_j}\tilde{\Pi}_{ij}(\tau,\vec y) = - y_i \tilde{n}(\tau,\vec y)$ in the harmonic-trap frame. This furnishes another detector definition for the conformal charge $C$
\be
\cD_{3}(\vec v)= - v_i\ptl_{v_j}\tilde{\Pi}_{ij}(\tfrac{\pi}{2},\vec v) = v_i \ptl_{\tau}\tilde{j}_i(\tfrac{\pi}{2},\vec v) + v^2 \tilde{n}(\tfrac{\pi}{2},\vec v), \qquad \int\! d^3\vec v \,\cD_{3}(\vec v) = 2C.
\ee
One advantage that this definition provides is that the detectors are now expressed in terms of operators whose commutation relations we already know, bypassing the explicit construction of the stress tensor in order to derive the action of Schr\"odinger generators on the detector.}
\be
H(\tau) = H \cos^2\!\tau - D \sin \tau \cos \tau + C \sin^2\! \tau.
\ee
Setting $\tau=\frac{\pi}{2}$, and using $2H(\tau)=\int_\vy \tilde{\Pi}_{ii}(\tau, \vec y)$ \cite{Nakayama:2009ww}, we have
\be
\int d^3 \vec v\, \cD_{2ii}(\vec v) = 2C.
\ee

In this paper, we focus only on explicit calculations of the energy detector built from $\cD(\vec v) = \tilde{n}(\frac{\pi}{2},\vec v)$. However, it would be extremely interesting to explicitly calculate correlators of these other detector operators in known theories as well as understand more general symmetry constraints (\`a la eq.\ \eqref{constraints_symm}), as it may provide a tractable setup for understanding ``active detectors.''

\section{Calculations of Detector Correlators in NRCFTs}\label{sec:calcs}

In this section we perform a number of explicit calculations of one-point correlators of detector operators in NRCFTs. In section \ref{sec:free_calc} we begin with calculations of detector operators in states with two and three free fermions. In section \ref{sec:unitarity_calc} we extend this to the case of an interacting theory, namely fermions at unitarity, where we compute detector correlators in two and three-body states, as well as in large-charge states. Additionally, we discuss a generalization of the Hofman-Maldacena conformal collider bounds  \cite{Hofman:2008ar} to nonrelativistic theories, and show that it is satisfied for all the cases we considered.

\subsection{Free Theory}\label{sec:free_calc}

In this section, we  explicitly compute the collider observable measuring energy distribution for two and three-particle states in the theory of free fermions. We present two complementary derivations: the first employs the definition \eqref{En_def} directly, while the second demonstrates its equivalence to the detector formulated as the limiting form of the number density \eqref{En_def_ji}. In both approaches, the central ingredients are the $n$-particle state and the local operator responsible for creating this state from the vacuum.

\subsubsection*{Two-particle states:}
We consider the situation where the final state consists of two non-interacting fermions, which we assume to be produced by a local two-body scalar operator. In this case, the detector distribution is completely fixed by kinematics. In the center of mass frame where the momentum of the source, $\vec k=0$, we have
\begin{align}
\<\cE_v(\hat n)\>_{\cO_2(\omega)} = \frac{v^3}{2\pi}\de(\omega-v^2)\,.
\end{align}
Here we show how this can be obtained from correlators in the NRCFT. Additionally, we compute the detector correlator for non-zero $\vec k$ to verify our understanding of symmetry constraints on the correlator.

\begin{figure}
\begin{center}
\includegraphics[width=0.5\linewidth]{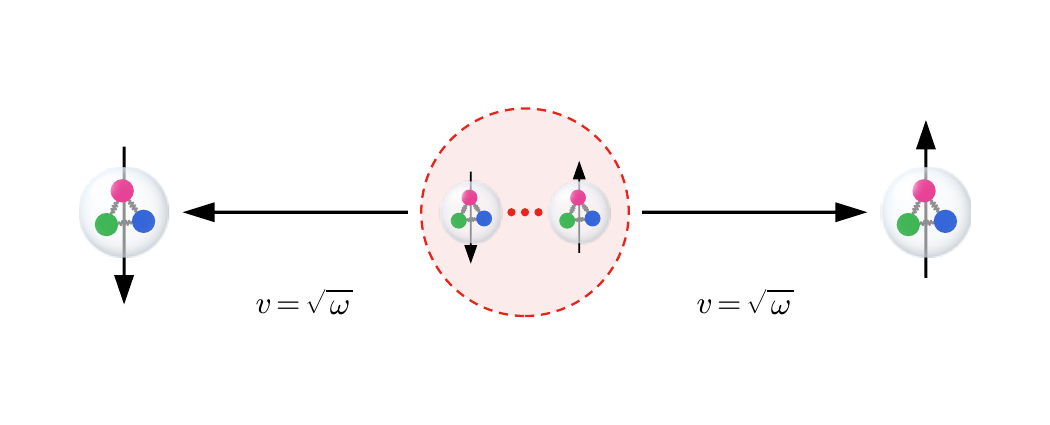}
\end{center}
\caption{Two-body states in the center of momentum frame, $\vec k=0$, produce a simple detector distribution $\mathcal{E}_v\sim \de(\omega-v^2)$.}
\label{two_body}
\end{figure}

The relevant charge-two sector has one spin-up and one spin-down fermion and the corresponding state is given by 
\begin{equation}\label{free2body}
    |\vk_1,\vk_2\rangle =\int_{\vec x, \vec y}
    \Psi^{\rm{free}}_{\vk_1,\vk_2} (\Vec{x},\Vec{y})  \psi^\dagger_\downarrow(\Vec{y})\psi^\dagger_\uparrow(\Vec{x})|0\rangle, \quad \Psi^{\rm{free}}_{\vk_1,\vk_2} (\Vec{x},\Vec{y})=\frac{1}{V} e^{i (\vk_1\cdot \vx+ \vk_2\cdot\vy)},\,\,\,   \int_\vx\equiv \int d^3\vx,
\end{equation}
where we have chosen to normalize the individual wave-functions within a rectangular box of volume $V$ with unit normalization. Thus the momenta are discrete but  in the limit of large volume $V$,  we can take the continuum limit using 
 \begin{equation}\label{2bodycontlimit}
 \sum_{\vec k_{i}}\rightarrow \frac{V}{(2\pi)^3} \int\! d^3\vec k_{i}\,.
\end{equation} 
The primary operator which creates this state from vacuum is therefore particularly simple to construct
\begin{align}\label{O2_free}
\cO^{\rm free}_2(t, \vx)= \psi_\uparrow\psi_\downarrow(t,\vx)\,.
\end{align}

We first evaluate the energy observable in the two-particle state \eqref{free2body} by the explicit action of \eqref{En_def}. As discussed in subsection \ref{DigNRCFT}, a more fundamental object is the velocity-graded detector observable $\mathcal{E}_v({\hat n})$ 
\begin{align}
\mathcal{E}_v({\hat n})&= \frac{v^4}{2(2\pi)^3}\sum_{\sigma=\uparrow,\downarrow} a^\dagger_{\sigma}(v \hat n)a_{\sigma}(v \hat n)\,.
\end{align}
In order to compute its matrix element, we decompose our energy observable in the state created by eq.\ \eqref{O2_free} in the following manner:
\begin{align}\label{O_2_E_O_2_free}
&\langle 0|\cO_2(\omega,\vk) {\mathcal E}_v({\hat n}) \cO^\dagger_2(\omega',\vk')|0\rangle\nn \\
&=\sum_{\vk_1,\vk_2, \vk_3, \vk_4} \langle \vk_3, \vk_4| {\mathcal E}_v({\hat n}) | \vk_1, \vk_2 \rangle  \langle 0| \cO_2(\omega,\vk)|\vk_3, \vk_4 \rangle \langle \vk_1, \vk_2| \cO^\dagger_2(\omega',\vk')|0 \rangle\,.
\end{align}
In this equation, we have inserted two complete sets of states and have used the fact that the matrix element between a charge-$n$ operator and an $m$-particle state, $\langle 0|\cO_n(0,0)| m \rangle$ is non-zero only for $n=m$ due to particle number conservation.  The action of the energy operator on the 2 particle state of the free theory is given by
\begin{align}
\mathcal{E}_v({\hat n}) | \vk_1, \vk_2\rangle &=\frac{v^4}{2(2\pi)^3}\sum_{\sigma=\uparrow,\downarrow} a^\dagger_{\sigma}(v \hat n)a_{\sigma}(v \hat n)\left( \int_{\vec x, \vec y}
    \Psi^{\rm{free}}_{\vk_1,\vk_2} (\Vec{x},\Vec{y})  \psi^\dagger_\downarrow(\Vec{y})\psi^\dagger_\uparrow(\Vec{x})|0\rangle\right)\nonumber\\
    &=\frac{v^4}{2} \sum_{i} \delta^{(3)}(\vk_i-v{\hat n}) | \vk_1, \vk_2\rangle,
\end{align}
where we have used the anticommutation relations $\{a_\sigma(\vk),a^\dagger_{\sigma'}(\vk')\}=(2\pi)^3\delta_{\sigma, \sigma'}\delta^{(3)}( \vk-\vk' )$. The Fourier transform over the matrix element of $\cO_2$ in the 2-particle states is given by 
\begin{align}
\langle \vk_1, \vk_2 | \cO^\dagger_2(\omega,\vk)|0\rangle=\frac{(2\pi)^4}{V} \delta\left(\omega-\frac{\left( k^2_1+k^2_2\right)}{2}\right) \delta^{(3)}\left(\vk-\left( \vk_1+\vk_2\right)\right)\,.
\end{align} 
Putting everything together and normalizing by the two point function, we obtain
\begin{align}
&\langle\mathcal{E}_v({\hat n}) \rangle_{\cO^{\rm{free}}_2(\omega, \vk)}\nonumber\\
&=\frac{1}{(2\pi)\sqrt{\omega-\frac{k^2}{4}}}\int d^3\vk_1\, d^3 \vk_2\, \delta\left(\omega-\frac{ k^2_1+k^2_2}{2}\right) \delta^{(3)}\Bigl(\vk-\bigl( \vk_1+\vk_2\bigr)\Bigr) \biggl(\frac{v^4}{2} \sum_{i=1}^2\delta^{(3)}(\vk_i-v{\hat n})\biggr)\,.
\end{align}
The integrals are performed using delta function localization and finally detector correlator as a function of $v$ and ${\hat n}$ takes the form
\begin{equation}\label{eq:Ev_Charge2_Free}
\langle\mathcal{E}_v({\hat n}) \rangle_{\cO^{\rm{free}}_2(\omega, \vk)}=\frac{v^4}{(2\pi)\sqrt{\omega-\frac{k^2}{4}}} \,\delta\biggl(\omega-\frac{ v^2+\bigl(\vk-v {\hat n}\bigr)^2}{2}\biggr)\,.
\end{equation} 
The result admits a natural physical interpretation: it describes a two-particle state created by the operator, with one particle carrying momentum $v \hat n$ and the other carrying the recoil momentum $\vk - v \hat n$, consistent with overall momentum conservation. It can also be written as
\begin{equation}
\langle\mathcal{E}_v({\hat n}) \rangle_{\cO^{\rm{free}}_2(\omega, \vk)} =\frac{v^4}{(2\pi)\left(\omega-\frac{k^2}{4}\right)^{\frac{3}{2}}} \,\delta\Biggl(1-\frac{\Bigl(v \hat n- \frac{\vk}{2} \Bigr)^2}{\omega-\frac{k^2}{4}}\Biggr)\,,
\end{equation} 
which is consistent with the symmetry constraints of eq.\ \eqref{constraints_symm}, once we account for the normalization due to the momentum space two point function as well as the definition of $\mathcal{E}_v({\hat n})$ in terms of $\cD({\vec v})$ given in eq.\ \eqref{eq:Dv_nHT_def}. Setting $\vec k=0$, we obtain the expected simple result 
\begin{equation}\label{eq:E_O2_free_final_2}
\<\cE_v(\hat n)\>_{\cO_2(\omega)} = \frac{v^3}{2\pi}\de(\omega-v^2)\,,
\end{equation}
describing two particles recoiling back-to-back, as depicted in figure \ref{two_body}.

We can also compute the integrated energy detector
\begin{equation}\label{E_O2_free_final}
\langle\mathcal{E}({\hat n}) \rangle_{\cO^{\rm{free}}_2(\omega, \vk)} = \int\limits_0^\infty\! dv\, \langle\mathcal{E}_v({\hat n}) \rangle_{\cO^{\rm{free}}_2(\omega, \vk)}\,.
\end{equation}
Invoking rotational symmetry, we align the detector direction ${\hat n}$ so that $\vk \cdot {\hat n} = k\cos\theta$. The structure of the resulting integrand is governed by the support of the delta function, which enforces energy-momentum conservation and admits solutions only within specific kinematic regimes. A careful analysis reveals three distinct cases:
\begin{itemize}
    \item \textbf{Sub-threshold regime} $\left(\omega < \frac{k^2}{4}\right)$: The delta function possesses no solution for real $\theta$ at any positive $v$, so the energy correlator receives no contribution.
    \item \textbf{Intermediate regime} $\left(\frac{k^2}{4} < \omega < \frac{k^2}{2}\right)$: Two distinct solutions exist, subject to a non-trivial kinematic constraint on the angular variable $\theta\leq \cos^{-1}\left( \frac{\sqrt{2} \sqrt{k^2-2 \omega }}{k} \right)$. The two solutions can be labeled as    \begin{align}
    v^\pm&= \frac{1}{2} \left(k \cos \theta \pm\sqrt{k^2 \cos ^2\theta -2 k^2+4 \omega } \right).
    \end{align}
    \item \textbf{Super-threshold regime} $\left(\omega > \frac{k^2}{2}\right)$: $v^+$ is the unique solution which is positive for all values of $\theta$.
\end{itemize}
Collecting the contributions from each kinematic region, the energy correlator takes the form,
\begin{align}
\langle\mathcal{E}({\hat n}) \rangle_{\cO^{\rm{free}}_2(\omega, \vk)}&=\begin{cases}
{\frac{\left(\sqrt{k^2 \cos ^2\theta -2 k^2+4 \omega }+k \cos \theta \right)^4}{16 \pi  \sqrt{4 \omega -k^2} \sqrt{k^2 \cos ^2\theta -2 k^2+4 \omega }}}, \qquad \omega>\frac{k^2}{2},\nonumber\\
{\frac{2 k^4 \cos ^4\theta -4 k^2 \cos ^2\theta  \left(k^2-2 \omega \right)+\left(k^2-2 \omega \right)^2}{2 \pi  \sqrt{4 \omega -k^2} \sqrt{k^2 \cos ^2\theta -2 k^2+4 \omega }}}\quad \frac{k^2}{2}>\omega>\frac{k^2}{4}, \quad \theta\leq\cos^{-1}\left( \frac{\sqrt{2} \sqrt{k^2-2 \omega }}{k} \right)\nonumber\\
\end{cases}
\end{align}
As a consistency check, one can integrate this expression to verify the Ward identity eq.\ \eqref{eq:Ward_E_generalk}, and more generally we have also verified the identities eq.\ \eqref{eq:Ward_P_N_generalk} using eq.\ \eqref{eq:Ev_Charge2_Free}.

We now arrive at the same result using the definition of the energy detector in terms of the number density in eq.\ \eqref{ji_def}. In order to do so, it is convenient to perform the Fourier transform over the external states at the end, after taking the detector limit. At an intermediate stage, we get
\begin{align}
&\langle 0| \cO_{2}(\omega, \vk) n\left(t_2, \vx_2 \right) \cO^\dagger_{2}(\omega', \vk')|0\rangle \\
&=\frac{-2}{(2\pi)^6}\int_{t_1,t_3,\vx_1,\vx_3} \frac{e^{i \left(\omega t_1-\omega' t_3- \vk \cdot \vx_1+\vk' \cdot \vx_3\right)}}{(t_{12}t_{23})^3}\int_{\vec{z}_1} \left( e^{\frac{i \left(\vx_1-\vec{z}_1\right)^2}{2t_{12}}+\frac{i \left(\vx_1-\vx_2\right)^2}{2t_{12}}+\frac{i \left(\vx_2-\vx_3\right)^2}{2t_{23}}+\frac{i \left(\vz_1-\vx_3\right)^2}{2t_{23}}} \right). \nn
\end{align}  
We now take the detector limit of the coordinates $(t_2, \vec{x}_2)$ and apply the scaling $\vec{z}_1 \rightarrow \frac{r}{v} \vec{z}_1$. The scaling is necessary in order to exchange the detector limit with the integral over $\vec z_1$, without it, the $\vec{z}_1$ dependence would drop out of the integrand. Finally, upon taking the Fourier transform of the external states, the integrals localize, and accounting for the normalization of the two-point function, we obtain
\begin{align}
\langle\cE_v(\hat n)\rangle_{\cO_2(\omega,\vec k)}&=\frac{\frac{v}{2}\lim_{r \rightarrow \infty} r^3 \langle 0| \cO_{2}(\omega, \vk) \,n\!\left(\frac{r}{v}, r {\hat n} \right) \cO^\dagger_{2}(\omega, \vk)|0\rangle}{\langle 0| \cO_{2}(\omega, \vk) \cO^\dagger_{2}(\omega, \vk)|0\rangle} \nonumber\\[6pt]
&=\frac{v^4}{2\pi \sqrt{\omega -\frac{\vk^2}{4}}}\, \delta \!\left(\omega-\frac{(\vk-{\hat n}\, v)^2+v^2}{2} \right),
\end{align}
in agreement with eq.\ \eqref{eq:Ev_Charge2_Free}. We note that the $\vec{z}_1$ integral may equivalently be performed prior to taking the detector limit without any scaling, leaving the final result unchanged. This calculation is an explicit verification of the equivalence between the two definitions of the energy detector, eqs.\ \eqref{En_def} and \eqref{En_def_ji}.

\subsubsection*{Three-particle States:}\label{sec:free_3particle}

We now turn our attention to the kinematically richer case in which the final state consists of three non-interacting fermions. We will compute the distribution of $\mathcal{E}_v$ in states produced by both spin 0 and spin 1 primary operators. Throughout this section, we always work in the center of momentum frame, $\vec k=0$, for simplicity.

We begin by considering a state created by the $\ell=0$ primary operator. The lowest dimensional $\ell=0$ primary operator necessarily involves at least $2$ derivatives, by rotational symmetry and Pauli exclusion. The independent operators at this level consist of $\psi_\uparrow \psi_\downarrow \nabla^2 \psi_\downarrow$ and $\nabla_i \psi_\uparrow \psi_\downarrow \nabla^i \psi_\downarrow$. The relative coefficient between them can be fixed by requiring that the operator transforms as a primary, which gives
\begin{align}
\cO^{ \rm{free}}_{3, \ell=0}=\psi_\uparrow \psi_\downarrow \nabla^2 \psi_\downarrow-2\nabla_i \psi_\uparrow \psi_\downarrow \nabla^i \psi_\downarrow.
\end{align}

\begin{figure}
    \centering
    \includegraphics[width=0.5\textwidth]{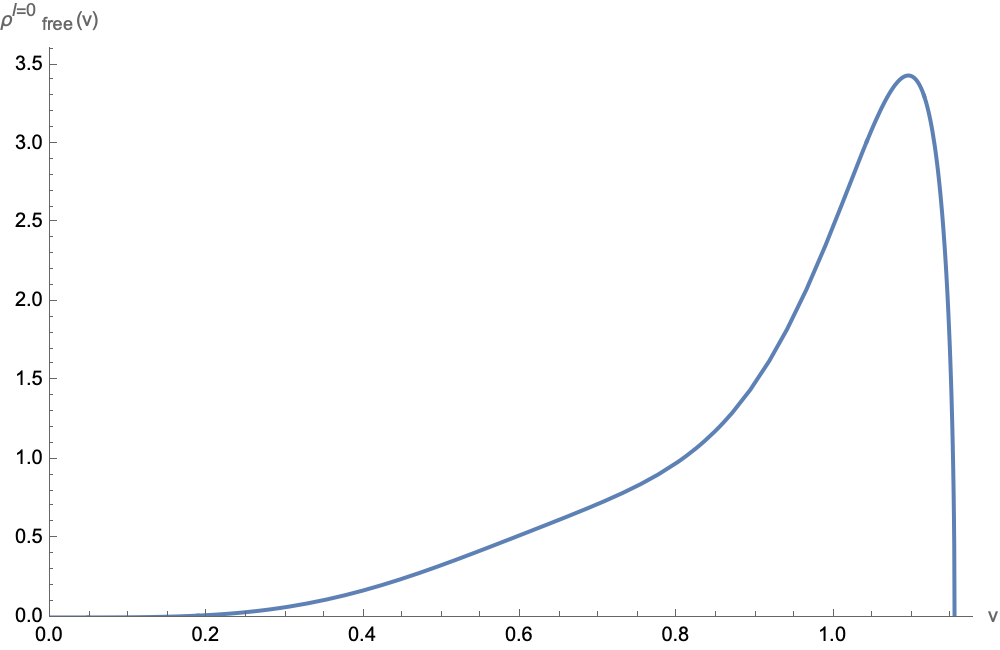}
    \caption{Velocity distribution for three free fermions in the $\ell=0$ state as a function of $v$, with $\omega=1$ and $\vk=0$.}
    \label{rhov_l0_free}
\end{figure}

\begin{figure}
\begin{center}
\includegraphics[width=0.5\linewidth]{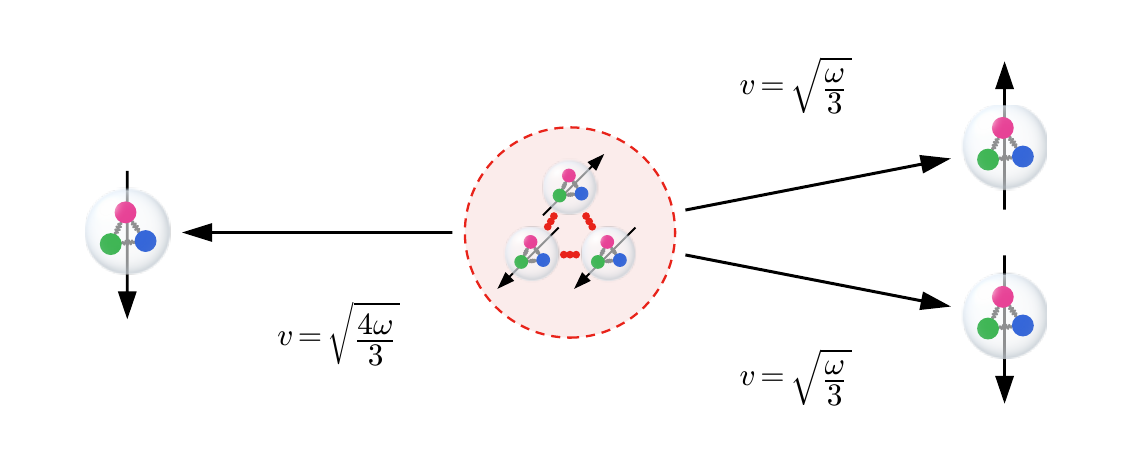}
\end{center}
\caption{Kinematic configuration for the maximal velocity in figure \ref{rhov_l0_free} and figure \ref{rhov_l1_free}: two collinear particles recoil against a third.}
\label{rhov_l1_free_kin}
\end{figure}

The detector distribution can be computed following the identical approach used
for the two-body state above.  We find
\begin{align}\label{eq:O3_l0_free}
\langle\mathcal{E}_v({\hat n}) \rangle_{\cO^{ \rm{free}}_{3, \ell=0}(\omega)}&=\frac{9 v^4 \sqrt{12 \omega -9 v^2} \left(3 v^4-4 v^2 \omega +2 \omega ^2\right)}{8 \pi ^2 \omega ^4} \theta(\sqrt{4\omega/3}-v)\,.
\end{align}
The distribution is plotted in figure \ref{rhov_l0_free}, where we have normalized it by defining $\rho^{\ell=0}_{\,\mathrm{free}}(v) = 4 \pi \langle\mathcal{E}_v({\hat n}) \rangle^{\mathrm{free}}_{\cO^{ \rm{free}}_{3, \ell=0}(\omega)}$.
The  distribution is independent of $\hat n$ due to the spin 0 nature of the source, and has support up to the maximal speed $v = \sqrt{4\omega/3}$. The corresponding kinematic configuration, shown in figure \ref{rhov_l1_free_kin}, consists of two particles moving collinearly in one direction with speed $\sqrt{\omega/3}$, while the third recoils in the opposite direction with speed $\sqrt{4\omega/3}$, as required by energy and momentum conservation. For the case of free fermions, the distribution vanishes at $v = \sqrt{4\omega/3}$. For interacting fermions, we will see that the velocity distribution differs markedly from the free case, while the allowed velocity range remains unchanged, as dictated by three-body kinematics. One can similarly obtain the charge and momenta distribution satisfying the Ward identities eq.\ \eqref{eq:Ward_P_N_generalk}.

We can also compute the distribution for a spin 1 source.  We denote the operator creating the state from vacuum by $\epsilon \cdot \cO_{3}(t, \vx)$, where $\epsilon^i$ is a normalized polarization vector. By rotational symmetry, the energy correlator can only depend on $\epsilon\.\hat n = \cos\th$
\begin{align}
\langle{\mathcal E}_v({\hat n})\rangle_{\epsilon\cdot \cO^{\rm free}_3(\omega)} &=  {\mathcal E}_{\perp}(v, \omega) \left( 1- |\epsilon\.\hat n|^2 \right) + {\mathcal E}_{\parallel} (v, \omega) |\epsilon\.\hat n|^2 \nonumber\\
&\equiv\frac{\sqrt{\omega}}{4\pi}\beta\left(\frac{v}{\sqrt{\omega}}\right) \left(1+ \alpha_2\left(\frac{v}{\sqrt{\omega}}\right)\left( \cos^2 \theta-\frac{1}{3} \right) \right)\,.
\end{align}
Famously, Hofman and Maldacena \cite{Hofman:2008ar} argued that positivity of the energy flux in spin 1 states places constraints on the value of $\alpha_2(v)$  (in this case assuming $\beta(v) $ is positive, which we will later show). In the nonrelativistic case, we extend positivity of $\mathcal{E}$ to positivity of $\mathcal{E}_v$, which translates to the bound
\begin{align}\label{eq:MH_positivity}
-\frac{3}{2}\leq \alpha_2(v)\leq 3 \,.
\end{align}
We will see that this is indeed true for the free field case, and much more interestingly, for the case of interacting fermions in section \ref{sec:NR_bounds}.

\begin{figure}
    \centering
    \includegraphics[width=0.5\textwidth]{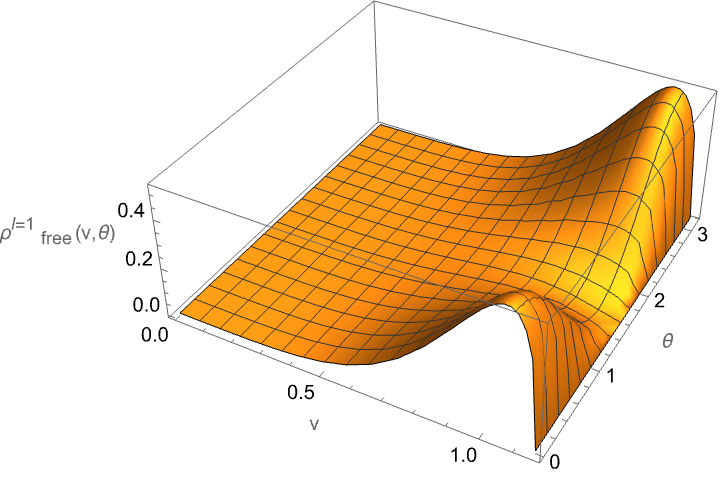}
    \caption{Velocity distribution for three free fermions in the $\ell=1$ state created by $\psi_\uparrow \psi_\downarrow \ptl_i \psi_\uparrow$ as a function of $v$ and $\theta$, with $\omega=1$ and $\vk=0$.}
    \label{rhov_l1_free}
\end{figure}

The free field realization of the lowest dimensional charge-three $l=1$ operator  is given in ref.\ \cite{Hammer:2021zxb} 
\begin{equation}
    \cO^{\rm free}_{3, i}(t,\vx) = \psi_\uparrow \psi_\downarrow \partial_i \psi_\uparrow (t,\vx) .
\end{equation}
The detector one-point function of this operator is given by
\begin{align}\label{En_free_O3_l1}
\langle\mathcal{E}_v({\hat n}) \rangle_{\epsilon \cdot \cO^{\rm{free}}_3(\omega)}&=\frac{9 v^4 \sqrt{12 \omega -9 v^2} }{8 \pi ^2 \omega ^2}  \left(1+\frac{9 v^2}{4\omega} \left( \cos^2\theta-\frac13 \right) \right),
\end{align}
where $\epsilon \cdot \hat n= \cos \theta$ and we have kept the theta function for velocity analogous to eq.~\eqref{eq:O3_l0_free} implicit. One can verify that this result satisfies the Ward identities eqs.~\eqref{eq:Ward_E_generalk} and \eqref{eq:Ward_P_N_generalk}. We define the velocity distribution $\rho$ as
\begin{align}
    \rho^{\ell=1}_{\,\mathrm{free}}(v, \theta) &= \langle\mathcal{E}_v({\hat n}) \rangle_{\epsilon \cdot \cO^{\rm{free}}_3(\omega)},
\end{align}
and we plot $\rho^{\ell=1}_{\,\mathrm{free}}$ in figure \ref{rhov_l1_free}. We see that the generalized Hofman-Maldacena bound eq.\ \eqref{eq:MH_positivity} is indeed satisfied for $\alpha^{\rm free}_2(v)=\frac{9 v^2}{4 \omega}$, and the upper bound is saturated for the maximal speed $v =\sqrt{4\omega/3}$.

\subsection{Fermions at unitarity}\label{sec:unitarity_calc}

Having understood the behavior of detector operators for two and three-body states in the case of free fermions, we now consider the case of an interacting fermionic NRCFT, namely fermions at unitarity. This system is commonly described by an effective Lagrangian for spin-half fermions with a point-like interaction, which is then renormalized and fine-tuned to the strongly interacting fixed point \cite{Nishida:2006eu}. The fine-tuning is characterized by an infinite scattering length, or equivalently by a contact interaction whose normalized coupling is tuned to the unitary limit. We will see that this infinite scattering length leads to non-trivial consequences for the correlators of detector operators.

In this section, we compute one-point function of the velocity-resolved detector $\cE_v(\hat n)$ in a variety of two- and three-body states in the theory of fermions at unitarity. This will be achieved using the wave-function approach to computing correlators in NRCFTs, discussed in section \ref{sec:wave-function}, combined with the non-perturbative detector definition in eq.\ \eqref{En_def_ji_v}. Analytic wave-function solutions are not known in closed form beyond the three-particle sector. However, for large enough particle numbers, such states can be approximated using the superfluid EFT. To investigate the behavior of detector correlators at large multiplicity, we therefore also compute the one-point detector correlator $\cE_v(\hat n)$ in many-body states using this effective theory. For each calculation, we review the structure of the corresponding wave-function to make the presentation self-contained.

\subsubsection{Two-Particle States}\label{sec:unitarity_2body}

In this subsection, we consider  one-point function of $\cE_v(\hat n)$ in the state created by the lowest dimensional charge-$2$ operator $\cO_2^\dagger$. As for the case of free fermions, the one-point correlator in a two-body state is completely fixed by kinematics, again corresponding to two recoiling back-to-back particles, as illustrated in figure \ref{two_body}.  We will see how this emerges from the two-body wave-function for fermions at unitarity.

Let us first review some essential ingredients of the charge-two sector of fermions at unitarity needed for this calculation \cite{Chowdhury:2023oas}. The charge-two state in fermions at unitarity can be written as
\be
|\Psi_{\vec P_{\mathrm{cm}},k}\> = \int_{\vec x,\vec y} \Psi_{\vec P_{\mathrm{cm}},k}(\vec x,\vec y) \psi^\dagger_{\downarrow}(\vec y)\psi^\dagger_{\uparrow}(\vec x)|0\>,
\ee
where the wave-function $\Psi_{\vec P_{\mathrm{cm}},k}(\vec x,\vec y)$ obeys the Bethe-Peierls boundary condition \eqref{eq:Bethe_Peierls}, and is given by\footnote{\label{foot:VRmax_convention}As explained in ref.\ \cite{Chowdhury:2023oas}, the appropriate way to describe the wavefunction is to assume that $\vec R_{\text{cm}}$ lives in a box of volume $V$, and $\vec r$ lives inside a sphere with radius $R_{\text{max}}$. After taking the cotinuum limit $V \to \oo, R_{\text{max}} \to \oo$, the discrete sum over states becomes integral over $\vec P_{\text{cm}}$ and $k$. In the end, all the correlation functions are independent of $V$ and $R_{\text{max}}$ in the continuum limit. Because of this, in this paper we neglect all the dependences on $V$ and $R_{\text{max}}$ (i.e., implicitly set $V=R_{\text{max}}=1$) to make the equations simpler. If the correct $V$ and $R_{\text{max}}$ dependences are restored, some of the equations in the intermediate steps will have different overall factors, but all our final results remain the same.} 
\be\label{eq:charge2_wave-function}
\Psi_{\vec P_{\mathrm{cm}},k}(\vec x,\vec y)= \frac{e^{i\vec P_{\mathrm{cm}}\.\vec R_{\mathrm{cm}}}}{\sqrt{2\pi }}\frac{\cos(kr)}{r},\qquad \vec R_{\mathrm{cm}}=\frac{\vec x+\vec y}{2},\quad \vec r=\vec x-\vec y.
\ee
The total energy of the state is $\frac{P_{\mathrm{cm}}^2}{4}+k^2$. The phase space measure for the charge-two sector is 
\begin{align}
 \frac{1}{(2\pi)^3\pi}\int d^3 \vec P_{\rm cm}\int\limits_0^\infty dk.
\end{align}
The lowest dimensional charge-2 operator which creates the initial state and its matrix element in the charge-two sector is given in ref.\ \cite{Nishida:2007pj}
\be
\cO_2(t,\vec X)= \lim_{\substack{\vec x \to \vec X \\ \vec y \to \vec X}}|\vx-\vy|\psi_\uparrow(t,\vx) \psi_{\downarrow}(t,\vy),\qquad
\<0|\cO_2(t,\vec X)|\Psi_{\vec P_{\mathrm{cm}},k}\> &=
 \frac{e^{i\vec P_{\mathrm{cm}}\.\vec X-i \left( \frac{P_{\rm cm}^2}{4}+k^2 \right) t}}{\sqrt{2\pi}}.
\ee
It is easy to check that $\Delta_{\cO_2}=2$. For our purposes, it is sufficient to consider the operator at zero spatial momentum giving us the following form factor of the charge-2 operator
\be\label{eq:O2_FormFactor_momentum}
\<0|\cO_2(\omega)|\Psi_{\vec P_{\mathrm{cm}},k}\> = (2\pi)^\frac{7}{2}\de^{(3)}(\vec P_{\textrm{cm}})\de(\omega-\tfrac{P_{\textrm{cm}}^2}{4}-k^2).
\ee

Analogous to the free theory case, we can compute the one-point function of the energy detector by inserting complete sets of states
\begin{align}\label{O_2_E_O_2_int}
&\langle 0| \cO_2(\omega){\mathcal E}_v({\hat n})\cO^\dagger_2(\omega') |0\rangle\\
&=\frac{1}{\left(2\pi \right)^6 \pi^2}\int d^3 \vec P_{\rm cm}\,d^3\vec P'_{\rm cm} \int\limits_0^\infty dk\, dk'\, \langle \Psi_{\vec P_{\mathrm{cm}},k}| {\mathcal E}_v({\hat n}) |\Psi_{\vec P'_{\mathrm{cm}},k'} \rangle  \langle 0| \cO_2(\omega)|\Psi_{\vec P_{\mathrm{cm}},k} \rangle \langle \Psi_{\vec P'_{\mathrm{cm}},k'}| \cO^\dagger_2(\omega')|0 \rangle.\nn
\end{align}
The matrix element of $\cE_v(\hat n)$ in the charge-two states can be obtained by taking the detector limit of the density operator $n$. We compute this in similar manner as the free theory case, but now using the explicit two body unitarity wave-function eq.\ \eqref{eq:charge2_wave-function} and the operator definition of $\cE_v(\hat n)$ from eq.\ \eqref{En_def_ji_v}, resulting in the following form:
\be
&\<\Psi_{\vec P_{\textrm{cm}},k}|\cE_v(\hat n)|\Psi_{\vec P'_{\textrm{cm}},k'}\>  \\
&=\frac{v}{2\pi}\lim_{r\to \oo}r^4 e^{i\p{\frac{P_{\mathrm{cm}}^2-P_{\mathrm{cm}}^{\prime2}}{4}+k^2-k^{\prime2}}\frac{r}{v}}\int_{\vec y_0}e^{-\frac{ir}{2}(\vec P_{\textrm{cm}}-\vec P'_{\textrm{cm}})\.(\hat n+\vec y_0)}\frac{\cos(k r|\hat n-\vec y_0|)\cos(k' r |\hat n-\vec y_0|)}{|\hat n-\vec y_0|^2},\nn
\ee
where we have rescaled the integration variable $\vec y \to r \vec y_0$. Recall that $\cE_v(\hat n)$ commutes with $H$ and $P_i$ (see eq.\ \eqref{eq:H_C_Dv_comm}). One therefore expects this matrix element to contain delta functions that enforce energy and momentum conservation. However, the delta functions are not manifest from the above expression, and it is not immediately obvious how such delta functions emerge after taking the $r\to \oo$ limit. 

To appropriately take the $r \to \oo$ limit, the expression should be interpreted as a distribution in $\vec P_{\textrm{cm}},\vec P'_{\textrm{cm}},k$ and $k'$. The reason is that the large-$r$ limit is highly oscillatory: away from stationary points of the phase, the oscillations lead to destructive interference, while neighborhoods of the stationary points give the leading contribution. Thus the appropriate way to extract the limit is to integrate the expression against smooth test functions and then apply stationary-phase analysis. For example, if we integrate the $\vec P_{\textrm{cm}}$-dependent part against a test function $f(\vec P_{\textrm{cm}})$ and take the $r\to \oo$ limit, using saddle-point analysis we get
\be
&\int\! d^3\! \vec P_{\textrm{cm}}\, e^{-\frac{ir}{2}\vec P_{\textrm{cm}}\.(\hat n + \vec y_0)+\frac{ir}{v}\frac{P_{\textrm{cm}}^2}{4}}f\bigl(\vec P_{\textrm{cm}}\bigr)= \p{2e^{\frac{i\pi}{4}}\sqrt{\frac{\pi v}{r}}}^3 e^{-i \frac{vr}{4}|\hat n + \vec y_0|^2} f\bigl(\vec P_{\textrm{cm}} = v(\hat n + \vec y_0)\bigr) +\cdots,
\ee
where $\cdots$ denote subleading terms in the $r\to \oo$ limit. In this distributional sense, the large-$r$ limit localizes $\vec P_{\textrm{cm}}$ onto the delta function $\de^{(3)}(\vec P_{\textrm{cm}} - v(\hat n + \vec y_0))$ at leading order.  Equivalently, after keeping track of the stationary-phase prefactor, the oscillatory kernel acts as a delta function supported at this value of $\vec P_{\textrm{cm}}$. One obtains a similar localization for the $k$-dependent part
\be
\int\limits_0^\infty\! dk\, e^{i\frac{r}{v}k^2} \cos(k r |\hat n -\vec y_0|) f(k)&= \frac{1}{2}e^{-i\frac{v r}{4}|\hat n - \vec y_0|^2}\sqrt{\frac{\pi v}{r}}e^{\frac{i\pi}{4}} f\left(k= \tfrac{v}{2}|\hat n - \vec y_0|\right) + \cdots.
\ee

In summary, the detector limit ensures the following distributional identities, 
\begin{align}\label{eq:distribution_identity}
e^{-\frac{ir}{2}\vec P_{\textrm{cm}}\.(\hat n + \vec y_0)+\frac{ir}{v}\frac{P_{\textrm{cm}}^2}{4}} &\sim \p{2e^{\frac{i\pi}{4}}\sqrt{\frac{\pi v}{r}}}^3 e^{-i \frac{vr}{4}|\hat n + \vec y_0|^2} \delta^{(3)}( \vec P_{\textrm{cm}} -v(\hat n + \vec y_0) ),\quad r \to \oo, \nonumber\\
e^{i\frac{r}{v}k^2} \cos(k r |\hat n -\vec y_0|)  &\sim \frac{1}{2}e^{-i\frac{v r}{4}|\hat n - \vec y_0|^2}\sqrt{\frac{\pi v}{r}}e^{\frac{i\pi}{4}} \delta\left( k-\tfrac{v}{2}|\hat n - \vec y_0|\right),\qquad\qquad\quad  r \to \oo.
\end{align}
These formulae should be understood in the distributional sense: when integrated against a smooth test function, the right-hand side reproduces the leading large-$r$ stationary-phase contribution of the corresponding oscillatory integral.

Putting everything together and using the fact that the integrals over $P'_{\rm cm}$ and $k'$ also lead to similar localizations, we obtain 
\be
&\<\Psi_{\vec P_{\textrm{cm}},k}|\cE_v(\hat n)|\Psi_{\vec P'_{\textrm{cm}},k'}\> \nn \\
&=\frac{v}{2\pi } 16 \pi^4 v^4\int_{\vec y_0}\frac{1}{|\hat n-\vec y_0|^2}\de^{(3)}\left(\vec P_{\textrm{cm}} - v(\hat n + \vec y_0)\right)\de^{(3)}\left(\vec P_{\textrm{cm}} - \vec P'_{\textrm{cm}}\right)\de\left(k-\tfrac{v}{2}|\hat n - \vec y_0|\right)\de\left(k-k'\right) \nn \\
&=(2\pi)^3\de^{(3)}(\vec P_{\textrm{cm}} - \vec P'_{\textrm{cm}})\pi \de\left(k-k'\right)\left(\frac{v^4}{4 \pi k^2}\de(k-|v\hat n - \tfrac{\vec P_{\textrm{cm}}}{2}|)\right).
\ee
We see that the matrix element is indeed proportional to $\de(\vec P_{\textrm{cm}} - \vec P'_{\textrm{cm}})\de(k-k')$, consistent with the fact that $\cE_v(\hat n)$ preserves energy and momentum. Combining this with eqs.\ \eqref{O_2_E_O_2_int} and  \eqref{eq:O2_FormFactor_momentum}, we find
\be
&\<0|\cO_2(\omega)\cE_v(\hat n)\cO^\dagger_2(\omega)|0\> =  \frac{v^2}{(2\pi)^2}\de(\omega-v^2).
\ee

Finally, normalizing by the two-point function of $\cO_2$
\be\label{eq:2pt_O2_interacting}
\<0|\cO_2(\omega)\cO^\dagger_2(\omega)|0\> = \frac{1}{2\pi\sqrt{\omega}},
\ee
we have
\be
\<\cE_v(\hat n)\>_{\cO_2(\omega)} = \frac{v^3}{2\pi}\de(\omega-v^2).
\ee
This result agrees with the free-theory case, eq.\ \eqref{eq:Ev_Charge2_Free} (for $\vec k=0$). This is not a coincidence-as we will explain later in section \ref{sec:general_properties}, the one-point function of $\cE_v(\hat n)$ in a scalar charge-two state is completely fixed by symmetry and the Ward identities.

\subsubsection{Three-Particle States}\label{sec:unitarity_3body}

We now consider the one-point function of $\cE_v(\hat n)$ in three-particle states. In particular, we will focus on states with two spin-up fermions and one spin-down fermion. We review the details of the state following refs.\ \cite{werner:tel-00285587, Chowdhury:2023oas}.\footnote{The state where all three fermions have the same spin is not affected by the Bethe-Peierls boundary condition, and thus has the same wave-function as the free theory.} It can be expressed as
\be\label{3_state_int}
|\Psi^{l,m,s}_{\vec P_{\textrm{cm}},k}\> = \frac{1}{\sqrt{2}}\int d^3\vec x_1 d^3\vec x_2 d^3\vec x_3\, \Psi^{l,m,s}_{\vec P_{\textrm{cm}},k}(\vec x_i)\psi^{\dagger}_{\uparrow}(\vec x_1)\psi^{\dagger}_{\downarrow}(\vec x_2)\psi^{\dagger}_{\uparrow}(\vec x_3)|0\>.
\ee
The state is labeled by quantum numbers $l$, $m$, and $s$, where $l$ and $m$ are angular momentum quantum numbers. The state depends further on the center of mass momentum $\vec P_{\textrm{cm}}$ and the relative momentum $k$ with energy  $\frac{P_{\textrm{cm}}^2}{6}+k^2$. As explained in section \ref{sec:wave-function}, the wave-function is anti-symmetric under $\vx_1 \leftrightarrow \vx_3$ and obeys the Bethe-Peierls boundary condition when $\vx_1 \to \vx_2$ and $\vx_3 \to \vx_2$. The quantum number $s$ is constrained by the Bethe-Peierls condition as we review below.

To describe the wave-function, let us introduce the coordinates
\be\label{jacobi_coord}
\vec R_{\textrm{cm}} &= \frac{\vec x_1 + \vec x_2 + \vec x_3}{3}, \quad \vec r = \vec x_2 - \vec x_1, \quad \vec \r =\frac{2}{\sqrt{3}}\p{\vec x_3 - \frac{\vec x_1 + \vec x_2}{2}}.
\ee
We further define the hyperradius $R$ and the hyperangular coordinates $\Omega$ as
\be\label{efimov_coord}
R = \sqrt{\frac{r^2 + \r^2}{2}},\quad \Omega=\p{\a,\hat r,\hat \r},\quad \a=\textrm{arctan}\frac{r}{\r}.
\ee
In these coordinates, the wave-function at unitarity $\Psi(\vec x_i)$ is obtained by solving the three-body Schr\"odinger equation via separation of variables leading to the normalized three-body wave-function
\be\label{eq:Charge3_Wave-function_Interacting}
\Psi^{l,m,s}_{\vec P_{\textrm{cm}},k}(\vec x_i) = \p{\frac{2}{\sqrt{3}}}^{3/2}N^l_{k} e^{i\vec P_{\textrm{cm}}\.\vec R_{\textrm{cm}}} \frac{J_s(\sqrt{2}k R)}{R^2}\Phi^{l,m}_s(\Omega), \qquad |N_k^l|^2 = \frac{\sqrt{2}\pi k}{f_s^l},
\ee
where $J_s(\sqrt{2}k R)$ is the Bessel function of the first kind and denotes the hyperradial wave-function, while $\Phi^{l,m}_s(\Omega)$ is the hyperangular wave-function, which is an eigenfunction of the hyperangular Laplacian with eigenvalue $4-s^2$. In this expression, $f_s^l$ denotes the overlap of the hyperangular wave-function and is given by
\be\label{eq:fs_def}
f_s^l\equiv \int d\bar \Omega |\Phi^{l,m}_s(\Omega)|^2, \qquad d\bar \Omega=2\sin^2 (2\a)d\a d^2\hat r d^2 \hat \r.
\ee
We choose the normalization constant $N_k^l$ such that $\int_{\vec x_1,\vec x_2,\vec x_3} |\Psi(\vec x_i)|^2 = 1$. (Note that we have also followed the convention explained in footnote \ref{foot:VRmax_convention}.) The phase space measure for the charge-three sector is given by
\begin{align}\label{cont_lim_3body}
\frac{\sqrt{2}}{(2\pi)^3\pi}\int d^3 \vec P_{\rm cm}\int\limits_0^\infty dk.
\end{align}

Intuitively, the hyperradial wave-function characterizes how fast the wave-function falls off as a function of the hyperradius $R$ and the quantum number $s$ ($\sim R^{s-2}$) as all three coordinates $\vx_i$ are contracted towards $R_{\rm cm}$ with fixed hyperangles, consistent with the Pauli principle. 

The Bethe–Peierls boundary condition constrains the hyperangular part of the wave-function, which may be written compactly as
\be\label{eq:charge3_hyperangular_wavefunc}
\Phi^{l,m}_s(\Omega) = (1-P_{13})\frac{\f_s^l(\a)}{\sin(2\a)}Y_l^m(\hat \r),
\ee
where $Y_l^m$ is the spherical harmonics\footnote{We use the normalization of \cite{Chowdhury:2023oas} for the spherical harmonics, 
\begin{equation}
 Y_l^m(\theta, \varphi) =  e^{i m \varphi } \sqrt{\frac{(l-m)!}{(l+m)!}} P_l^m(\cos (\theta )), \qquad
 \int\! d\Omega\, |Y_l^m(\theta, \varphi)|^2 = \frac{4\pi}{2 l +1} \,,
\end{equation}}, and $P_{13}$ is an operator that exchanges particles $1$ and $3$.  The expressions of $\f_s^l$ for $l=0,1$ are
\be\label{eq:charge3_alpha_wavefunc}
\f_s^{l=0}(\a)=\sin\p{s\left(\tfrac{\pi}{2}-\a\right)},\quad \f_s^{l=1}(\a) = -s\cos\p{s(\tfrac{\pi}{2}-\a)} + \tan \a \sin\p{s(\tfrac{\pi}{2}-\a)}.
\ee
Imposing the boundary condition yields a transcendental equation for the allowed values of $s$. The remaining ingredient is the overlap function $f_s^l$. While the closed form expression for $l=0$ is known, we rely on numerical integration for the higher $l$ values. We summarize the first few $s$ and corresponding $f_s^l$ values for $l=0,1$ in table \ref{table:allowed_s}. 

The charge-three primary operators form an infinite family labeled by the quantum numbers $s, l$ and $m$. They are defined through their matrix elements between the vacuum and the charge-three state
 \begin{equation}\label{O_3_def}
  \< 0 | \cO_3^{l, m, s}(0,\vec R_{\rm cm}) | \Psi^{l,m,s}_{\vec P_{\textrm{cm}},k} \> = \lim_{R\to0} R^{2-s} \int\!d\bar\Omega\, \Psi^{l,m,s}_{\vec P_{\textrm{cm}},k}(\vec R_{\rm cm}, R, \Omega) \Phi_s^{l, m*}(\Omega).
\end{equation}
These primary operators have dimensions $\Delta_{\cO^{l, m, s}_3}= s+\frac{5}{2}$. The matrix elements at an arbitrary time $t$ are obtained by using $\cO(t)= e^{i H t} \cO(0) e^{-i Ht}$. We now proceed to evaluate our detector operator in states created by the charge-three primary operator.
 \begin{table}
	\begin{center}
		\begin{tabular}{|c|c|}
 \hline
  $s$& $f_s^{l=0}$ \\		
			\hline
			$~2.16622~$ & $~15.9415~$ \\
			\hline
			 5.12735 & 554.016 \\
			\hline
			 7.11448 & 443.969 \\
			\hline
          8.83225 & 457.113 \\
          \hline
		\end{tabular}
  \qquad\qquad
  \begin{tabular}{|c|c|}
  \hline
  $s$& $f_s^{l=1}$ \\
  \hline
       $~1.77272~$ & $~345.377~$ \\
			\hline
			 4.35825 & 2174.79\\
    \hline
			 5.71643 & 3915.43\\
			\hline
			 8.05319 & 11514.6\\
    \hline
  \end{tabular}
  \end{center}
\caption{Values of $s$ and $f_s^l$ for $l=0,1$.}
\label{table:allowed_s}
 \end{table}

The expectation value of the energy detector operator in the charge-three primary states takes the form
\begin{align}\label{O_3_E_O_3_int}
&\langle 0| \cO^{l,m,s}_3(\omega){\mathcal E}_v({\hat n})\cO^{l,m,s~\dagger}_3(\omega') |0\rangle\nonumber\\
&=\frac{2}{\left(2\pi \right)^6 \pi^2}\int d^3 \vec P_{\rm cm}\,d^3\vec P'_{\rm cm} \int\limits_0^\infty dk\, dk'\,\nonumber\\
&\times \langle \Psi^{l,m,s}_{\vec P_{\mathrm{cm}},k}| {\mathcal E}_v({\hat n}) |\Psi^{l,m,s}_{\vec P'_{\mathrm{cm}},k'} \rangle  \langle 0| \cO^{l,m,s}_3(\omega)|\Psi^{l,m,s}_{\vec P_{\mathrm{cm}},k} \rangle \langle \Psi^{l,m,s}_{\vec P'_{\mathrm{cm}},k'}| \cO^{l,m,s,~\dagger}_3(\omega')|0 \rangle,
\end{align}
where the form factor is given by
\be\label{eq:O3_int_formfactor}
\<0|\cO^{l,m,s}_3(\omega)|\Psi^{l,m,s}_{\vec P_{\textrm{cm}},k}\> = \p{\frac{2}{\sqrt{3}}}^{\frac{3}{2}}N_k^l f_s^l \frac{k^s}{2^{\frac{s}{2}}\G(s+1)}(2\pi)^4\de^{(3)}(\vec P_{\textrm{cm}})\de(\omega-( \tfrac{P_{\textrm{cm}}^2}{6}+ k^2)), 
\ee
with the normalization $N_k^l$  defined in eq.\ \eqref{eq:Charge3_Wave-function_Interacting}. Similar to the two-particle case, the matrix elements of $\cE_v$ between interacting three-particle states $|\Psi^{l,m,s}_{\vec P_{\text{cm}},k}\>$ can be evaluated using eq.\ \eqref{3_state_int},
\be
&\<\Psi^{l,m,s}_{\vec P_{\textrm{cm}},k}|{\mathcal E}_v(\hat n)|\Psi^{l',m',s'}_{\vec P'_{\textrm{cm}},k'}\>  \\
&=\frac{v}{2} \lim_{r\to \oo} r^9 e^{i\p{\frac{P_{\mathrm{cm}}^2-P_{\mathrm{cm}}^{\prime2}}{6}+k^2-k^{\prime2}}\frac{r}{v}} \nn \\
&\times \int_{\vec y,\vec z} \left(2\Psi^{l,m,s*}_{\vec P_{\textrm{cm}},k}(r \vec y, r \vec z, r \hat n)\Psi^{l',m',s'}_{\vec P'_{\textrm{cm}},k'}(r \vec y, r \vec z, r \hat n)+\Psi^{l,m,s *}_{\vec P_{\textrm{cm}},k}(r \vec y, r \hat n,r \vec z)\Psi^{l',m',s'}_{\vec P'_{\textrm{cm}},k'}(r \vec y, r \hat n,r \vec z)\right).\nn
\ee
The two terms correspond to the two ways a particle can reach the detector at $r \hat n$. The first term, carrying a relative factor of two, accounts for either of the two spin-up particles arriving at the detector, while the second accounts for the single spin-down particle. 

Next, we plug in the explicit form of the wave-function eq.\ \eqref{eq:Charge3_Wave-function_Interacting} and take the detector limit. In this limit, the hyperangular wave-function $\Phi_s^{l,m}(\Omega)$ remains unchanged, which is evident from the explicit expressions of the hyperangular coordinates defined in eqs.\ \eqref{jacobi_coord} and \eqref{efimov_coord}. As in the two-particle case, the matrix element can be evaluated using the distributional identities eq.\ \eqref{eq:distribution_identity}.\footnote{In this case, one also needs an additional identity
\be
e^{ik^2\frac{r}{v}}\sin(k x_0 r) \sim -\frac{1}{2i}e^{-\frac{iv r}{4}x_0^2}\sqrt{\frac{\pi v}{r}}e^{i\frac{\pi}{4}}\de(k-\tfrac{v x_0}{2}),\qquad r \to \oo.
\ee
} This gives
\be\label{eq:Ev_Charge3_matrix_expr0}
\<\Psi^{l,m,s}_{\vec P_{\textrm{cm}},k}|\cE_v(\hat n)|\Psi^{l',m',s'}_{\vec P'_{\textrm{cm}},k'}\>  =&\frac{6\sqrt{3}\pi^3v^{10}}{k^6}N^{l*}_kN^{l'}_{k'}e^{\frac{i\pi}{2}(s-s')}\de^{(3)}(\vec P_{\textrm{cm}}-\vec P'_{\textrm{cm}})\de\left(k-k'\right) \\
&\times\int_{\vec y,\vec z}\de^{(3)}\left(\vec P_{\textrm{cm}}-v(\vec y+\vec z+\hat n)\right)\de\left(k-\tfrac{v R_{\vec y,\vec z,\hat n}}{\sqrt{2}}\right)\nn \\
&\times\p{2\Phi^{l,m*}_s(\Omega_{\vec y,\vec z,\hat n})\Phi^{l',m'}_{s'}(\Omega_{\vec y,\vec z,\hat n})+\Phi^{l,m*}_s(\Omega_{\vec y,\hat n,\vec z})\Phi^{l',m'}_{s'}(\Omega_{\vec y,\hat n,\vec z})}. \nn
\ee
where $R_{\vec y,\vec z,\hat n}$ and $R_{\vec y,\hat n,\vec z}$ denote the hyperradius evaluated from eq.\ \eqref{efimov_coord} with the identifications $(\vx_1,\vx_2,\vx_3)=(\vec y,\vec z,\hat n)$ and $(\vx_1,\vx_2,\vx_3)=(\vec y,\hat n,\vec z)$, respectively, and the hyperangular coordinates are assigned analogously. We also recover the delta functions preserving energy and momentum. 

To simplify the integral over $\vec y,\vec z$, we make a change of variables
\be
\vec p = \vec y + \vec z + \hat n,\quad \vec q = \vec y + \frac{1}{2}\hat n - \frac{1}{2v}\vec P_{\textrm{cm}}.
\ee
The delta functions in eq.\ \eqref{eq:Ev_Charge3_matrix_expr0} localize $\vec p$ and $|\vec q|$ to be
\be\label{eq:localisation_p_|q|}
\vec p =\frac{\vec P_{\textrm{cm}}}{v},\quad |\vec q| = \frac{1}{v}\sqrt{k^2 - \frac{3}{4}v^2+\frac{v \hat n\.\vec P_{\textrm{cm}}}{2} - \frac{\vec P_{\textrm{cm}}^2}{12}},
\ee
and we are left with an angular integral over $\hat q$. Then, eq.\ \eqref{eq:Ev_Charge3_matrix_expr0} becomes
\be
&\<\Psi^{l,m,s}_{\vec P_{\textrm{cm}},k}|\cE_v(\hat n)|\Psi^{l',m',s'}_{\vec P'_{\textrm{cm}},k'}\>  \\
&=\frac{6\sqrt{3}\pi^3 v^4}{k^4}\sqrt{1-\frac{3v^2}{4k^2}+\frac{v \hat n\.\vec P_{\textrm{cm}}}{2k^2} - \frac{\vec P_{\textrm{cm}}^2}{12k^2}}N^{l*}_kN^{l'}_{k'}e^{\frac{i\pi}{2}(s-s')}\de^{(3)}(\vec P_{\textrm{cm}}-\vec P'_{\textrm{cm}})\de(k-k') \nn \\
& \left.\int d\Omega_{\hat q}\p{2\Phi^{l,m*}_s(\Omega_{\vec y,\vec z,v\hat n})\Phi^{l',m'}_{s'}(\Omega_{\vec y,\vec z,v\hat n})+\Phi^{l,m*}_s(\Omega_{\vec y,v\hat n,\vec z})\Phi^{l',m'}_{s'}(\Omega_{\vec y,v\hat n,\vec z})}\right|_{\substack{\vec y =\vec q + \frac{\vec P_{\textrm{cm}}}{2}-\frac{v\hat n}{2},~ \vec z = -\vec q + \frac{\vec P_{\textrm{cm}}}{2}-\frac{v\hat n}{2} \\|\vec q| =\sqrt{k^2 - \frac{3}{4}v^2+\frac{v \hat n\.\vec P_{\textrm{cm}}}{2} - \frac{\vec P_{\textrm{cm}}^2}{12}}}}.\nn
\ee

Together with the form factor eq.\ \eqref{eq:O3_int_formfactor} and performing the $\vec P_{\rm cm}, \vec P'_{\rm cm}, k$ and $k'$ integrals in eq.\ \eqref{O_3_E_O_3_int}, the normalized one-point function of $\cE_v(\hat n)$ in the state created by $\cO^{l,m,s}_3$ is given by
\be\label{eq:Ev_generaO3_normalized}
&\<\cE_v(\hat n)\>_{\cO_3^{l,m,s}(\omega)}  \nn \\
&=\frac{3\sqrt{3}v^4}{2f^l_s\omega^{\frac{3}{2}}}\sqrt{1-\frac{3v^2}{4\omega}}  \left.\int d\Omega_{\hat q}\p{2\left|\Phi^{l,m}_s(\Omega_{\vec y,\vec z,v\hat n})\right|^2+\left|\Phi^{l,m}_s(\Omega_{\vec y,v\hat n,\vec z})\right|^2}\right|_{\substack{\vec y =\vec q -\frac{v\hat n}{2},~ \vec z = -\vec q -\frac{v\hat n}{2} \\|\vec q| =\sqrt{\omega - \frac{3}{4}v^2}}}.
\ee
where we use the normalization of the two-point function of the $\cO^{l,m,s}_3$ operator
\be
&\< 0 |\cO_3^{l,m,s}(\omega)\cO_3^{l,m,s \dagger}(\omega)| 0 \> =\frac{\pi}{3\sqrt{3}}\frac{2^{4-s} f^l_s}{\Gamma(s+1)^2}\omega^s.
\ee
The localization conditions eq.~\eqref{eq:localisation_p_|q|} require $|\vec q|$ to be real, yielding the same velocity bound as in the free case eq.~\eqref{eq:O3_l0_free}. The maximal velocity thus also follows directly from the interacting calculation. For future convenience we also note the expression of the hyperangle $\alpha$ for both the cases in terms of $ v, \hat n$, $\omega$ and $\hat q$.
\be\label{eq:O3_int_alpha}
\alpha_{\vec y,\vec z,v\hat n}&= \tan^{-1}\left[ \frac{\sqrt{4\omega-3v^2}}{\sqrt{3}v} \right], \quad P_{13} \alpha_{\vec y,\vec z,v\hat n}= \tan^{-1}\left[ \frac{|3 v \hat n + \hat q \sqrt{4\omega - 3 v^2}|}{\sqrt{3}| v \hat n - \hat q \sqrt{4\omega - 3 v^2}|}\right], \\
\alpha_{\vec y,v\hat n,\vec z}&= \tan^{-1}\left[ \frac{|3 v \hat n - \hat q \sqrt{4\omega - 3 v^2}|}{\sqrt{3}| v \hat n + \hat q \sqrt{4\omega - 3 v^2}|}\right] , \quad P_{13} \alpha_{\vec y,v\hat n,\vec z}= \tan^{-1}\left[ \frac{|3 v \hat n + \hat q \sqrt{4 \omega - 3 v^2}|}{\sqrt{3}| v \hat n - \hat q \sqrt{4 \omega - 3 v^2}|}\right].\nn
\ee

The $\cE_v(\hat n)$ one-point function can be computed once we plug in the explicit form of the hyperangular wave-function.

\subsubsection*{$l=0$ state}

We start with the simplest case, where the operator is the leading charge-three scalar operator. This operator has quantum numbers $l=m=0$ and $s=s_0\approx 2.16622$ \cite{Chowdhury:2023oas}. Using the symmetry constraints eq.\ \eqref{constraints_symm} and the fact that the Fourier transform of the two-point function scales as $\sim \omega^{\Delta_O- \frac{5}{2}}$, the  normalized one point function of the detector in the states created by $\cO_3^{0,0,s_0}(\omega) \equiv \cO_3^{s_0}(\omega)$ is given by
\be
\<\cE_v(\hat n)\>_{\cO_3^{s_0}(\omega)} &= \frac{\sqrt{\omega}}{4\pi}\r\biggl(\frac{v}{\sqrt{\omega}}\biggr).
\ee
Since the $\omega$-scaling is fixed by symmetry, in what follows, we simply set $\omega=1$. 

\begin{figure}
	\centering
	\includegraphics[scale=0.7]{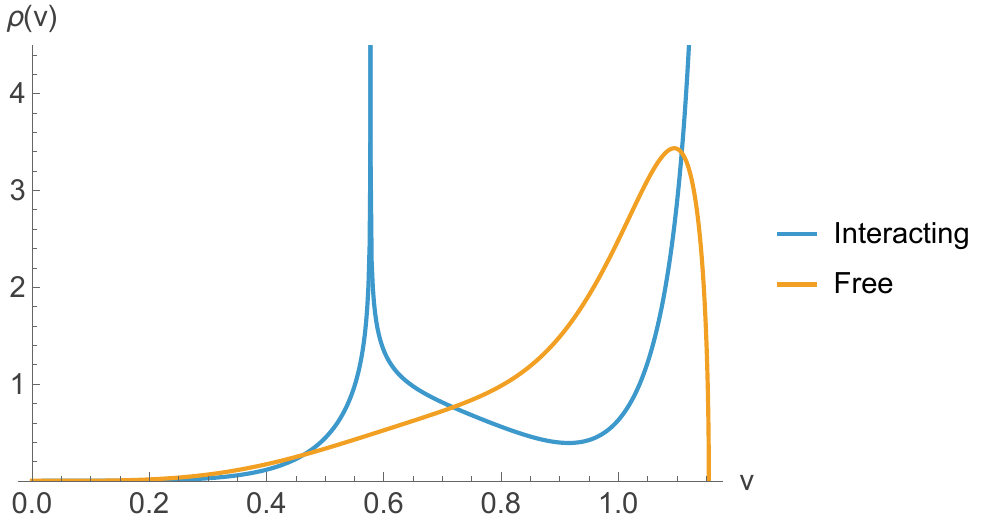}
	\caption{Distribution of velocity in the state created by leading charge-three scalar operator $\cO_3^{s_0}(\omega)$ in the interacting theory and free theory. Both curves satisfy the Ward identities $\int dv \r(v) = 1$ and $\int dv \frac{2}{v^2}\r(v) = 3$. The $\r(v)$ for interacting theory has divergences at $v=\sqrt{\frac{1}{3}}$ and $v=\sqrt{\frac{4}{3}}$.}
	\label{fig:charge3_interacting_rhov}
\end{figure}

\begin{figure}
\begin{center}
\includegraphics[width=0.8\linewidth]{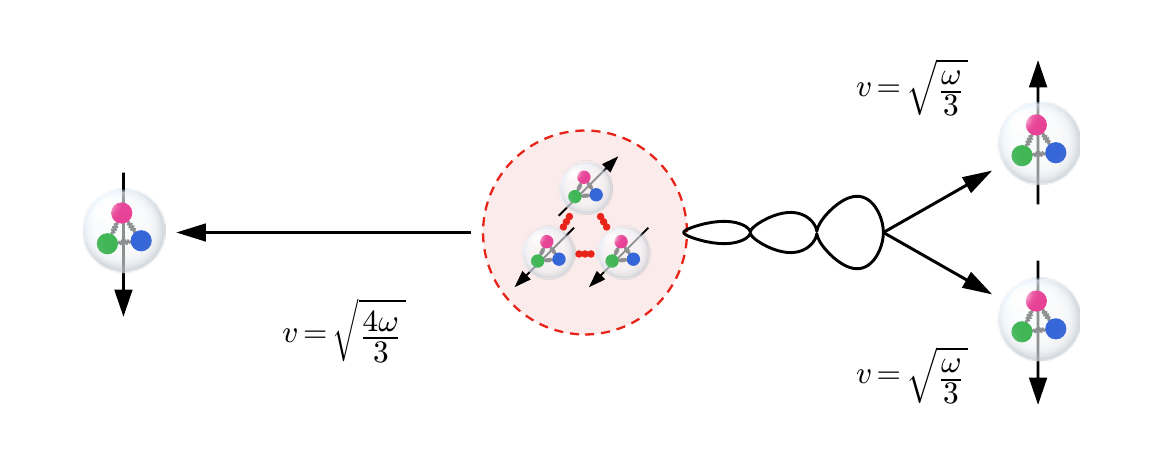}
\caption{As compared to the case of free fermions, due to resonant interactions, collinear fermions interact, producing a non-trivial distribution for $\mathcal{E}_v$, with enhancements at $v=\sqrt{\frac{1}{3}}$ and $v=\sqrt{\frac{4}{3}}$. }
\label{interacting_fermions_configuration}
\end{center}
\end{figure}

By eq.\ \eqref{eq:Ev_generaO3_normalized}, the function $\r(v)$ for the leading charge-three scalar operator is given by
\be
\r(v)&=\frac{6\sqrt{3} \pi\,v^4}{f^l_s}\sqrt{1-\frac{3v^2}{4}}  \left.\int d\Omega_{\hat q}\p{2\left|\Phi^{0}_s(\Omega_{\vec y,\vec z,v\hat n})\right|^2+\left|\Phi^{0}_s(\Omega_{\vec y,v\hat n,\vec z})\right|^2}\right|_{\substack{\vec y =\vec q -\frac{v\hat n}{2},~\vec z = -\vec q  -\frac{v\hat n}{2},\\ |\vec q| = \sqrt{1 - \tfrac{3}{4}v^2}}}.
\ee
The remaining angular integrals can be evaluated numerically. We plot the result in figure \ref{fig:charge3_interacting_rhov}. For comparison, we also plot the corresponding result for the $l=0$ charge-three operator in the free theory obtained in section \ref{sec:free_3particle} (see eq.\ \eqref{eq:O3_l0_free}). For the charge-$3$ state, the $\r(v)$ functions in the two theories exhibit very different behavior, unlike in the charge-$2$ state where they are the same.

We have also checked that the result obeys the Ward identities eqs.\ \eqref{eq:Ward_E_generalk} and \eqref{eq:Ward_P_N_generalk}, which imply
\be
\int dv \r(v) = 1,\qquad \int dv \frac{2}{v^2}\r(v) = 3.
\ee

One crucial difference between interacting case and the free theory is that the $\r(v)$ function for the interacting case has integrable divergences at $v=\sqrt{\frac{1}{3}}$ and at $v=\sqrt{\frac{4}{3}}$. Both the divergences originate from the kinematic configuration shown in figure \ref{interacting_fermions_configuration}, where one of the spin-up fermions gets close to the spin-down fermion, and they both go in the opposite direction of the other spin-up fermion. This kinematic configuration is the same as the maximal velocity configuration depicted in figure \ref{rhov_l1_free_kin}.

Let us analyze the divergences near the two peaks. Consider the divergence near $v=\sqrt{\frac{4}{3}}$, the maximal velocity of the particle reaching our detector for a three-body initial state. For this, the only possible configuration of the three particles is depicted in figure \ref{interacting_fermions_configuration} where particles having opposite spins approach each other.  The source of the divergence, therefore, must stem from the Bethe-Peierls boundary condition that we have imposed on the hyperangular wave-function. Suppressing the angular momentum and $s$ labels, we can write the hyperangular wave-function in eq.\ \eqref{eq:charge3_hyperangular_wavefunc} as
\be
\Phi(\Omega) = \Phi_{12}(\Omega) - \Phi_{32}(\Omega),
\ee
where the subscript denotes the positions of the pair of particles on which the unitarity condition is imposed. The two contributions are related by $\Phi_{32}(\Omega)=P_{13}\Phi_{12}(\Omega)$. When the hyperangular coordinate $\a$ becomes small (i.e., $\vec x_1 \to \vec x_2$), the hyperangular wave-function $\Phi_{12}(\Omega)$ has the behavior
\be
\Phi_{12}(\Omega) \sim \frac{1}{\a},
\ee
due to the Bethe-Peierls boundary condition. 

As $v$  approaches the maximal velocity $\sqrt{\frac{4}{3}}$, it follows from eq.\ \eqref{eq:O3_int_alpha} that the hyperangular coordinate $\a_{\vec y, \vec z, v\hat n}$ in the integrand of eq.\ \eqref{eq:Ev_generaO3_normalized} goes like $\a_{\vec y, \vec z, v\hat n} \sim \sqrt{\frac{4}{3}-v^2}$, consistent with the kinematics of figure \ref{interacting_fermions_configuration}. Using this fact, we see that in eq.\ \eqref{eq:Ev_generaO3_normalized}, we have the leading divergent piece
\be
\sqrt{1-\frac{3}{4}v^2}\, \left|\Phi_{12}(\Omega_{\vec y,\vec z,v\hat n})\right|^2 \sim \frac{1}{\sqrt{\frac{4}{3}-v^2}},\qquad v \to \sqrt{\frac{4}{3}}.
\ee
Note that the other terms in the integrand of eq.\ \eqref{eq:Ev_generaO3_normalized} give finite contributions in this limit. In summary, near the peak at $v=\sqrt{\frac{4}{3}}$, $\r(v)$ has the divergent behavior
\be
\r(v) \sim \frac{1}{\sqrt{\frac{4}{3}-v^2}},\qquad v\to \sqrt{\frac{4}{3}}.
\ee
This endpoint singularity is the Watson-Migdal enhancement due to final-state interactions~\cite{Watson:1952ji, Migdal:1955}.

On the other hand, when the velocity of the particle reaching the detector is $\sqrt{\frac{1}{3}}$, the collinear kinematics shown in figure \ref{interacting_fermions_configuration} is not the only kinematically allowed possibility, and one has to integrate over all possible configurations. In particular, consider the $\Phi_{32}(\Omega)$ wave-function and the region where $\hat q$ is anti-parallel to $\hat n$ in the angular integral in eq.\ \eqref{eq:Ev_generaO3_normalized}. In this configuration, let us look at the behavior of the hyperangular coordinate $\a_{v\hat n,\vec z,\vec y}=P_{13}\a_{\vec y,\vec z,v\hat n}$ given by eq.\ \eqref{eq:O3_int_alpha}, which can be written as
\be
P_{13}\a_{\vec y,\vec z,v\hat n}=\textrm{tan}^{-1}\p{\sqrt{\frac{3v^2+2+3v\sqrt{4-3v^2}\cos\th_q}{-3v^2+6-3v\sqrt{4-3v^2}\cos\th_q}}},
\ee
where $\cos\th_q=\hat q\.\hat n$. In the limit $v\to \sqrt{\frac{1}{3}}$ and $\th_q \to \pi$, the hyperangular coordinate $P_{13}\a_{\vec y,\vec z,v\hat n}$ and the hyperangular wave-function $\Phi_{32}(\Omega)$ behave like
\be
P_{13}\a_{\vec y,\vec z,v\hat n} \sim \sqrt{\frac{3}{2}(\pi-\th_q)^2 + 6(\tfrac{1}{3}-v^2)^2 + \ldots}\, , \qquad \Phi_{32}(\Omega) \sim \frac{1}{P_{13}\a}.
\ee 
As a result, the angular integral near $\th_q = \pi$ gives
\be
\int d\th_q \sin\th_q \left| \Phi_{32}(\Omega_{\vec y,\vec z,v\hat n})\right|^2 &\sim\int d\th_q \sin\th_q\p{\frac{1}{\sqrt{\frac{3}{2}(\pi-\th_q)^2 + 6(\frac{1}{3}-v^2)^2 + \ldots}}}^2, \nonumber\\
&\sim  \log{\frac{1}{\left|\frac{1}{3}-v^2\right|}}\,,
\ee
as $v\to \sqrt{\frac{1}{3}}$. Therefore, this divergence again results from the kinematic configuration depicted in figure \ref{rhov_l1_free_kin}, where the particle reaching the detector has velocity $v=\frac{1}{\sqrt{3}}$ instead of the maximal velocity. We obtain similar divergences from the other permutation terms from $\left|\Phi(\Omega_{\vec y,v \hat n,\vec z})\right|^2$ in eq.\ \eqref{eq:Ev_generaO3_normalized} as well. In particular, the term $\left|\Phi_{12}(\Omega_{\vec y,v \hat n,\vec z})\right|^2$ gives rise to similar logarithmic divergences from the $\th_q = 0$ region, and the term $\left|\Phi_{32}(\Omega_{\vec y,v \hat n,\vec z})\right|^2$ gives the same divergences near $\th_q=\pi$. For example, for the $\left|\Phi_{12}(\Omega_{\vec y,v \hat n,\vec z})\right|^2$ term we have, as $v\to \sqrt{\frac{1}{3}}$,
\begin{align}
\int d\th_q \sin\th_q \left| \Phi_{12}(\Omega_{\vec y, v\hat n,\vec z})\right|^2 &\sim\int d\th_q \sin\th_q\p{\frac{1}{\sqrt{\frac{3}{2}\th_q^2 + 6(\frac{1}{3}-v^2)^2 + \ldots}}}^2, \nonumber\\
&\sim  \log{\frac{1}{\left|\frac{1}{3}-v^2\right|}}.
\end{align}
In summary, near the peak at $v=\sqrt{\frac{1}{3}}$, $\r(v)$ has the divergent behavior
\be
\r(v) \sim  \log{\frac{1}{\left|\frac{1}{3}-v^2\right|}},\qquad v\to \sqrt{\frac{1}{3}}\,.
\ee

It would be interesting to investigate the behavior of the detector correlator near these singular regions in more detail. An interesting feature of the one-point correlator $\mathcal{E}_v$, as compared to $\mathcal{E}$, is that the additional measurement of $v$ introduces an additional scale. While we focus in this paper on the case of a genuine NRCFT, it is expected that scattering length corrections would become important precisely in the regions $v\to \sqrt{4/3}$ and $v\to \sqrt{1/3}$, ultimately smoothing out the divergence. These corrections might play an important role in measurements of $\mathcal{E}_v$ using slow neutrons.

\subsubsection{Nonrelativistic Hofman-Maldacena Bounds}\label{sec:NR_bounds}

One of the most important theoretical consequences of the study of conformal collider physics \cite{Hofman:2008ar} was the derivation of the conformal collider bounds, obtained by imposing positivity of the ANE in states created by spinning operators, in particular conserved currents, $J$, and the stress tensor, $T$. These constraints have been generalized in a number of different directions, as reviewed in the introduction.

In this section, we compute the one-point correlator $\mathcal{E}_v(\hat n)$ in a variety of different spin 1 states in the theory of fermions at unitarity. We will see that the nonrelativistic analog of the conformal collider bounds are indeed satisfied pointwise in velocity. This suggests that these bounds could provide useful constraints on the space of NRCFTs.

\begin{figure}[t]
  \centering
  \begin{minipage}{0.48\textwidth}
    \centering
    \includegraphics[width=\linewidth]{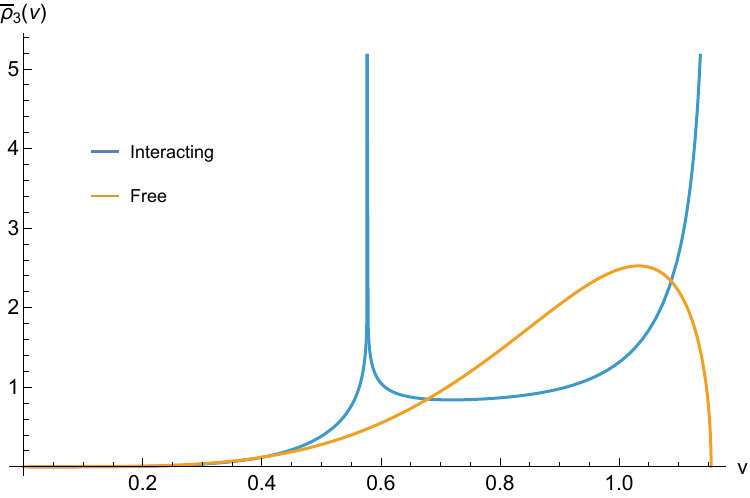}
  \end{minipage}
  \hspace{2.5mm}
  \begin{minipage}{0.48\textwidth}
    \centering
    \includegraphics[width=\linewidth]{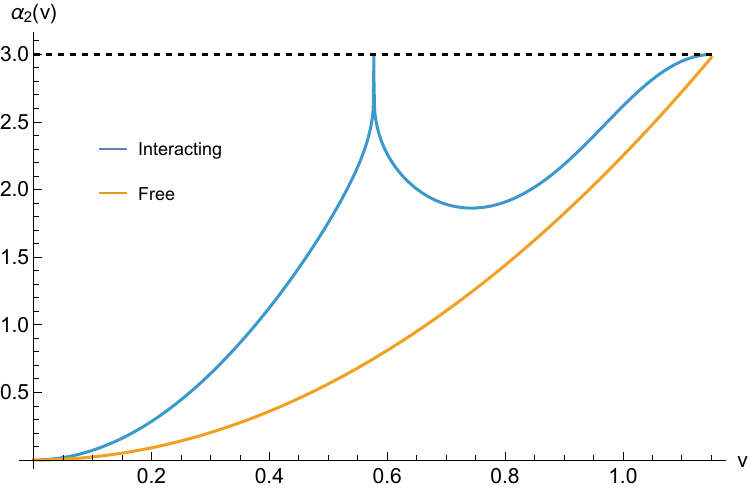}
  \end{minipage}
  \caption{$\bar \r_3(v)$ and $\a_2(v)$ functions for the leading $l=1$ charge-three operator in fermions at unitarity and free fermion theory. For both theories, we have $\int dv \bar \r_3(v) = 1, \int dv \frac{2}{v^2}\bar \r_3(v) = 3$ due to the Ward identities. The black dashed line on the right plot is the upper bound of the generalized Hofman-Maldacena bound.}
  \label{fig:rhobar_withalpha2plot}
\end{figure}
The leading charge-three, $l=1$, operator $\cO_{3}^{i}$ carries $s=s_1\approx 1.77272$. We will study a state created by $\e \cdot  \cO^{s_1}_{3}(\omega) $, where $\e_i$ is a normalized polarization vector. By rotational invariance, $\<\cE_v(\hat n)\>_{\e\.\cO^{s_1}_{3}(\omega)}$ is a function of the angle between $\e$ and $\hat n$, which we denote as $\th$. We can fix $\hat n=\hat z$, and the $\th$-dependence should be given by
\be\label{eq:Ev_O3_indifferentm}
\<\cE_v(\hat n)\>_{\e\.\cO^{s_1}_{3}(\omega)} = \cos^2\th \<\cE_v(\hat n)\>_{\cO^{0, s_1}_3(\omega)} + \sin^2\th \<\cE_v(\hat n)\>_{\cO^{\pm, s_1}_3(\omega)},
\ee
where  for notational convenience, $\cO^{ l=1, m=0, s_1}_{3} \equiv \cO^{0, s_1}_3$ and $\cO^{ l=1, m=\pm 1, s_1}_{3} \equiv \cO^{\pm, s_1}_3$  denote the different harmonics of the spin 1 operator. In arriving at this result, we have also used the fact that the matrix elements $\<0|\cO^{l=1,m}_{3}\cE_v(\hat n)\cO^{l=1,m'\dagger}_{3}|0\>$ vanish for $m\neq m'$, and the matrix element for $m=-1$ is the same as $m=1$.

Similar to the $l=0$ case, we can use the symmetry constraints to write the one-point function as 
\be
\<\cE_v(\hat n)\>_{\e\.\cO^{s_1}_{3}(\omega)} = \frac{\sqrt{\omega}}{4\pi}\bar \r_3\biggl(\frac{v}{\sqrt{\omega}}\biggr)\p{1+\a_2\biggl(\frac{v}{\sqrt{\omega}}\biggr)\p{\cos^2\th-\frac{1}{3}}}.
\ee
The two functions $\bar \r_3\left(\frac{v}{\sqrt{\omega}}\right)$ and $\a_2\left(\frac{v}{\sqrt{\omega}}\right)$ are
\be\label{eq:rho3_alpha2_expr0}
\bar \r_3\biggl(\frac{v}{\sqrt{\omega}}\biggr) &=  \frac{4\pi}{3 \sqrt{\omega}}\p{\<\cE_v(\hat n)\>_{\cO^{0, s_1}_{3}(\omega)}+2\<\cE_v(\hat n)\>_{\cO^{+, s_1}_{3}(\omega)}},\nn\\
\a_2\biggl(\frac{v}{\sqrt{\omega}}\biggr) &= \frac{3\left(\<\cE_v(\hat n)\>_{\cO^{0, s_1}_{3}(\omega)}-\<\cE_v(\hat n)\>_{\cO^{+, s_1}_{3}(\omega)}\right)}{\<\cE_v(\hat n)\>_{\cO^{0,s_1}_{3}(\omega)}+2\<\cE_v(\hat n)\>_{\cO^{+,s_1}_{3}(\omega)}}.
\ee
Each $\<\cE_v(\hat n)\>_{\cO^{l,m, s_1}_{3}(\omega)}$ can be computed using eq.\ \eqref{eq:Ev_generaO3_normalized} and the $l=1$ hyperangular wave-function eqs.\ \eqref{eq:charge3_hyperangular_wavefunc} and \eqref{eq:charge3_alpha_wavefunc}. The constraints from the energy and particle number Ward identities are
\be 
\int\! dv\, \bar \r_3(v) = 1,\qquad  \int\! dv\, \frac{2}{v^2}\bar \r_3(v) = 3.
\ee
We plot the $\bar \r_3(v)$ and $\a_2(v)$ functions for the leading $l=1$ operator in figure \ref{fig:rhobar_withalpha2plot}, together with the corresponding functions in free fermion theory (see eq.\ \eqref{En_free_O3_l1}). We observe that $\bar \r_3(v)$ has divergences at $v=\sqrt{\frac{1}{3}}, \sqrt{\frac{4}{3}}$ similar to the $l=0$ case. More precisely, we find that these divergences only show up in the $m=0$ term in eq.\ \eqref{eq:rho3_alpha2_expr0}, and the $m=1$ term is finite for all allowed values of $v$.

The positivity condition $\cE_v(\hat n)\geq 0$ proposed in section \ref{sec:positivity} implies that the $\a_2(v)$ function satisfies a generalized Hofman-Maldacena bound
\be
-\frac{3}{2}\leq \a_2(v)\leq 3,
\ee
and we find that it is indeed satisfied for the leading $l=1$ operator for all $v$. The upper bound is saturated at $v=\sqrt{\frac{1}{3}}$ and $v=\sqrt{\frac{4}{3}}$. As one can see from eq.\ \eqref{eq:rho3_alpha2_expr0}, this is due to the divergences of the $m=0$ term.

Finally, we can integrate over $v$ to get the familiar expression of the one-point energy correlator
\be\label{eq:spin1_EC_witha2}
\<\cE(\hat n)\>_{\e\.\cO_{3}(\omega)} = \frac{\omega}{4\pi}\p{1+a_2(\cos^2\th-\frac{1}{3})},
\ee
and we find
\be
a_2=\int\! dv\, \bar\r_3(v)\a_2(v) \approx 2.512.
\ee
For comparison, in free theory we have $a_2=15/8=1.875$.

\begin{figure}[t]
  \centering
  \begin{minipage}{0.48\textwidth}
    \centering
    \includegraphics[width=\linewidth]{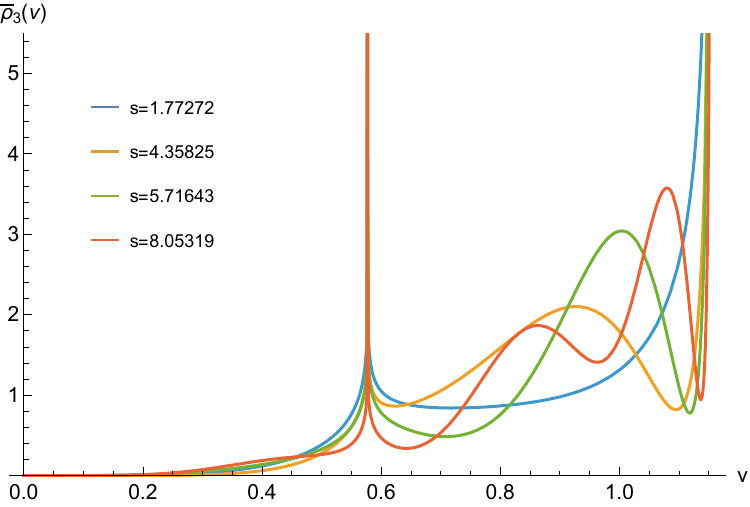}
  \end{minipage}
  \hspace{2.5mm}
  \begin{minipage}{0.48\textwidth}
    \centering
    \includegraphics[width=\linewidth]{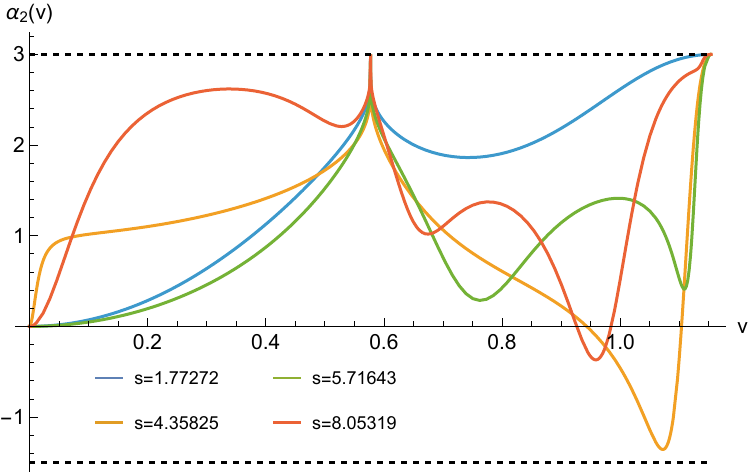}
  \end{minipage}
  \caption{$\bar \r_3(v)$ and $\a_2(v)$ functions for the four lowest-dimension $l=1$ charge-three operators in fermions at unitarity. All four operators satisfy $\int dv \bar \r_3(v) = 1, \int dv \frac{2}{v^2}\bar \r_3(v) = 3$. The black dashed lines on the right plot correspond to the generalized Hofman-Maldacena bound.}
  \label{fig:rhobar_withalpha2_highers}
\end{figure}

It is straightforward to generalize the calculation to other $l=1$ charge-three operators with higher $s$. In particular, the next three values of $s$ are approximately given by $s\approx 4.35825$, 5.71643, and 8.05319. In figure \ref{fig:rhobar_withalpha2_highers}, we show the $\bar \r_3(v)$ and $\alpha_2(v)$ for the first four $l=1$ operators (including the leading one). At higher $s$, both the $\bar \r_3(v)$ and $\a_2(v)$ exhibit interesting oscillatory behaviors between the two divergences at $v=\sqrt{\frac{1}{3}}$ and $v=\sqrt{\frac{4}{3}}$. It would be very interesting to establish a more precise connection between the oscillations and the corresponding hyperangular wave-function.

Interestingly, we find that for the operator with $s \approx 4.35825$, the $\a_2(v)$ function can get quite close to the lower bound $-3/2$. This is because one gets $\a_2(v)=-3/2$ when the $m=0$ contribution vanishes, and we find that for $s\approx 4.35825$, the $m=0$ term becomes much smaller than the $m=1$ term near $v\sim 1.07$, which leads to $\a_2(v)$ being close to $-3/2$. It would be interesting to better understand why we get such a minimum, and if we can find examples which saturate the lower bound. Finally, we can compute the $a_2$ coefficient for the integrated energy detector defined in eq.\ \eqref{eq:spin1_EC_witha2}. We find
\be
a_2^{s=4.35825}\approx 0.826,\quad a_2^{s=5.71643}\approx 1.447,\quad a_2^{s=8.05319}\approx 1.618.
\ee

It would be interesting to find NRCFTs which saturate the given bounds, or which exhibit some form of special behavior. Another interesting class of theories to investigate are nonrelativistic anyons. These have well studied supersymmetric analogs \cite{Nakayama:2008qz,Nakayama:2009cz,Lee:2009mm,Doroud:2015fsz,Doroud:2016mfv}, which might also provide a simplified setting to improve the understanding of the structure of the bounds.

In this section we have simply illustrated that there are analogs of the conformal collider bounds in NRCFTs, and that they are satisfied in a highly non-trivial manner in the particular case of fermions at unitarity. It would be particularly interesting to explore the implications of the nonrelativistic conformal collider bounds on general NRCFTs. For example, by generalizing the calculation performed in appendix \ref{app:ward_id_flat} to spin-1 states, we can translate the bounds on $\a_2(v)$ into bounds on the integral of NRCFT three-point functions over the cross ratio with a velocity-dependent kernel. The main power of the relativistic conformal collider bounds rests on the ability to interpret them as constraints on coefficients in the three-point functions of currents or stress tensors, which in many cases can be interpreted in terms of anomaly coefficients. It would be interesting to similarly classify the different structures of three-point functions of spin-1 and spin-2 currents in NRCFTs, and their relation to anomalies. It would also be interesting to generalize the discussion here to the case of multi-point positivity \cite{Mecaj:2026kji,Mecaj:2025ecl,Belin:2026wkc,Dempsey:2025yiv}. We leave these directions to future work.

\subsubsection{Large-Charge States}

We now consider the situation when we have a many-body initial state. In particular, we will compute the one-point function $\<\cE_v(\hat n)\>$ in a state created by the leading charge-$Q$ operator $\cO_Q$, and take the $Q \to \oo$ limit. We will evaluate the distribution using the large-charge EFT \cite{Beane:2024kld}, which was reviewed in section \ref{sec:large_charge}. 

When computing detector correlators, a common step in both nonrelativistic and relativistic CFTs is performing the Fourier transform of the correlation function. For the relativistic case, the authors of ref.\ \cite{Cuomo:2025pjp} pointed out that in the large-charge limit, the Fourier transform gets localized at a saddle point at imaginary time. The main idea is that in the large $Q$ limit, the integral is dominated by the $\<0|\cO_Q \cO^\dagger_Q|0\>$ two-point function, whose Fourier transform can be computed using a saddle point approximation. We now show that the same argument also works in the nonrelativistic case. More precisely, let us consider the Fourier transform of the two point function of two primary operators of dimension $\Delta_{Q}$ and charge $N_{\cO}=-Q$. The integral can be evaluated exactly and the leading answer at large $Q$ is given by
\begin{align}
 \int\! dt\, d^{3}\vec x\, e^{i\omega t}e^{-i\vec p\.\vec x} \frac{1}{(t-i\e)^{\De_Q}}\exp\p{\frac{i Q \vec x^2}{2t}}&=\frac{4 \sqrt{2} \pi ^{5/2} i^{\Delta_Q } \left(\omega -\frac{p^2}{2 Q}\right)^{\Delta_Q -\frac{5}{2}}}{Q^{3/2} \Gamma \left(\Delta_Q -\frac{3}{2}\right)}\nonumber\\
 &\underset{Q\to\infty}\simeq i^{\De_Q} \frac{4\pi^2 e^{\De_Q}\De_Q^{2-\De_Q}}{Q^{\frac{3}{2}}}\p{\omega-\tfrac{p^2}{2Q}}^{\De_Q-\frac{5}{2}}\,,
\end{align}
where we have ignored overall normalization for the two-point function. 

In the limit $Q \to \oo$, this integral can also be evaluated using the saddle point approximation. The saddles are at
\be\label{FT_saddles}
t^*=-i\frac{\De_Q}{\omega-\frac{p^2}{2Q}},\quad \vec x^* = -i\frac{\De_Q}{\omega-\frac{p^2}{2Q}}\frac{\vec p}{Q}.
\ee
Performing the Gaussian integral around this saddle point,\footnote{Because the saddle point location is complex, we must deform the contour so that it passes through the saddle point. In particular, the saddle point in $t$ lies in the lower half plane, and thus the deformation does not cross the branch cut in the upper half plane due to the $i\epsilon$-prescription. In addition the integrals along the real direction have vanishing contributions at infinity. So, we can simply shift the contour in the imaginary direction without rotating it.} we obtain
\be
&\int\! dt\, d^{3}\vec x\, e^{i\omega t}e^{-i\vec p\.\vec x} \frac{1}{(t-i\e)^{\De_Q}}\exp\p{\frac{i Q \vec x^2}{2t}} \underset{Q\to\infty}{\longrightarrow} i^{\De_Q} \frac{4\pi^2 e^{\De_Q}\De_Q^{2-\De_Q}}{Q^{\frac{3}{2}}}\p{\omega-\tfrac{p^2}{2Q}}^{\De_Q-\frac{5}{2}}\,.
\ee
In summary, the Fourier transform of the two-point function of two large-charge primary operators is equivalent to evaluating their position space two-point function at the spacetime point eq.\ \eqref{FT_saddles}.

\begin{figure}
\includegraphics[width=0.5\linewidth]{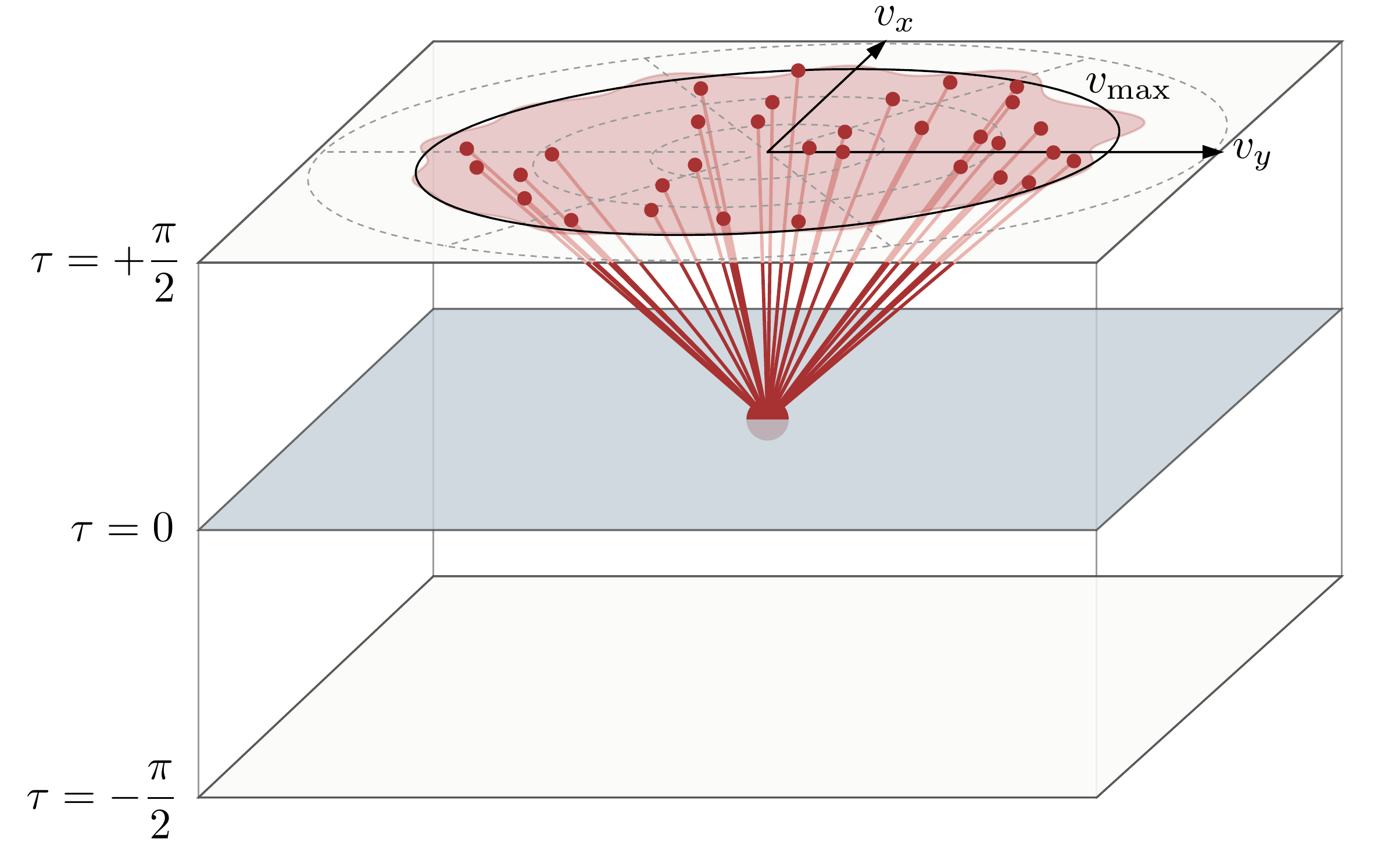}\qquad
\includegraphics[width=0.5\linewidth]{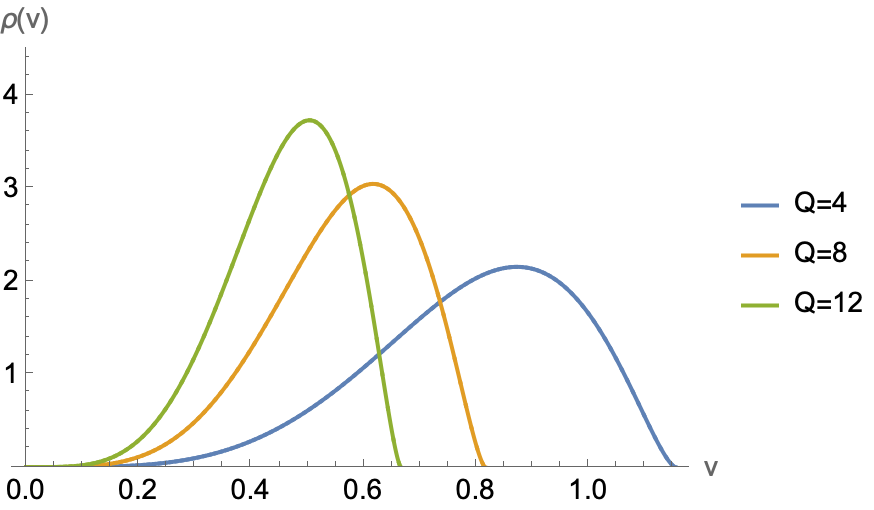}
\caption{An illustration of a large-charge state, and plot of the distribution of the one-point function $\mathcal{E}_v$, for $Q=4,8,12$. The large-charge calculation using a superfluid state provides a good approximation for small velocities, but suffers from edge effects, at large velocities, near the edge of the droplet.}
\label{fig:large_charge}
\end{figure}

The energy correlation entails the calculation of the $\<0|\cO_Q n \cO^\dagger_Q|0\>$ correlation function which can be computed from the large-charge path integral
\be\label{eq:PathInt_largeQ}
\<0|\cO_Q n \cO^\dagger_Q|0\>=\int\! \cD \th\, \cO_Q(t_{E1},\vec x_1)n(t_{E},\vec x)\cO_Q^{\dag}(t_{E2},\vec x_2)e^{-\int dt_E d^3 \vec x \cL},
\ee
where $\theta$ is the Nambu-Goldstone field. Note that, however, the large-charge path integral involves the Euclidean time $t_E$ and for our calculation we should Wick rotate it by $t_E =it$ to get the Lorentzian time. Our calculations generically involve three-point functions of the form $\langle \mathcal{O}_Q \cO_q \mathcal{O}_{Q+q}^{\dagger}\rangle$, involving two large-charge operators and a light operator $\cO_q$. In the limit $Q\to\infty$ with $\cO_q$ held fixed, the light insertion can be treated as a probe of the background sourced by the heavy operators. The leading semiclassical contribution is therefore obtained by evaluating the operator insertions on the two-point saddle reviewed in eq.\ \eqref{eq:saddle_theta}, with the backreaction of the light insertion neglected at this order \cite{Beane:2024kld}. Combining this with the Fourier transform saddle in eq.\ \eqref{FT_saddles}, one has
\be\label{eq:LargeCharge_FT_Result}
&\frac{\int dtd^3\vec x\, e^{i\omega t-i\vec p \.\vec x}\<\cO_Q(t,\vec x) n(t_0,\vec x_0) \cO^{\dag}_Q(0,\vec 0)\>}{\int dtd^3\vec x\, e^{i\omega t-i\vec p \.\vec x}\<\cO_Q(t,\vec x)\cO_Q^{\dag}(0,\vec 0)\>} \underset{Q\to\infty}{\longrightarrow} \frac{\<\cO_Q(t^*,\vec x^*) n(t_0,\vec x_0) \cO_Q^{\dag}(0,\vec 0)\>}{\<\cO_Q(t^*,\vec x^* )\cO_Q^{\dag}(0,\vec 0)\>}\bigg|_{\theta= \theta_s}\,,
\ee
where, the solution eq.\ \eqref{eq:saddle_theta} at Lorentzian time $t$ and position $\vec x$ can be written as
\be\label{eq:theta_solution_Lorentzian}
\th_s(t,\vec x) = \frac{i}{2}\g \log\p{\frac{\frac{\De_Q}{\omega}-it}{it}}-\frac{i}{4}\left[\frac{\vec x^2}{it}-\frac{\vec x^2}{\frac{\De_Q}{\omega}-it}\right],
\ee
for
\be
t_{E1}=\frac{\De_Q}{\omega},\quad t_{E2}=0,\quad \vec x_1=\vec x_2=\vec 0.
\ee

\begin{figure}
\begin{center}
\includegraphics[width=0.8\linewidth]{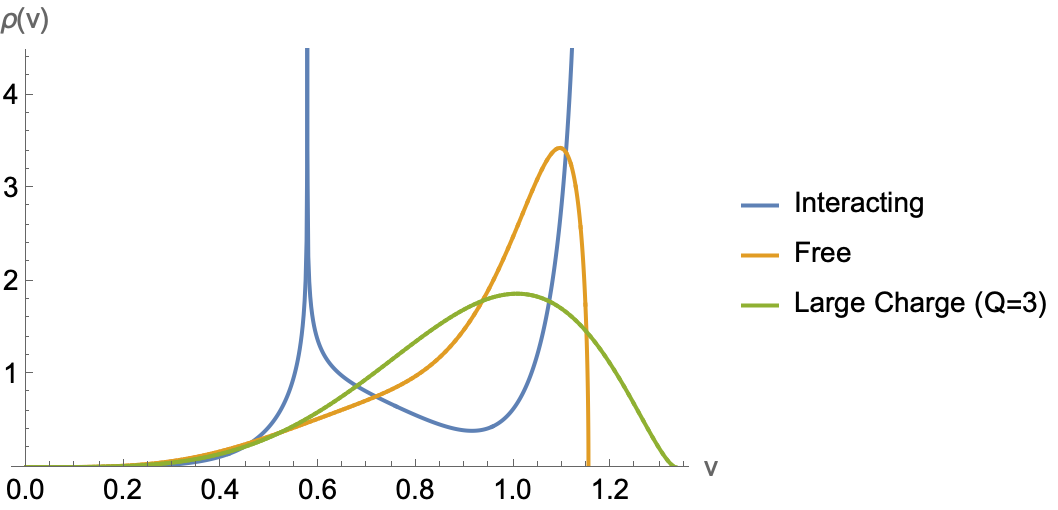}
\end{center}
\caption{A comparison of the $\rho(v)$ distribution for three body final states of fermions, in the non-interacting theory, the exact result in fermions at unitarity, and the large-charge approximation for fermions at unitarity. Extrapolated to $Q=3$, the large-charge result shows approximate agreement at small velocities but misses the resonant features associated with three-body kinematics and extends beyond the exact kinematic endpoint.
}
\label{fig:large_charge_comparison}
\end{figure}

Next, we evaluate the leading order superfluid EFT representation of the density operator \eqref{eq:EFT_nOp} on  the saddle solution \eqref{eq:theta_solution_Lorentzian} and take the detector limit. After using eq.\ \eqref{eq:LargeCharge_FT_Result}, we obtain the $\cE_v(\hat n)$ one-point function in the leading large-charge state,
\be\label{eq:Ev_largeQ_result}
\<\cE_v(\hat n)\>_{\cO_Q(\omega)}= \frac{27Q^{4}}{1024\pi^2\omega^3}v^4 \p{\frac{16\omega}{3Q}-v^2}^{\frac{3}{2}}.
\ee
The result has the scaling behavior $\sim v^4$ at small $v$, and $\sim (v_{\text{max}}^2-v^2)^{\frac{3}{2}}$ near the endpoint, which agrees with the prediction made in ref.\ \cite{Son:2021kkx} from a completely different perspective. One can also easily verify that it obeys the Ward identities eqs.\ \eqref{eq:Ward_E_generalk} and \eqref{eq:Ward_P_N_generalk}.

It is interesting to understand the validity of the large-charge expansion for detector distributions. This can be understood intuitively from figure \ref{fig:large_charge}. The superfluid EFT state can be viewed as a droplet on the spatial slice at $\tau=\pi/2$. For small velocities, the detector is deep inside the droplet, and will be well described by the superfluid EFT, but for larger velocities, it will be near the edge of the droplet, and suffer from edge effects. These have been discussed in ref.\ \cite{Hellerman:2020eff}. We therefore expect that our large-charge calculation of the $\mathcal{E}_v$ distribution should be valid at small $v$. 

In figure \ref{fig:large_charge_comparison}, we show a comparison of the distributions of the $\mathcal{E}_v$ detector on three fermion final states in the free theory, the exact result in the theory of fermions at unitarity, and the large-charge approximation evaluated at $Q=3$. The states are created by the leading $l=0$ operator. We indeed see agreement at small $v$, below the first singularity in the distribution for fermions at unitarity. This is expected, as this singularity is strongly related to the kinematics of the three-body state, and therefore will not be reproduced by the large-charge approximation. 

It would be extremely interesting to numerically evaluate the four-body result using either exact numerics for the four body wave-function, or approximations (see, e.g., refs.\ \cite{Platter:2004he,Hammer:2006ct}). It is common in the relativistic case that kinematic singularities are smoothed out in higher body states (see, e.g., ref.\ \cite{Chicherin:2023gxt}), and it would be interesting to understand the extent to which this occurs in the nonrelativistic case, and at which $Q$, the large $Q$ expansion begins to become a good approximation.

An interesting feature of the solution is that the allowed value of $v$ has a cutoff at $v\sim O(1) \times \sqrt{\frac{\omega}{Q}}$. Naïvely, one can imagine a kinematic configuration in which only an $O(1)$ number of particles carry most of the energy $\omega$, while the remaining $O(Q)$ particles carry almost no energy. This would lead to a nonzero contribution for $v \sim O(1)\times \sqrt{\omega}$. The EFT calculation above suggests that the contributions from such configurations are suppressed at large $Q$.

\section{Detector Correlators and Celestial Wave-functions}\label{sec:celestial}

The correlators of detector operators, $\mathcal{E}_v$, at future infinity exhibit a number of structural similarities to the case of cosmological correlators, which are also evaluated on a spacelike slice at future infinity (see, e.g., refs.\ \cite{Maldacena:2002vr,Arkani-Hamed:2015bza,Baumann:2021fxj,Baumann:2020dch,Arkani-Hamed:2018kmz,Baumann:2019oyu} for recent discussions). In this context, it is often convenient to work with a wave-function, from which the correlator can then ultimately be computed. More ambitiously, the conjectured dS/CFT correspondence \cite{Strominger:2001pn} suggests that the calculation of cosmological correlators is equivalent to the calculation of correlators in a CFT. In the case of asymptotic observables in asymptotically flat spacetimes, a similar program is being pursued in the celestial holography program \cite{Pasterski:2021raf}.

In this section, we argue that the correlation functions of detector operators in NRCFT can be interpreted as wave-functions at the fixed time slice at $\tau=\frac{\pi}{2}$ in the harmonic-trap frame. We first discuss some general properties of detector correlators in section \ref{sec:general_properties}. Then, in sections \ref{sec:wave-function_free} and \ref{sec:wave-function_interacting}, we explain how they can be viewed as wave-functions in free theories and fermions at unitarity. We believe that this could provide a simple toy model in which to study celestial holography.

\subsection{General Properties}\label{sec:general_properties}

We will study the $\cD(\vec v)$ detector defined in eq.\ \eqref{eq:Dv_nHT_def}. First, let us write down how the generators of the Schrödinger group act on the detector. Using eqs.\ \eqref{eq:generators_general} and \eqref{eq:Dop_conformalgenerators}, we find
\be
[D,\cD(\vec v)] &=  -i (\vec v\.\ptl_{\vec v} + d)\cD(\vec v), \nn \\
[K_i,\cD(\vec v)] & = -i \ptl_{v_i} \cD(\vec v), \nn \\
[M_{ij},\cD(\vec v)] & = i(v_i\ptl_{v_j}-v_j\ptl_{v_i})\cD(\vec v), \nn \\
[H,\cD(\vec v)] &= [P_i,\cD(\vec v)] = 0.
\ee
From these actions, we observe that the detector operators $\cD(\vec v)$ in a $(d+1)$-dimensional NRCFT behave like local operators in a scale-invariant theory on $\R^d$, where $D$ acts as dilatation, $K_i$ acts as translations, and $M_{ij}$ acts as rotations. In particular, the $\R^d$ can be thought of as the velocity space of the original NRCFT. From the point of view of this fictitious boundary theory, $\cD(\vec v)$ transforms like a scalar local operator with scaling dimension $d$.

Let us consider the general $n$-point function of $\cD(\vec v)$ in the state created by a local operator $\cO(\omega,\vec k)$ with $M$ particles (i.e., $M=-N_{\cO}$)
\be
\<0|\cO(\omega,\vec k)\cD(\vec v_1)\ldots \cD(\vec v_n)\cO^\dagger(\omega,\vec k)|0\>.
\ee
The symmetry constraints on this $n$-point function are given by eq.\ \eqref{eq:npoint_symmetry}. Additionally, due to the Ward identities \eqref{eq:Ward_forDv}, the $(n+1)$-point function and the $n$-point function are related by
\be\label{eq:Ward_general_npoint}
\int d^d\vec v_{n+1}\<0|\cO(\omega,\vec k)\cD(\vec v_1)\ldots \cD(\vec v_{n+1})\cO^\dagger(\omega,\vec k)|0\>&=M\<0|\cO(\omega,\vec k)\cD(\vec v_1)\ldots \cD(\vec v_n)\cO^\dagger(\omega,\vec k)|0\>, \nn \\
\int d^d\vec v_{n+1}\vec v_{n+1}\<0|\cO(\omega,\vec k)\cD(\vec v_1)\ldots \cD(\vec v_{n+1})\cO^\dagger(\omega,\vec k)|0\>&=\vec k \<0|\cO(\omega,\vec k)\cD(\vec v_1)\ldots \cD(\vec v_n)\cO^\dagger(\omega,\vec k)|0\>, \nn \\
\int d^d\vec v_{n+1}\frac{\vec v_{n+1}^2}{2}\<0|\cO(\omega,\vec k)\cD(\vec v_1)\ldots \cD(\vec v_{n+1})\cO^\dagger(\omega,\vec k)|0\>&=\omega\<0|\cO(\omega,\vec k)\cD(\vec v_1)\ldots \cD(\vec v_n)\cO^\dagger(\omega,\vec k)|0\>.
\ee
Also, we recall that $\cD(\vec v)$ can be obtained by evolving the number density operator $n(\vec v)$ from $\tau=0$ to $\tau=\frac{\pi}{2}$ under the harmonic-trap potential $H+C$
\be
\cD(\vec v) = e^{i(H+C)\frac{\pi}{2}}n(\vec v)e^{-i(H+C)\frac{\pi}{2}},
\ee
where $n(\vec v)\equiv n(t=0, \vec v)$. As a result, the $n$-point detector correlator in a state created by $\cO^\dagger(\omega,\vec k)$ can be written as
\be\label{eq:N-point_generalform}
&\<\cO(\omega,\vec k)|\cD(\vec v_1)\ldots \cD(\vec v_n)|\cO^{\dagger}(\omega,\vec k)\> \nn \\
&= \<\cO(\omega,\vec k)|e^{i(H+C)\frac{\pi}{2}}n(\vec v_1)\ldots n(\vec v_n)e^{-i(H+C)\frac{\pi}{2}}|\cO^{\dagger}(\omega,\vec k)\>.
\ee

From eq.\ \eqref{eq:N-point_generalform}, we see that there are two equivalent ways to compute detector correlators. We can choose to evolve each number density operator $n(\vec v)$ to $\tau=\pi/2$. This was the approach used in the previous sections, and it was achieved by taking the detector limit. Alternatively, one can also evolve the state $|\cO^\dagger(\omega,\vec k)\>$ using $e^{-i(H+C)\frac{\pi}{2}}$, and compute the expectation value of $n(\vec v_1)\ldots n(\vec v_n)$ in the evolved state. In this section, we will use the latter approach.

So far, the statements we have made are true in any general NRCFTs. On the other hand, it turns out that for the NRCFTs we consider in this paper (free theories and fermions at unitarity), their detector correlators satisfy stronger conditions. In particular, given an operator $\cO(\omega,\vec k)$ creating an $M$-particle state, it can always be expressed as $M$ $\psi^\dagger$'s acting on the vacuum integrated against a wave-function (which is an eigenfunction of the Hamiltonian $H$). We can then follow eq.\ \eqref{eq:N-point_generalform} and evolve it to $\tau=\pi/2$ using the harmonic-trap Hamiltonian $H+C$. This gives us a ``celestial wave-function,'' from which we can compute all the detector correlators. In other words, we will show that there exists a corresponding function $\Phi_{\cO}$ depending on $M$ positions $\vec v_i$, energy $\omega$ and momenta $\vk$
\be
\Phi_{\cO}(\vec v_1,\ldots,\vec v_M;\omega,\vec k),
\ee
such that the $n$-point detector correlators (for $n\leq M$) can be written as
\be\label{eq:D_npt_inPhi}
&\<0|\cO(\omega,\vec k)\cD(\vec v_1)\ldots \cD(\vec v_n)\cO^\dagger(\omega,\vec k)|0\> \nn \\
&= \int\! d^d\vec u_1\ldots d^d\vec u_M\p{\sum_{i=1}^{M}\de^{(d)}(\vec u_i-\vec v_1)}\ldots \p{\sum_{i=1}^{M}\de^{(d)}(\vec u_i-\vec v_n)}\Phi_{\cO}(\vec u_1,\ldots ,\vec u_{M}; \omega, \vec k) \nn \\
&=\frac{M!}{(M-n)!} \Phi^{(n)}(\vec v_1,\ldots ,\vec v_{n}; \omega, \vec k) + \textrm{contact terms},
\ee
where the contact terms are terms with $\de^{(d)}(\vec v_i - \vec v_j)$, and they can be written in terms of $\Phi_{\cO}^{(m<n)}$. Here, $\Phi_{\cO}^{(n)}$ is defined as
\be\label{eq:wavefunc_nth_def}
\Phi_{\cO}^{(n)}(\vec v_1,\ldots,\vec v_n;\omega,\vec k)\equiv \int\! d^d \vec v_{n+1}\ldots d^d\vec v_{M}\, \Phi_{\cO}(\vec v_1,\ldots ,\vec v_{M}; \omega,\vec k).
\ee
For example, for $n=2$ we have
\be
&\<0|\cO(\omega,\vec k)\cD(\vec v_1)\cD(\vec v_2)\cO^{\dagger}(\omega,\vec k)|0\> \nn \\
&=M(M-1)\Phi^{(2)}(\vec v_1,\vec v_2;\omega,\vec k) + M \de^{(d)}(\vec v_1 - \vec v_2)\Phi^{(1)}(\vec v_1;\omega,\vec k).
\ee

Using eq.\ \eqref{eq:D_npt_inPhi}, the Ward identity constraints eq.\ \eqref{eq:Ward_general_npoint} can be simplified further. For example, the particle number Ward identity gives
\be\label{eq:PhiO_Ward_0}
\int d^d\vec v_{n+1}\, \Phi_{\cO}^{(n+1)}(\vec v_1,\ldots,\vec v_{n+1};\omega,\vec k) = \Phi_{\cO}^{(n)}(\vec v_1,\ldots,\vec v_{n};\omega,\vec k),
\ee
which is consistent with eq.\ \eqref{eq:wavefunc_nth_def}. The other two Ward identities give
\be\label{eq:PhiO_Ward_1}
\int d^d\vec v_{n+1}\, \vec v_{n+1} \Phi_{\cO}^{(n+1)}(\vec v_1,\ldots,\vec v_{n+1};\omega,\vec k) = \frac{\vec k-\sum_{r=1}^n \vec v_r}{M-n}\Phi_{\cO}^{(n)}(\vec v_1,\ldots,\vec v_{n};\omega,\vec k), \nn \\
\int d^d\vec v_{n+1}\, \frac{\vec v_{n+1}^2}{2} \Phi_{\cO}^{(n+1)}(\vec v_1,\ldots,\vec v_{n+1};\omega,\vec k) = \frac{\omega-\sum_{r=1}^{n}\frac{\vec v_r^2}{2}}{M-n} \Phi_{\cO}^{(n)}(\vec v_1,\ldots,\vec v_{n};\omega,\vec k).
\ee
The additional subtraction terms $\sum_{r=1}^n \vec v_r$, $\sum_{r=1}^{n}\frac{\vec v_r^2}{2}$ are consequences of the contact terms in eq.\ \eqref{eq:D_npt_inPhi}. 

Furthermore, due to energy and momentum conservation, the full $\Phi_{\cO}$ function is always proportional to the energy-conserving and momentum-conserving delta functions. This implies that $\Phi_{\cO}$ is fixed to be
\be
\Phi_{\cO}(\vec v_1,\ldots ,\vec v_{M}; \omega, \vec k) = \Phi^{(M-1)}_{\cO}(\vec v_1,\ldots ,\vec v_{M-1}; \omega, \vec k)\de^{(d)}(\vec k - \vec v_1-\ldots - \vec v_M),
\ee
and $\Phi_{\cO}^{(M-1)}$ is proportional to the energy-conserving delta function
\be
&\Phi^{(M-1)}_{\cO}(\vec v_1,\ldots ,\vec v_{M-1}; \omega, \vec k) \nn \\
&= \tilde{\Phi}^{(M-1)}_{\cO}(\vec v_1,\ldots ,\vec v_{M-1}; \omega, \vec k)\de(\omega-\tfrac{\vec v_1^2}{2}-\ldots-\tfrac{\vec v_{M-1}^2}{2} -\tfrac{(\vec k - \vec v_1-\ldots - \vec v_{M-1})^2}{2}).
\ee
As a result, all the detector correlators in the state created by $\cO(\omega,\vec k)$ are fixed by the function $\tilde{\Phi}_{\cO}^{(M-1)}$. In the next two subsections, we will show that $\tilde{\Phi}^{(M-1)}_{\cO}$ can be written as the square of a ``celestial wave-function."

In general, $\tilde{\Phi}^{(M-1)}_{\cO}$ is not completely fixed. One exception is when $\cO$ is a charge-2 scalar operator ($M=2$). In this case, after imposing the Ward identities eqs.\ \eqref{eq:PhiO_Ward_0}, \eqref{eq:PhiO_Ward_1}, and the symmetry constraints eq.\ \eqref{eq:npoint_symmetry}, we obtain
\be\label{eq:charge2_scalar_general}
\Phi^{(1)}_{\cO_2} (\vec v;\omega,\vec k)&=\frac{2\Phi^{(0)}_{\cO_2}(\omega,\vec k)}{\vol(S^{d-1})\p{\omega-\tfrac{\vec k^2}{4}}^{\frac{d}{2}-1}}\de(\omega-\tfrac{\vec k^2}{4}-(\vec v - \tfrac{\vec k}{2})^2),
\ee
where $\Phi_{\cO_2}^{(0)}$ is simply the two-point function $\<0|\cO(\omega,\vec k)\cO^\dagger(\omega,\vec k)|0\>$. Note that the $\vec v$-dependence only shows up in the energy-conserving delta function. In other words, energy conservation forces the $\Phi^{(1)}(\vec v;\omega,\vec k)$ for a charge-2 scalar state to be localized on a sphere centered at $\frac{\vec k}{2}$ with radius $\sqrt{\omega-\tfrac{\vec k^2}{4}}$. This result explains why we obtain the same $\<\cE_v(\hat n)\>$ for the leading charge-2 scalar state in free theories and fermions at unitarity in section \ref{sec:calcs}.

\subsection{Free Theories}\label{sec:wave-function_free}
We now explain how to compute the squared wave-function $\tilde{\Phi}_{\cO}^{(M-1)}$ in free theories in $d=3$. We will consider an operator $\cO$ which creates an $M$-particle state with $M_\uparrow$ spin-up fermions and $M_\downarrow=M-M_\uparrow$ spin-down fermions.

The main idea is to compute the detector correlators using eq.\ \eqref{eq:N-point_generalform}. We start with the free theory $M$-particle state, which can be written as
\be\label{eq:Mparticle_free_state}
|\{\vec k_i\}\>=\int \left[\prod_{i=1}^Md^3 \vec x_i \right] \Psi^{(M)}_{\text{free}}(\{\vec x_i\}; \{\vec k_i\}) \left[\prod_{i=1}^M \psi^\dagger_{\s_i}(\vec x_i)\right]|0\>.
\ee
Here, the wave-function $\Psi^{(M)}_{\text{free}}(\{\vec x_i\}; \{\vec k_i\})$ is a sum of products of plane waves $e^{i\vec k_i\.\vec x_j}$ given by eq.\ \eqref{free_wvfn_p_q}. Following eq.\ \eqref{eq:N-point_generalform}, we would like to compute the action of $e^{-i(H+C)\frac{\pi}{2}}$ on this state. Using eqs.~\eqref{eq:free_fermions}, \eqref{ji_def} and \eqref{eq:nj_algebra}, the action of the harmonic-trap Hamiltonian $H+C$ on this state is given by
\be
(H+C)|\{\vec k_i\}\>=\int \left[\prod_{i=1}^M d^3 \vec x_i \right] \left[\sum_{i=1}^M\p{-\frac{\partial^2_{\vec x_i}}{2}+\frac{\vec x_i^2}{2}}\Psi^{(M)}_{\text{free}}(\{\vec x_i\}; \{\vec k_i\})\right]\left[\prod_{i=1}^M \psi^\dagger_{\s_i}(\vec x_i)\right]|0\>.
\ee

Next, recall that the eigenfunctions of the 1D harmonic oscillator potential $-\frac{\partial^2_{x}}{2}+\frac{x^2}{2}$ are the Hermite functions $h_n(x) = \frac{1}{\pi^{1/4}\sqrt{2^n n!}}H_n(x) e^{-\frac{x^2}{2}}$, where $H_n(x)$ is the Hermite polynomial, and $h_n(x)$ has eigenvalue $n+\frac{1}{2}$. Thus, the action of $e^{-i(-\frac{\ptl_x^2}{2}+\frac{x^2}{2})\frac{\pi}{2}}$ on the eigenfunction $h_n(x)$ is
\be\label{eq:Free_eigenfunc_evolution_FT}
e^{-i(-\frac{\ptl_x^2}{2}+\frac{x^2}{2})\frac{\pi}{2}}h_n(x) =  (-i)^n e^{-\frac{i \pi}{4}}h_n(x) = \frac{e^{-\frac{i \pi}{4}}}{\sqrt{2\pi}} \int dy\, e^{-ixy}h_n(y).
\ee
We see that the action of $e^{-i(-\frac{\ptl_x^2}{2}+\frac{x^2}{2})\frac{\pi}{2}}$ is simply the Fourier transform. Since Hermite functions form a complete orthonormal  basis for square-integrable wave-functions, it then follows that the action of $e^{-i(H+C)\frac{\pi}{2}}$ on the $M$-particle state \eqref{eq:Mparticle_free_state} is given by
\be\label{eq:HplusC_free_eq}
e^{-i(H+C)\frac{\pi}{2}}|\{\vec k_i\}\>=\frac{e^{-\frac{3 i\pi M }{4}}}{(2\pi)^{\frac{3M}{2}}}\int \left[\prod_{i=1}^M d^3 \vec x_i \right] \widetilde{\Psi}^{(M)}_{\text{free}}(\{\vec x_i\}; \{\vec k_i\})\left[\prod_{i=1}^M \psi^\dagger_{\s_i}(\vec x_i)\right]|0\>,
\ee
where
\be
\widetilde{\Psi}^{(M)}_{\text{free}}(\{\vec x_i\}; \{\vec k_i\}) \equiv \int \left[\prod_{i=1}^M d^3 \vec y_i \right] e^{-i \sum_{j=1}^M \vec x_j\.\vec y_j}\Psi^{(M)}_{\text{free}}(\{\vec y_i\}; \{\vec k_i\}),
\ee
which is just a sum of products of delta functions $\de^{(3)}(\vec x_i -\vec k_j)$.

Meanwhile, the action of a local operator $\cO(\omega,\vec k)$ on the state $|\{\vec k_i\}\>$ is characterized by its form factor $\cF_\cO$,
\be
\<0|\cO(\omega,\vec k)|\{\vec k_i\}\> = \cF_{\cO}(\{\vec k_i\};\omega,\vec k)(2\pi)^4 \de(\omega-\sum_i \tfrac{\vec k_i^2}{2})\de^{(3)}(\vec k-\sum_i \vec k_i).
\ee
Combining this with eq.\ \eqref{eq:HplusC_free_eq}, we have
\be
&\<0|\cO(\omega,\vec k)\cD(\vec v_1)\ldots \cD(\vec v_n)\cO^{\dagger}(\omega,\vec k)|0\> \nn \\
&=\int \frac{\left[\prod_{i=1}^M d^3 \vec k_i\right]\left[\prod_{i'=1}^M d^3 \vec k'_{i'}\right]}{(M_{\uparrow}!)^2(M_{\downarrow}!)^2(2\pi)^{6M}}\cF^*_{\cO}(\{\vec k_i\}; \omega,\vec k)\cF_{\cO}(\{\vec k'_i\}; \omega,\vec k)  (2\pi)^4 \de(\omega-\sum_i \tfrac{\vec k_i^2}{2})\de^{(3)}(\vec k-\sum_i \vec k_i) \nn \\
&\times \frac{1}{(2\pi)^{3M}}\int \left[\prod_{i=1}^M d^3 \vec x'_i \right]\left[\prod_{i=1}^M d^3 \vec x_i \right]\left(\widetilde{\Psi}^{(M)}_{\text{free}}(\{\vec x'_i\}; \{\vec k'_i\})\right)^*\widetilde{\Psi}^{(M)}_{\text{free}}(\{\vec x_i\}; \{\vec k_i\}) \nn \\
&\times \<0|\left[\prod_{i=1}^M \psi^\dagger_{\s_i}(\vec x'_i)\right]^\dagger n(\vec v_1)\ldots n(\vec v_n)\left[\prod_{i=1}^M \psi^\dagger_{\s_i}(\vec x_i)\right]|0\>.
\ee
Finally, after plugging in the expression of $\widetilde{\Psi}^{(M)}_{\text{free}}$ and comparing with eq.\ \eqref{eq:D_npt_inPhi}, we arrive at a remarkably simple result for the squared celestial wave-function $\tilde{\Phi}^{(M-1)}_{\cO}$:\footnote{Note the identity
\be
n(\vec v)\left[\prod_{i=1}^M \psi^\dagger_{\s_i}(\vec x_i)\right]|0\> = \p{\sum_i \de^{(3)}(\vec v- \vec x_i)}\left[\prod_{i=1}^M \psi^\dagger_{\s_i}(\vec x_i)\right]|0\>.
\ee
}
\be
&\tilde{\Phi}^{(M-1)}_{\cO}(\vec v_1,\ldots,\vec v_{M-1}; \omega,\vec k) \nn \\
&= \frac{1}{(2\pi)^{3M-4}(M!)(M_\uparrow !)(M_\downarrow !)}\sum_{\s \in S_M} |\cF_\cO(\vec v_{\s(1)},\ldots , \vec v_{\s(M)};\omega, \vec k)|^2,
\ee
where $\vec v_M=\vec k-\vec v_1-\ldots-\vec v_{M-1}$. We see that in free theories, the celestial wave-function of a state created by operator $\cO$ is completely fixed by the form factor of the operator. This is consistent with our expectations, since by using the free theory definition of the detector, one can compute all the detector correlators straightforwardly provided we know the form factors.

\subsection{Fermions at Unitarity}\label{sec:wave-function_interacting}
Now, we consider the celestial wave-functions for fermions at unitarity. We will consider the wave-functions for $M=2$ and $M=3$. We emphasize that unlike the free theory case, in the interacting case it is not yet fully understood how to determine the celestial wave-function for $M>3$, but we will make some comments on what form we expect it to take at the end of the section.

The analysis in the previous subsection does not work in interacting theories. In particular, we expanded the free plane-waves in terms of the Hermite functions. In the interacting theory, this construction requires eigenfunctions satisfying the Bethe–Peierls boundary condition, which the free oscillator eigenfunctions do not obey. We therefore construct the appropriate trapped eigenbasis and determine the action of $e^{-i(H+C)\pi/2}$ on it.

\subsubsection*{Two-particle states}
Let us first consider the $M=2$ case with one spin-up and one spin-down fermion. We introduce the coordinates
\be
\vec R_{\textrm{cm}} = \frac{\vec x_1 + \vec x_2}{2},\quad \vec r =\vec x_1 - \vec x_2,
\ee
in which the harmonic potential can be written as
\be
&\sum_{i=1}^2\p{-\frac{\partial^2_{\vec x_i}}{2}+\frac{\vec x_i^2}{2}} =-\frac{\ptl^2_{\vec R_{\text{cm}}}}{4} + \vec R^2_{\text{cm}} - \ptl^2_{\vec r} + \frac{\vec r^2}{4}.
\ee
The $\vec R_{\text{cm}}$-dependence is not affected by the Bethe-Peierls condition, and therefore should be the same as the free theory case. However, for the $\vec r$-dependent part of the wave-function, the eigenfunctions satisfying the Bethe-Peierls boundary condition are given by
\be
\psi^{\text{int}}_n(r) =C_n \frac{e^{-\frac{r^2}{4}}H_{2n}(\tfrac{r}{\sqrt{2}})}{r},\quad \p{-\ptl^2_{\vec r} + \frac{\vec r^2}{4}}\psi^{\text{int}}_n(r) = (2n+\frac{1}{2})\psi^{\text{int}}_n(r).
\ee
This implies
\be
e^{-i(-\ptl^2_{\vec r} + \frac{\vec r^2}{4})\frac{\pi}{2}}\psi^{\text{int}}_n(r)  = e^{-i\frac{\pi}{4}}(-1)^n\psi^{\text{int}}_n(r) = \frac{e^{-i\frac{\pi}{4}}}{\sqrt{\pi}r}\int\limits_0^{\oo}\! du\cos\p{\frac{u r}{2}} (u \psi^{\text{int}}_n(u)).
\ee
We again see that the action of $e^{-i(H+C)\frac{\pi}{2}}$ on $\psi^{\text{int}}_n(r)$ leads to a linear functional that is independent of $n$, and thus we can apply the functional to the full interacting wave-function. Note that this is different from the free theory case, where the integral is a Fourier transform, which essentially replaces the $\cos\p{\frac{u r}{2}}$ factor with $\sin\p{\frac{u r}{2}}$.

As reviewed in section \ref{sec:unitarity_2body}, the charge-2 wave-function for fermions at unitarity is given by
\be
\Psi^{\text{int}}_{M=2}(\vec x_i) = \frac{e^{i\vec P_{\text{cm}}\.\vec R_{\text{cm}}}}{\sqrt{2\pi}}\frac{\cos(k r)}{r}.
\ee
As a result, the action of $e^{-i(H+C)\frac{\pi}{2}}$ on the charge-2 state is given by
\be
&e^{-i(H+C)\frac{\pi}{2}} |\Psi_{\vec{P}_{\text{cm}},k}\> = \frac{e^{-i\pi}}{(2\pi)^{\frac{3}{2}}\sqrt{\pi}\sqrt{2\pi}} \nn \\
&\qquad \times \int\! d^3\!\vec R_{\text{cm}}\,d^3\vec r\, (2\pi)^3\de^{(3)}\Bigl(\sqrt{2}\vec R_{\text{cm}}-\tfrac{\vec P_{\text{cm}}}{\sqrt{2}}\Bigr) 
\biggl({\frac{1}{r}\int\limits_0^{\oo}\! du \cos(\tfrac{u r}{2})\cos(k u)}\biggr)
\psi^\dagger_{\downarrow}(\vec x_1)\psi^\dagger_{\uparrow}(\vec x_2)|0\>,
\ee
where the $\de^{(3)}(\sqrt{2}\vec R_{\text{cm}}-\tfrac{\vec P_{\text{cm}}}{\sqrt{2}})$ comes from the Fourier transform of $e^{i\vec P_{\text{cm}}\.\vec R_{\text{cm}}}$. The $u$-integral also gives another delta function $\de(k-\tfrac{r}{2})$. In the end, we get
\be\label{eq:HplusC_charge2_state_interacting}
&e^{-i(H+C)\frac{\pi}{2}}|\Psi_{\vec P_{\text{cm}},k}\> = e^{-i\pi}\sqrt{2}\pi^{\frac{3}{2}} k \int\! d\Omega_{\hat n}\, \psi^\dagger_{\downarrow}(\tfrac{\vec P_{\text{cm}}}{2}+k \hat n)\psi^\dagger_{\uparrow}(\tfrac{\vec P_{\text{cm}}}{2}-k \hat n)|0\>.
\ee
We see that the position of the fermions are replaced by the velocity determined by the center of mass momentum $\vec P_{\text{cm}}$ and relative momentum $k$.

Finally, combining eq.\ \eqref{eq:HplusC_charge2_state_interacting} with the form factor of the leading charge-$2$ local operator $\cO_2$ given by eq.\ \eqref{eq:O2_FormFactor_momentum}, we find that the squared celestial wave-function $\tilde{\Phi}^{(1)}$ for the charge-2 interacting state is given by
\be
&\tilde{\Phi}^{(1)}_{\cO_2}(\vec v;\omega,\vec k) = \frac{1}{(2\pi)^2(\omega-\frac{\vec k^2}{4})}.
\ee
One can verify that this is consistent with the general form \eqref{eq:charge2_scalar_general} and the two-point function \eqref{eq:2pt_O2_interacting}.

\subsubsection*{Three-particle states}
We now consider the same calculation for the charge-3 states, whose unitary wave-function in the harmonic trap was derived in \cite{PhysRevLett.97.150401}. To consider the action of $e^{-i(H+C)\frac{\pi}{2}}$ on the charge-3 wave-functions, we use Efimov coordinates given by eqs. \eqref{jacobi_coord} and \eqref{efimov_coord}. Again, the $\vec R_{\text{cm}}$-dependence is the same as free theory, but the hyperradial and hyperangular part must be treated differently. Note that the hyperangular part of the harmonic-trap Hamiltonian $H+C$ is the same as $H$. Consequently, for a function of the form $f(R)\Phi_s(\Omega)$, where $\Phi_s(\Omega)$ is an eigenfunction of the hyperangular Laplacian with eigenvalue $4-s^2$, the action of $H+C$ can be written as
\be
-\frac{1}{2}\p{\ptl^2_R+\frac{5}{R}\ptl_R + \frac{4-s^2}{R^2}} + \frac{R^2}{2}.
\ee
The eigenfunctions of this differential operator are given by 
\be
\psi^{(3)}_{n,s}(R) &= R^{s-2}e^{-\frac{R^2}{2}}L^s_n(R^2),
\ee
where $L_n^s$ is the generalized Laguerre polynomial, and the eigenvalue of $\psi^{(3)}_{n,s}(R)$ is $s+1+2n$. Thus, the action of $e^{-i(H+C)\frac{\pi}{2}}$ becomes
\be\label{eq:HplusC_eigenfunc_charge3_interacting}
e^{-i(H+C)\frac{\pi}{2}}\psi^{(3)}_{n,s}(R) =e^{-i(s+1+2n)\frac{\pi}{2}}\psi^{(3)}_{n,s}(R)  \equiv \frac{e^{-(s+1)\frac{i\pi}{2}}}{R^{\frac{5}{2}}}\int\limits_0^\oo\! du\, \sqrt{u R}J_s(u R) u^{\frac{5}{2}}\psi^{(3)}_{n,s}(u).
\ee
Recall that the charge-3 interacting wave-function is given by eq.\ \eqref{eq:Charge3_Wave-function_Interacting}, where the hyperradial and hyperangular part is $\frac{J_s(\sqrt{2}k R)}{R^2}\Phi_s(\Omega)$. Using eq.\ \eqref{eq:HplusC_eigenfunc_charge3_interacting}, we find that acting $e^{-i(H+C)\frac{\pi}{2}}$ gives\footnote{We have used the Bessel function orthogonality relation,
\be
\int_0^\oo\! dx\, x J_s(ax)J_s(bx) = \frac1{a} {\de(a-b)}.
\ee
}
\be
\frac{e^{-(s+1)\frac{i\pi}{2}}}{R^{\frac{5}{2}}}\int\limits_0^\oo\! du\, \sqrt{u R}J_s(u R) u^{\frac{5}{2}}\frac{J_s(\sqrt{2}k u)}{u^2}\Phi_s(\Omega) = \frac{e^{-(s+1)\frac{i\pi}{2}}}{R^3}\de(R-\sqrt{2}k)\Phi_s(\Omega).
\ee

We can now write down the action of $e^{-i(H+C)\frac{\pi}{2}}$ on the full three-body state. Recall that the Jacobian from the $\vec x_i$ coordinates to $\vec R_{\text{cm}}, \vec r,\vec \r$ is $(\sqrt{3}/2)^3$. We also have
\be
d^3 \vec R_{\text{cm}}\,d^3 \vec r\, d^3 \vec \r = d^3 \vec R_{\text{cm}}\, dR\, R^5\, d\bar \Omega,
\ee
where $d\bar \Omega=2\sin^2(2\a)d\a d\hat r d\hat \r$ denotes the hyperangular measure. Combining everything, we obtain
\begin{align}
&e^{-i(H+C)\frac{\pi}{2}}|\Psi^{l,m,s}_{\vec P_{\text{cm}},k}\> \nn \\
&=e^{-(2s+5)\frac{i\pi}{4}}N_{k}^l \p{\frac{\pi}{\sqrt{3}}}^{\frac{3}{2}}\sqrt{2} k^2\!\int\! d\bar \Omega\, \Phi_s^{l,m}(\Omega) \psi^\dagger_\uparrow(\vec x_1)\psi^\dagger_\downarrow(\vec x_2)\psi^\dagger_\uparrow(\vec x_3)|0\>\Bigg|_{\substack{\sum\vec x_i=\vec P_{\text{cm}} \\ R_{\vec x_1,\vec x_2,\vec x_3}=\sqrt{2}k}},
\end{align}
where $R_{\vec x_1,\vec x_2,\vec x_3}=\sqrt{\frac{\vec x_{12}^2 + \vec x_{13}^2+\vec x_{23}^2}{3}}$. We see that the positions of the fermions again become velocities at $\tau=\frac{\pi}{2}$. The momentum and energy of the state fix the center of mass position and the hyperradius. 

To find the full celestial wave-function, we must also include the form factor of the charge-$3$ local operator $\cO_3^{l,m,s}$ given by eq.\ \eqref{eq:O3_int_formfactor}. Our final result is
\be\label{eq:Phi2_charge3_final}
\tilde{\Phi}^{(2)}_{\cO_3^{l,m,s}}(\vec v_1,\vec v_2;\omega,\vec k) =\frac{2^{4-s}\pi (\omega-\frac{\vec k^2}{6})^{s-2}}{3\G(s+1)^2} \sum_{\s \in S_3}\left|\Phi_s^{l,m}(\Omega_{\vec v_{\s(1)},\vec v_{\s(2)},\vec v_{\s(3)}})\right|^2,
\ee
where $\vec v_3=\vec k-\vec v_1-\vec v_2$.

\begin{figure}
	\centering
	\includegraphics[scale=0.7]{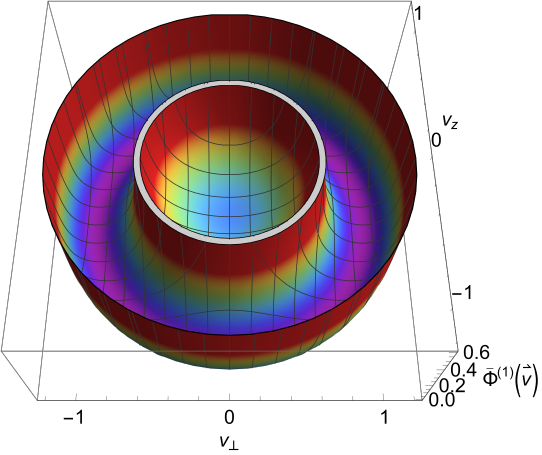}
	\caption{The integrated squared celestial wave-function, $\bar{\Phi}^{(1)}(\vec v)$ for the leading $l=0$ charge-3 state ($s=2.16622$).}
	\label{fig:3D_spin0}
\end{figure}

\begin{figure}
    \centering
    \begin{minipage}[t]{0.4\textwidth}
        \centering
        \vspace{0pt}
        {\small 
         \hspace{-3mm}
        \[
          s=1.77272
        \]
         \hspace{3mm}
        }
        \vspace{-2mm}
        \includegraphics[width=\linewidth]{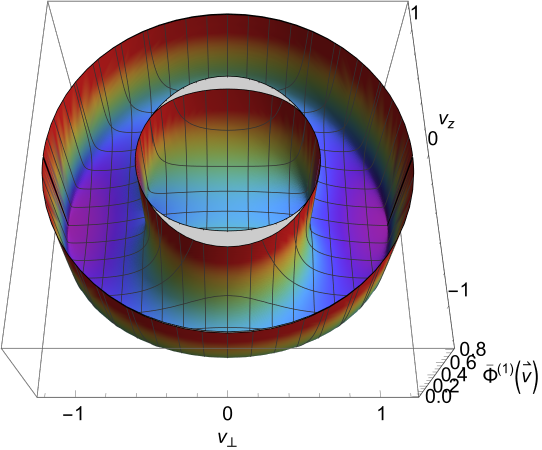}
    \end{minipage}
    \hspace{2em}
     \begin{minipage}[t]{0.4\textwidth}
        \centering
        \vspace{0pt}
        {\small 
         \hspace{-3mm}
        \[
          s=4.35825
        \]
         \hspace{3mm}
        }
        \vspace{-2mm}
        \includegraphics[width=\linewidth]{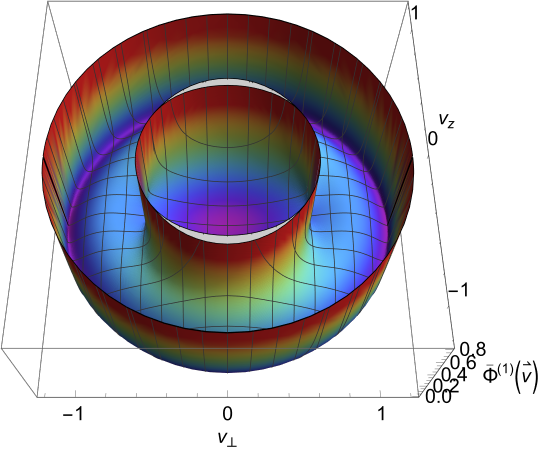}
    \end{minipage}
    
    \vspace{1em}

    \begin{minipage}[t]{0.4\textwidth}
        \centering
        \vspace{0pt}
        {\small 
         \hspace{-3mm}
        \[
          s=5.71643
        \]
         \hspace{3mm}
        }
        \vspace{-2mm}
        \includegraphics[width=\linewidth]{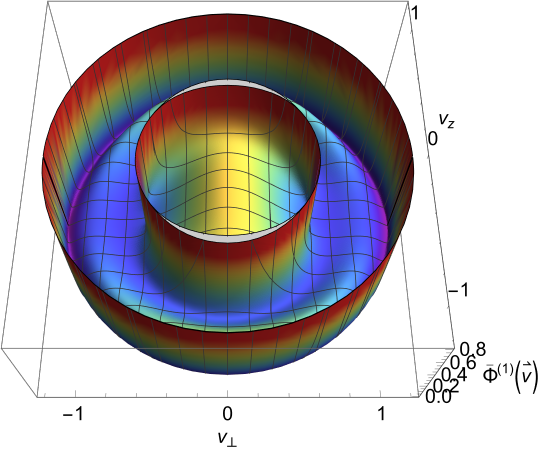}
    \end{minipage}
    \hspace{2em}
    \begin{minipage}[t]{0.4\textwidth}
        \centering
        \vspace{0pt}
        {\small 
        \hspace{-3mm}
        \[
          s=8.05319
        \]
        \hspace{3mm}
        }
        \vspace{-2mm}
        \includegraphics[width=\linewidth]{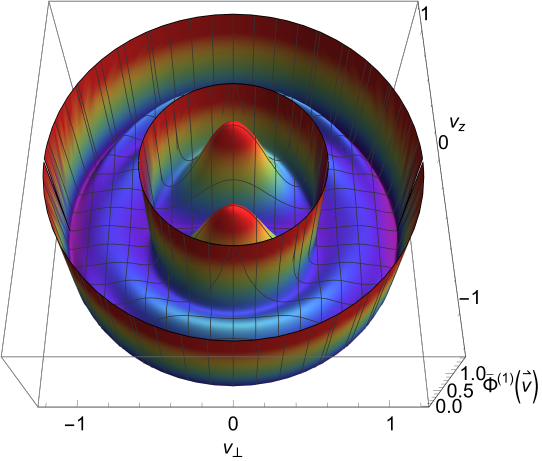}
    \end{minipage}
    
    \caption{The integrated squared celestial wave-function, $\bar{\Phi}^{(1)}(\vec v)$ for $l=1$ charge-3 states with $s=1.77272$, 4.35825, 5.71643, 8.05319 and polarization $\hat \e = \hat z$.}
    \label{fig:3D_spin1}
\end{figure}

From the squared celestial wave-function, we can further integrate over $\vec v_2$ to get the one-point detector correlator $\Phi^{(1)}_{\cO}(\vec v;\omega,\vec k)$. The celestial wave-function approach then gives an alternative way to obtain the same results we computed in section \ref{sec:unitarity_calc} by taking the detector limit of the number density operator. Let us define
\be
\bar{\Phi}^{(1)}(\vec v) \equiv \frac{\Phi^{(1)}(\vec v)}{\Phi^{(0)}}.
\ee
We now consider $\bar{\Phi}^{(1)}(\vec v)$ for operators with $l=0$ and $l=1$. For the $l=1$ operator, by rotational invariance, the one-point function $\bar{\Phi}^{(1)}(\vec v)$ is a function of $|\vec v|$ and $\theta$, the angle between $\vec v$ and the polarization vector. So, we can make a 3D plot of $\bar{\Phi}^{(1)}(\vec v)$ on the $v_{\perp}$-$v_z$ plane, where $v_z=|\vec v|\cos\theta$ and $v_{\perp} = |\vec v|\sin\theta$. In figure \ref{fig:3D_spin0}, we show the 3D plot of the $\bar{\Phi}^{(1)}(\vec v)$ for the leading scalar charge-$3$ operator, which is automatically isotropic on the $v_{\perp}$-$v_z$ plane. In figure \ref{fig:3D_spin1}, we show the $\bar{\Phi}^{(1)}(\vec v)$ for the first four $l=1$ charge-$3$ operators.

For states with $M>3$ in fermions at unitarity, we cannot perform the same calculation, since their wave-functions are not known exactly. However, we note that one can also define similar hyperspherical coordinates even for higher-body states, and the wave-function can be written as a product $e^{i\vec R_{\text{cm}}\. P_{\text{cm}}}f(R)\Phi_s(\Omega)$ for some unknown hyperangular wave-function $\Phi_s(\Omega)$ and $s$ (see  Appendix C of \cite{2006PhRvA..74e3604W}). Then, we expect that the analysis we did for the $M=3$ case should also work, and the celestial wave-function for $M>3$ is fully determined by the hyperangular wave-function similar to eq.\ \eqref{eq:Phi2_charge3_final}.

It will be interesting to explore this ``celestial wave-function" for more general NRCFTs. We have focused on the case of two and three body states in the free theory, and in fermions at unitarity. The picture will extend to other theories with a wave-function description. For example it should also apply for the case of nonrelativistic anyons \cite{Nishida:2007pj,Nishida:2007de}. A potential obstruction would be the case of an NRCFT with a continuous spectrum. Recently, ref.\ \cite{Belin:2026wkc} has shown that in the relativistic case, the fundamental properties of energy correlators imply that they admit an event distribution formula. In the nonrelativistic case, it would be extremely interesting to study whether a similar argument can lead to eq.\ \eqref{eq:D_npt_inPhi}, but we leave this to future work.

\section{Conclusions and Future Directions}\label{sec:conc}

In this paper we have initiated the study of ``nonrelativistic conformal collider physics'' by formalizing the concept of detector operators in nonrelativistic theories. In the case of NRCFTs, we found that the space of asymptotic detectors was naturally graded by velocity, leading us to define detectors, such as energy flux detectors, $\mathcal{E}_v (\hat n)$, which measure energy deposited in calorimeters differentially in velocity. We have also constructed similar velocity graded detectors for measuring momenta and charge flux. These detectors conserve energy and momentum and are examples of ``passive" detectors. We have also constructed ``active" detectors which do not do so, for e.g., detectors which measure the other conserved charges of the Schr\"odinger group. Using the harmonic-trap geometry, the analog of the Lorentzian cylinder, we studied the operator definitions of these detectors, their symmetries, and Ward identities. We then performed a number of explicit calculations of one-point functions of the passive detectors in states with different charges and spins corresponding to different NRCFTs. Particularly for  three body states in the NRCFT of fermions at unitarity, we found that they exhibit a remarkably rich structure, reflective of the underlying three-body wave-functions. For the energy detector, we also proposed a generalization of the Hofman-Maldacena conformal collider bound, namely that $\langle \mathcal{E}_v (\hat n) \rangle \geq 0$, as a function of velocity, and showed that it was satisfied for a number of non-trivial states in fermions at unitarity.

Since this paper represents the first exploration of detector operators in NRCFTs, there are a large number of directions for future exploration, both in the further development of the theory of detector operators in NRCFTs and their application to constraining NRCFT data, as well as in the application to real world experiments. Here we summarize a number of different directions which we believe would be interesting to explore:
\smallskip

{\bf{Conformal Symmetry Breaking:}} In this paper we have studied detectors purely within the context of NRCFTs.  The most immediate open question, from both theoretical and experimental perspectives, is the effect of breaking of conformal (and scale) invariance.  The unitarity fermion fixed point has a unique relevant deformation at zero chemical potential
\begin{equation}\label{eq:DeltaL-relevant}
\Delta\mathcal{L}_\text{rel} = \frac{4\pi}{a} \cO_2^\dagger \cO_2,
\end{equation}
that triggers (for $a<0$) a renormalization group (RG) flow toward the free-fermion fixed point.
Along this RG flow, final-state observables are no longer independent of the total energy, but now depend, in a universal manner, on the dimensionless parameter $\omega a^2$. The velocity distribution of final particles  should approach the unitarity fermion shape at $\omega a^2\gg 1$ and the free fermion shape in the opposite limit $\omega a^2\ll 1$.
In particular, the singularities of the distribution of final particles should be smoothed out by finite $1/a$.

A lesson may be learned from the calculation, 
using conformal perturbation theory, of the effects of the deformation on the total rate of three-neutron production \cite{Chowdhury:2023oas, Beane:2024kld, Beane:2025tum}. 
Although formally the $1/a$ correction is expected to be small when the total energy is larger than $\hbar^2/(m_n a^2)\sim 0.1~ \text{MeV}$, it has been found to be numerically significant for the total rate of point production of neutrons \cite{Chowdhury:2023oas,Beane:2025tum,Higgins:2025dnz}.  Thus, for a quantitative comparison with experiment, it may be important to be able to resum all powers of $1/a$, or, in other words, to do the calculation along the whole RG flow.  

The interpolation between the unitary-fermion and free-fermion behaviors was studied in refs.~\cite{Backert:2026yjx,Higgins:2025dnz}. In this direction, it would be interesting to directly compute detector correlators in the pionless EFT \cite{Kaplan:1998tg,Kaplan:1998we}, which should enable their description along the flow from fermions at unitarity to free neutrons. See, e.g., refs.\ \cite{Hildenbrand:2019sgp, Dietz:2021haj} and \cite{Backert:2026yjx} for recent calculations of three-particle production. One can also try to approach the RG flow from the IR side, i.e., as an expansion over the scattering length $a$ around the free fixed point.  Finite scattering length corrections have been incorporated in the evaluation of the $K\to 3 \pi$ decay amplitudes in ref.\ \cite{Bissegger:2007yq}.

A particularly interesting aspect of detector correlators, which has been exploited in QCD \cite{Electron-PositronAlliance:2025fhk}, is that they map out the RG flow of the theory as a function of angle (and in this case velocity). The flow from fermions at unitarity provides an example of a tractable flow, where it may be possible to study this, as well as related questions of detector matching \cite{Chang:2025zib} in a solvable setting. 

The leading irrelevant deformation of the unitarity fermion fixed point is proportional to the $s$-wave effective range
\begin{equation}\label{eq:DeltaL}
\Delta\mathcal{L}_\text{irrel} = -\pi r_\text{eff} \cO_2^\dagger \biggl(i\overset\leftrightarrow{\partial_t} +\frac{\overleftarrow{\nabla}^2+\overrightarrow{\nabla}^2}{4} \biggr)\cO_2\,.
\end{equation}
Numerically $a\approx-19~\text{fm}$ and $r_\text{eff}\approx2.75~\text{fm}$.  Unlike the relevant deformation, one cannot hope to resum all $r_\text{eff}$ corrections, but only to compute them perturbatively \cite{Chowdhury:2023oas, Beane:2024kld, Beane:2025tum}. 
\smallskip

\textbf{Nonrelativistic complex CFTs:}  Another way to destroy conformality is through the mechanism of ref.\ \cite{Kaplan:2009kr}: two fixed points merge, and move to the complex plane.  These ``complex CFTs'' \cite{Gorbenko:2018ncu} have a nonrelativistic analog in the Efimov effect, realized in the most important example of unitarity bosons \cite{Braaten:2004rn}.  The physics of the latter is not scale invariant, but depends on the energy scale in a log-periodic manner (the ``limit cycle''), in which the phase of the log-periodic function is the only parameter needed to make predictions.  It would be interesting to extend our calculation to this case.

Experimentally, unitarity bosons are realized approximately by helium atoms (more precisely, the atoms of the $^4\text{He}$ isotope), which have an anomalously large scattering length of about $100~\mathring{\text{A}}$.  The $^4\text{He}$ atoms form bound clusters of arbitrary size \cite{Barranco2006}, as well as ion complexes such as $\text{He}_n\text{H}^+$ \cite{Grandinetti2004}. One can also consider related processes such as knockout on helium trimers \cite{Kunitski:2015Efimov}\footnote{We thank Hans-Werner Hammer for suggesting this possibility.}, which is analogous to performing a deep-inelastic scattering experiment on a helium trimer.  One could also, in principle, envision processes of dissociation of ionic helium complexes where the interesting kinematic regime is that of small relative energy of the final-state helium atoms.
\smallskip

\textbf{Nonrelativistic Anyons:} A particularly interesting example of an NRCFT is nonrelativistic anyons in 2+1d \cite{Hagen:1984mj,Jackiw:1990mb,Leblanc:1992wu,Jackiw:1992fg,Bergman:1993kq}. These were studied from the NRCFT perspective in refs.\ \cite{Nishida:2007de,Nishida:2007pj}. It will be particularly interesting to compute detector correlators in these systems. Few-anyon states have been extensively studied \cite{Wu:1984py,Mashkevich:1994me,Mashkevich:1995ep,Khare:1991mw,Sen:1992vz,Sporre:1991pm,Murthy:1992zm,Chou:1991rg}, see also  refs.\ \cite{Doroud:2016mfv,Doroud:2015fsz} for recent work.  In the fractional quantum Hall effect, as charged particles in magnetic field, anyons cannot have a dispersion relation, but with the recent discovery of the fractional quantum anomalous Hall effect \cite{Cai2023,Zeng2023,Park2023,Xu2023}, there is now a possibility of dispersive anyons, with dispersion relation quadratic near the bottom of an anyonic band \cite{Shi:2024cgb}.

It would be interesting to explore the extent to which such measurements could be performed experimentally.  Previously, ref.\ \cite{Morampudi:2016ler} proposed the threshold production of two and three-anyon states as a signature of their statistics. Analogously, one would expect that the detector OPE would depend on the statistics. (One complication that may invalidate the assumptions of NRCFT is the Coulomb interaction between the anyons.  This may need to be screened out by a gate to enable the approximate scale and Schr\"odinger invariance.)
\smallskip

\textbf{Point production of four and more fermions near unitarity:} The problem of $N=4$ and more unitarity fermions is not exactly solvable, unlike the case of $N=3$.  On the other hand, many current and future experiments will explore production of four \cite{Kisamori:2016jie,Duer:2022ehf} (for recent reviews, see refs.\ \cite{Nakamura2024,Faestermann:2025our}), or six neutrons \cite{Nakamura2024,Nasr:2025lmi}.   In order to interpret the results of these experiments, it will be necessary to develop approximate or numerical methods to deal with more than three neutrons.  It is possible also that the large-charge expansion is already useful at $N=6$ or even $N=4$.
\smallskip

\textbf{Extended initial state:} We have, as of now, assumed that the particles are produced locally at a single point by the action of a local operator on the vacuum.  In real nuclear processes, neutrons are emitted from an extended source. In fact, 
the correlation between the neutrons in the initial state has been argued to play an important role \cite{Lazauskas:2022mvq} in the $4n$ spectrum in the core knockout experiment with $^8\text{He}$ beam~\cite{Duer:2022ehf}. It seems that finite size effects of the source should be treatable using an OPE, and expanding over a complete set of primary operators of a given mass, and it would be interesting to explore this.
\smallskip

{\bf{Higher Point Functions:}} In this paper we have computed the simplest one-point detector correlator, $\langle \mathcal{E}_v (\hat n) \rangle$. As compared to the relativistic case where the one-point function is fixed by symmetries, the one-point function in an NRCFT has a non-trivial structure due to its velocity dependence. It will be extremely interesting to compute higher point functions, in particular $\langle \mathcal{E}_{v_1} (\hat n_1) \mathcal{E}_{v_2} (\hat n_2) \rangle$. For charge-$3$ states in fermions at unitarity, it should be a straightforward exercise to compute them using the celestial wave-function approach we describe in section \ref{sec:wave-function_interacting}. Such correlators should have non-trivial angular dependence already in the three-body states considered in this paper. In the relativistic case, these two-point functions have been extremely important experimentally, and we believe are also the natural observable in the nonrelativistic case.
\smallskip

{\bf{Detectors for Other Schr\"odinger Charges:}}
In this paper we focused on the particular case of energy flux. It would also be interesting to perform explicit calculations for the detectors of other charges of the Schr\"odinger group, which we constructed in section \ref{sec:general}. A particularly interesting candidate would be angular momentum\footnote{We thank Julio Parra-Martinez for this suggestion.}. Integrated angular momentum flux is of great relevance in the study of collisions of binary black holes, and has thus been studied extensively in classical gravity \cite{Jakobsen:2021smu,Mougiakakos:2021ckm,Damour:2020tta,Bini:2021gat,Gralla:2021qaf,Manohar:2022dea}. Detector operators for measuring the differential angular momentum flux have also been proposed in gravity \cite{Gonzo:2020xza}. In gravity and gauge theories, these detectors are sensitive to the infrared structure. In NRCFTs, such as fermions at unitarity, they should be well behaved, providing an interesting opportunity to study such observables in a non-perturbative setting. It would also be interesting to understand if they could be studied experimentally.
\smallskip

{\bf{``Higher Twist" Detector Operators:}} In this paper we have studied the simplest possible detector operator, which ``detects" a single particle state. There is a much broader class of detectors, which in the relativistic case are referred to as ``higher twist" detectors, and have the interpretation of detecting multi-particle states. In an NRCFT, it is natural for a true detector to have particle number zero. There is a wide class of such operators, which take the form of normal ordered operators $: (\phi^\dagger)^n \phi^n:$, as well as analogs including derivatives. It would be interesting to explore detectors formed from these more general detectors. As compared to the detectors considered in this paper, which had canonical operator dimensions, and were not renormalized, the operator $\phi^n$ is renormalized for $n>1$. These operators could therefore provide an interesting example of detector renormalization in a simplified setup. Many of their scaling dimensions are known explicitly \cite{Nishida:2007pj}.
\smallskip

{\bf{Detector OPE:}} Our detectors are formed from charge-$0$ operators in NRCFTs. There has recently been progress understanding such operators \cite{Boisvert:2025hex}, in particular illustrating that there is a state-operator correspondence in an NRCFT to a thermofield double Hilbert space. Such a construction illustrates the existence of the OPE for charge-0 operators, and therefore also the detector operators studied in this paper.  It will be interesting to develop this detector OPE explicitly. As compared to the case of a relativistic theory, this OPE predicts scaling laws in \emph{both} angle, and velocity, which could be observed experimentally. 
\smallskip

{\bf{Finite Temperature and Transport:}} One of the original reasons for the interest in cold atoms from the high energy physics community was the observation that they exhibit a low $\eta/s$ \cite{Enss:2010qh} similar to the case of the quark gluon plasma. Indeed, they come close to the holographic value \cite{Kovtun:2004de}. For a review on the interplay of the fields, see ref.\ \cite{Adams:2012th}. Energy correlator observables have recently been measured in heavy ion collisions, and there is a renewed interest in understanding how to extract hydrodynamic/transport properties directly from detector measurements. NRCFTs may be a useful theoretical laboratory to study this problem.
\smallskip

{\bf{Conformal Collider Bounds on Trace Anomalies and Transport:}} In this paper, we proposed analogs of the Hofman-Maldacena conformal collider bounds. In the relativistic context, these bounds are particularly appealing, because by studying $\langle TTT \rangle$ or $\langle JTJ \rangle$ three-point functions, they are able to bound anomaly coefficients \cite{Hofman:2008ar}, or transport coefficients 
\cite{Cordova:2017zej}. To make our bounds useful, it will be useful to relate them to anomaly coefficients in NRCFTs, as well as to extend them to higher spin operators. Anomaly coefficients have been classified in NRCFTs \cite{Pal:2016rpz,Auzzi:2016lxb,Auzzi:2017jry,Auzzi:2016lrq,Auzzi:2015fgg}. To obtain bounds on these coefficients will require the decomposition of three-point functions of spinning operators into tensor structures. In the relativistic context, this problem has effectively been solved \cite{Costa:2011mg}. To our knowledge, the only calculations of three and four-point functions in NRCFTs are for scalar operators \cite{Volovich:2009yh,Fuertes:2009ex}. It would therefore be important to further develop the understanding of spinning three-point structures in NRCFTs, and apply them to develop collider bounds.  
\smallskip

{\bf{Holographic Descriptions of Detector Operators:}} In holographic relativistic CFTs, detector operators have a particularly simple description in the bulk \cite{Hofman:2008ar}, where they are dual to shock waves
\cite{tHooft:1987vrq}. In addition to providing a simple prescription for calculating detector observables, this provides a clean interpretation of the conformal collider bounds as bulk causality. Holography for NRCFTs has been explored in \cite{Son:2008ye,Adams:2008wt,Goldberger:2008vg,Herzog:2008wg,Maldacena:2008wh,Guica:2010sw}. It would be interesting to explore the interpretation of detector operators in these setups.
\smallskip

{\bf{Nonrelativistic Celestial Holography:}} Detector operators in NRCFTs have a particularly simple geometric interpretation in the harmonic-trap geometry. The boundary at future infinity ($\tau=\pi/2$) is a Euclidean space in which the state is imprinted in velocity space, and the detectors, $\mathcal{E}_v (\hat n)$ are local correlators, while $\mathcal{E} (\hat n)$ is a line defect. We have shown that the Schr\"odinger symmetries act on this space as dilations, rotations and translations, giving correlators on this space the symmetry structure of a Euclidean scale invariant theory. This structure is extremely similar to recent proposals for celestial holography (see, e.g., refs.\ \cite{Pasterski:2021raf,Strominger:2017zoo} for reviews). In the standard approach to celestial holography, amplitudes are Mellin transformed \cite{Pasterski:2016qvg}, giving rise to operators indexed by a Mellin variable $\omega$, conjugate to energy. This is much in analogy with our detector variable, $v$. However, in our case, due to the causal structure of the harmonic trap, this variable $v$ is also geometrized. In some sense this causal structure is more similar to cosmological correlators \cite{Arkani-Hamed:2015bza} and dS/CFT \cite{Strominger:2001pn}. It would be extremely interesting to explore this analogy in more detail, as this may provide a simple example of celestial holography. In particular, while we have explored the detectors, it will be important to further explore the structure of the states in the ``boundary theory.'' Another appealing feature of this NR-celestial holography is that it could be potentially directly realized in cold atom experiments.

It would also be interesting to investigate the relation between our detectors and the nonrelativistic BMS group \cite{Batlle:2017ghk}. In the relativistic case, detector operators are localized versions of Lorentz charges, and therefore have a close connection with the BMS algebra \cite{Cordova:2018ygx}. It would be interesting to explore if this persists in the nonrelativistic case.
\smallskip

{\bf{Null reduction of light-ray operators:}} As has also been emphasized in \cite{Boisvert:2025hex}, detector operators in NRCFTs exhibit many of the properties of those in CFTs, albeit in a simplified setting. It will be interesting to explore this explicit connection further. In particular, many examples of NRCFTs can be obtained from null-reduction of CFTs. It would be interesting to explore the null reduction of light-ray operators more explicitly.

\section*{Acknowledgments}
The authors thank Clay Córdova, Gabriel Cuomo, Shehab Hossam Fadda, Antoine Georges, Hans-Werner Hammer, Justin Kulp, Julio Parra-Martinez, Jonathan Sorce, Zhiquan Sun, Jesse Thaler and F\'elix Werner for discussions. The authors thank the organizers of ``Bootstrap 2025'' where this work was initiated. The work of D.T.S. is supported, in part, by the U.S.\ DOE
Grant No.\ DE-FG02-13ER41958.  D.T.S. thanks the Institut des Hautes \'Etudes Scientifiques, where part of this work was completed, for hospitality.
S.D.C. is supported by ``Exotic High Energy Phenomenology'' (X-HEP), a project funded by the European Union - Grant Agreement n.~101039756 (PI: J.~Elias~Mir\'o). Views and opinions expressed are however those of the author(s) only and do not necessarily reflect those of the European Union or the ERC Executive Agency (ERCEA). Neither the European Union nor the granting authority can be held responsible for them.  I.M. is supported by the DOE Early Career Award DE-SC0025581, the Sloan Foundation, and the Simons Collaboration on Confinement and QCD Strings. The authors have used Claude Code and ChatGPT to make some figures and to check the results of analytic calculations.

\appendix
\section{Ward Identities from Three-Point Functions}\label{app:ward_id_flat}

In this appendix, we verify directly the Ward identities eqs.\ \eqref{eq:Ward_E_generalk} and \eqref{eq:Ward_P_N_generalk} for the detector one-point function in a state created by a general scalar $\cO$ with arbitrary particle number. Our strategy is to start with the local three-point functions $\<\cO n \cO^\dagger\>$ and $\<\cO \vec j \cO^\dagger\>$, and take the detector limit eq.\ \eqref{En_def_ji_v} to get the detector one-point function. Then, we directly evaluate the integrals of the Ward identities to verify them.

By eq.\ \eqref{eq:3ptfunc_general}, the $\<\cO n \cO^\dagger\>$ three-point function is given by
\be\label{eq:OnO_general}
&\<0|\cO(t_1,\vec x_1) n(t_2,\vec x_2) \cO^\dagger(t_3,\vec x_3)|0\> =\frac{e^{\frac{i M_\cO}{2}\frac{\vec x_{13}^2}{t_{13}}}}{t_{13}^{\frac{2\De_\cO-\De_n}{2}}t_{12}^{\frac{\De_n}{2}}t_{23}^{\frac{\De_n}{2}}}F_n(v_{123}),
\ee
where $t_{ij}=t_i - t_j$. The $i\e$-prescription is $t_i \to t_i -i\e_i$, and $\e_1 > \e_2 > \e_3$. $M_\cO=N_{\cO^\dagger}$ is the particle number of $\cO^\dagger$. The cross ratio $v_{123}$ is given by
\be
v_{123}\equiv \frac{1}{2}\p{\frac{\vec x_{12}^2}{t_{12}}+\frac{\vec x_{23}^2}{t_{23}}-\frac{\vec x_{13}^2}{t_{13}}}.
\ee
On the other hand, since $\vec j$ is not a primary, $\<\cO \vec j \cO^\dagger\>$ does not agree with eq.\ \eqref{eq:3ptfunc_general}. By imposing invariance under all the nonrelativistic conformal generators, we find
\be\label{eq:OjO_general_expr}
&\<\cO(t_1,\vec x_1) \vec j(t_2,\vec x_2) \cO^\dagger(t_3,\vec x_3)\> \nn \\
&= \frac{e^{\frac{i M_\cO}{2}\frac{\vec x_{13}^2}{t_{13}}}}{t_{13}^{\frac{2\De_\cO-(\De_j-1)}{2}}t_{12}^{\frac{\De_j-1}{2}}t_{23}^{\frac{\De_j-1}{2}}}\p{\p{\frac{\vec x_{12}}{2t_{12}}+\frac{\vec x_{23}}{2t_{23}}}F_n(v_{123})+\p{\frac{\vec x_{12}}{t_{12}}-\frac{\vec x_{23}}{t_{23}}}F_{\perp}(v_{123})}.
\ee
Here, $F_n$ is the same function that appears in $\<\cO n \cO^\dagger\>$. The reason why $F_n$ appears in both three-point functions is because the action of $\vec K$ and $C$ on $\vec j$ can give $n$ (see eq.\ \eqref{K_C_comm_ji}).

Note that we can form conformal generators by integrating $n(\vec x)$ or $\vec j(\vec x)$ at $t=0$. In particular, we have
\be
N = \int\! d^d \vec x\, n(\vec x),\quad D=\int\! d^d \vec x\, \vec x\.\vec j(\vec x),\quad M_{ij} = \int\! d^d \vec x\, (x_i j_j(\vec x) - x_j j_i(\vec x)).
\ee
These relations imply that the above $F_n, F_\perp$ functions satisfy nontrivial Ward identities from integrating eqs.\ \eqref{eq:OnO_general} and \eqref{eq:OjO_general_expr}. Let us assume that the two-point function $\<\cO \cO^\dagger\>$ is normalized such that $C=1$ in eq.\ \eqref{eq:2pt_convention}. Then, for the $N$ generator, we have
\be
\int\! d^d \vec x_2 \, \frac{e^{\frac{i M_\cO}{2}\frac{\vec x_{13}^2}{t_{13}}}}{t_{13}^{\frac{2\De_\cO-d}{2}}t_{1}^{\frac{d}{2}}(-t_{3})^{\frac{d}{2}}}F_n(-\tfrac{t_{13}}{2t_1 t_3}(\vec x_2 -\tfrac{t_1 \vec x_3 - t_3 \vec x_1}{t_{13}})^2) = M_{\cO} \frac{e^{\frac{i M_\cO}{2}\frac{\vec x_{13}^2}{t_{13}}}}{t_{13}^{\De_\cO}},
\ee
which gives
\be\label{eq:Fn_Ward_1}
\int\! d^d \vec y\, F_n(\tfrac{\vec y^2}{2}) = M_{\cO}.
\ee
The rotation generator $M_{ij}$ also gives the same relation as above (at least for scalar $\cO$). From the $D$ generator, we find two additional identities,
\be\label{eq:Fn_Ward_2}
\int\! d^d \vec y\, \vec y^2 F_n(\tfrac{\vec y^2}{2}) = 2 i \De_{\cO},\quad \int\! d^d \vec y\, \vec y^2 F_{\perp}(\tfrac{\vec y^2}{2}) = 0.
\ee

Now, let us compute $\<\cO(\omega,\vec k) \cE_v(\hat n) \cO^\dagger(\omega,\vec k)\>$. Taking the detector limit of eq.\ \eqref{eq:OjO_general_expr}, we have
\be\label{eq:Ev_OnePt_general}
\<\cO(\omega,\vec k) \cE_v(\hat n) \cO^\dagger(\omega,\vec k)\> =\frac{i^d v^{d+1}}{2} \! \int\! dt_{13} \, d^d \vec x_{13}\, e^{i\omega t_{13}-i\vec k\.\vec x_{13}} \frac{e^{\frac{i M_\cO}{2}\frac{\vec x_{13}^2}{t_{13}}}}{t_{13}^{\frac{2\De_\cO-d}{2}}}F_n(-\tfrac{(\vec x_{13}-t_{13} v \hat n)^2}{2t_{13}}).
\ee
To verify the Ward identities for $\cE_v$, we additionally perform the integral $\int d\Omega_{\hat n} \int dv$. Note that this integral can be turned into an integral over a $d$-dimensional vector $\vec v = v\hat n$. After some change of variables, we obtain
\be
&\int \! d\O_{\hat n} \! \int\limits_0^{\oo} \! dv \, \<\cO(\omega,\vec k) \cE_v(\hat n) \cO^\dagger(\omega,\vec k)\> \nn \\
=& \frac{i^d}{2} \! \int \! dt_{13} \, d^d \vec y_{13} \! \int \! d^d \vec v\, \vec v^2 \, e^{i\omega t_{13}-i\vec k\.(\vec y_{13}+t_{13} \vec v)} \frac{e^{\frac{i M_\cO}{2}\frac{(\vec y_{13}+t_{13} \vec v)^2}{t_{13}}}}{t_{13}^{\frac{2\De_\cO-d}{2}}}F_n(-\tfrac{\vec y_{13}^2}{2t_{13}}).
\ee
Computing the Gaussian integral over $\vec v$, we get
\be
&\frac{i^d}{2}  \int\! dt_{13} \,   \frac{e^{i(\omega-\frac{\vec k^2}{2M_\cO}) t_{13}}}{t_{13}^{\De_\cO+2}}\! \int\! d^d \vec y_{13} \p{\tfrac{2\pi i}{M_\cO}}^{\frac{d}{2}}\Bigl(\tfrac{t_{13}^2}{M_\cO^2} \vec k^2+\tfrac{(i d-2\vec k\.\vec y_{13} ) t_{13}}{M_\cO}+\vec y_{13}^2 \Bigr)F_n(-\tfrac{\vec y_{13}^2}{2t_{13}}).
\ee

Next, the important observation is that we can evaluate the $\vec y_{13}$-integral using the two identities for $F_n$ given by eqs.\ \eqref{eq:Fn_Ward_1} and \eqref{eq:Fn_Ward_2}. We obtain\footnote{One has to be careful with contour deformation since we are integrating $F_n(-\tfrac{\vec y_{13}^2}{2t_{13}})$, while in \eqref{eq:Fn_Ward_1} and \eqref{eq:Fn_Ward_2} the argument of $F_n$ is positive. Note that the cross ratio can be written as $v_{123}=\frac{\vec y_0^2}{2}$, where $\vec y_0 = \sqrt{\frac{t_{12}t_{23}}{t_{13}}}\left(\frac{\vec x_{12}}{t_{12}}-\frac{\vec x_{23}}{t_{23}}\right)$. When taking the detector limit, we have to analytically continue from $t_1>t_2>t_3$ to $t_2>t_1>t_3$. This gives $\vec y_0 \to i\frac{\vec y_{13}}{\sqrt{t_{13}}}$. As a result, we should rotate the $\vec y_{13}$-contour by $\vec y_{13} \to e^{-i\frac{\pi}{2}} \vec y_{13}$, which gives a $(-i)^d$ phase factor.}
\be
&\int\! d\O_{\hat n}\!\int\limits_0^{\oo} dv\, \<\cO(\omega,\vec k) \cE_v(\hat n) \cO^\dagger(\omega,\vec k)\> \nn \\
=&\frac{1}{2}\p{\tfrac{2\pi i}{M_\cO}}^{\frac{d}{2}} \! \int \! dt_{13} \,   \frac{e^{i(\omega-\frac{\vec k^2}{2M_\cO}) t_{13}}}{t_{13}^{\De_\cO+1-\frac{d}{2}}}\bigl(\tfrac{t_{13}}{M_\cO} \vec k^2+i d -2i\De_\cO \bigr).
\ee
Finally, the $t_{13}$-integral can be evaluated by deforming the contour to the upper half plane (which introduces a step function $\th(\omega-\tfrac{\vec k^2}{2M_\cO})$). In the end, we find
\be
&\int\! d\O_{\hat n}\!\int\limits_0^{\oo}\! dv\, \<\cO(\omega,\vec k) \cE_v(\hat n) \cO^\dagger(\omega,\vec k)\>  = \omega \<\cO(\omega,\vec k)\cO^\dagger(\omega,\vec k)\>,
\ee
in agreement with the detector energy Ward identity eq.\ \eqref{eq:Ward_E_generalk}. Note that the two-point function is given by
\be
&\<\cO(\omega,\vec k)\cO^\dagger(\omega,\vec k)\> \nn \\
&= 2i^{\De_\cO}\p{\tfrac{2\pi}{M_\cO}}^{\frac{d}{2}}\sin(\pi(\De_\cO-\tfrac{d}{2}))\G(1+\tfrac{d}{2}-\De_\cO)\p{\omega-\tfrac{\vec k^2}{2M_\cO}}^{\De_\cO-\frac{d}{2}-1}\th(\omega-\tfrac{\vec k^2}{2M_\cO}).
\ee

The two additional Ward identities for particle number and momentum can be verified in a similar way. They correspond to choosing different kernels for the $\vec v$-integral. We find that the integrals are also completely fixed after imposing eqs.\ \eqref{eq:Fn_Ward_1} and \eqref{eq:Fn_Ward_2}, and we have
\be
&\int\! d\O_{\hat n} \! \int\limits_0^{\oo}\! dv\, \frac{2\hat n}{v}\<\cO(\omega,\vec k) \cE_v(\hat n) \cO^\dagger(\omega,\vec k)\>  = \vec k \<\cO(\omega,\vec k)\cO^\dagger(\omega,\vec k)\>, \nn \\
&\int\! d\O_{\hat n}\!\int\limits_0^{\oo}\! dv\, \frac{2}{v^2} \<\cO(\omega,\vec k) \cE_v(\hat n) \cO^\dagger(\omega,\vec k)\>  = M_{\cO} \<\cO(\omega,\vec k)\cO^\dagger(\omega,\vec k)\>.
\ee

\bibliographystyle{JHEP}
\bibliography{refs_longpaper}

\end{document}